\documentclass{aa}
\usepackage{psfig}

\begin{document}

\title{Three-dimensional simulations of non-stationary accretion by
remnant black holes of compact object mergers}
\titlerunning{Post-merger black hole accretion}

\author{
S.~Setiawan \inst{1}\thanks{e-mail: {\tt S.Setiawan@ed.ac.uk}}
\and M.~Ruffert \inst{1}\thanks{e-mail: {\tt M.Ruffert@ed.ac.uk}}
\and H.-Th.~Janka \inst{2}\thanks{e-mail: {\tt thj@mpa-garching.mpg.de}}
}
\authorrunning{S.~Setiawan et.~al}
\institute{
School of Mathematics, University of Edinburgh, Edinburgh, EH9 3JZ,
Scotland, U.K.
\and
Max-Planck-Institut f\"ur Astrophysik, Postfach 1317, 85741
Garching, Germany
}

\offprints{M.~Ruffert}

\date{\today}

\abstract{
By means of three-dimensional hydrodynamic simulations with an
Eulerian PPM code we investigate the time-dependent evolution and
properties of accretion tori around nonrotating and rotating 
stellar-mass black holes, using a pseudo-Newtonian
(Paczy\'nski \& Wiita or Artemova-Bj\"ornsson-Novikov) potential
to approximate the effects of general relativity. The simulations
are performed with three nested Cartesian grids to ensure
sufficient resolution near the central black hole on the one
hand and a large computational volume on
the other. The black hole and torus are considered as the remnant
of a binary neutron star or neutron-star black-hole merger.
Referring to results from previous hydrodynamical simulations of
such events, we assume the initial configurations to consist of 
a black hole with a mass of about 4$\,M_\odot$ girded by a toroidal
accretion disk with a mass in the range from about 0.01~$M_\odot$ to
0.2~$M_\odot$. We simulate the 
torus evolution without and with physical shear viscosity, 
employing a simple $\alpha$-model for the gas viscosity.
As in our previous work on merging neutron star binaries and
neutron star/black hole binaries, we use the equation of state
of Lattimer and Swesty. The energy loss and lepton number change 
due to neutrino emission from the hot torus are treated by a 
neutrino-trapping scheme. The energy
deposition by neutrino-antineutrino annihilation around the disk
is evaluated in a post-processing step. The time-dependent 
efficiency of converting gravitational energy to neutrinos,
expressed by the ratio of neutrino luminosity to accretion rate of
rest-mass energy, can reach maximum values of up to about 10\%.
The corresponding efficiency of converting neutrino energy into
a pair-photon fireball by neutrino annihilation peaks at values
of several percent. Interestingly, 
we find that the rate of neutrino-antineutrino annihilation decays
with time much less steeply than the total neutrino luminosity does 
with the decreasing gas mass of the torus, because the ongoing 
protonization of the initially neutron-rich disk matter leads to
a rather stable product of neutrino and antineutrino luminosities.
The neutrino luminosity and total energy release
of the torus increase steeply with higher viscosity, larger torus
mass, and larger black hole spin in corotation with the disk,
in particular when the spin parameter is $a \ga 0.8$.
The latter dependence is moderated in case of a high disk viscosity.
For rotation rates as expected for post-merger black holes
($a\ga 0.5$) and reasonable values of the alpha viscosity of the torus
($\alpha \sim 0.1$), torus masses in the investigated range can
release sufficient energy in neutrinos to account for the 
energetics of the well-localized short gamma-ray bursts
recently detected by {\em Hete} and {\em Swift}, if collimation 
of the ultrarelativistic outflows into about 1\% of the sky 
is invoked, as predicted by recent hydrodynamic jet simulations.

}

\maketitle

\section{Introduction
\label{sec:introduction}} 

Merging double neutron stars or neutron-star black-hole
binaries have long been discussed as possible sources of 
cosmic gamma-ray bursts (GRBs) (e.g., Blinnikov et al.~\cite{bli84};
Paczy\'nski~\cite{pac86}; Goodman~\cite{goo86};
Goodman et al.~\cite{goo87}; Eichler et al.~\cite{eic89};
Paczy\'nski~\cite{pac91}; Narayan et al.~\cite{nar92};
M\'esz\'aros \& Rees~\cite{mes93};
Woosley~\cite{woo93a}; Jaroszy\'nski~\cite{jar93,jar96};
Mochkovitch et al.~\cite{moc93,moc95};
Thompson~\cite{tho94}; Witt et al.~\cite{wit94};
Janka \& Ruffert~\cite{jan96}; M\'esz\'aros \& Rees~\cite{mes97};
Popham et al.~\cite{pop99}; M\'esz\'aros et al.~\cite{mes99}).
After the discovery that long GRBs are linked to massive
star explosions, which confirmed predictions by Woosley~(\cite{woo93a})
and MacFadyen \& Woosley~(\cite{mac99}), it is in particular
the class of short, hard GRBs with durations of less than about 
2$\,$seconds, which is thought to originate from colliding 
compact objects in binary systems.
Recent observations of well-localized short GRBs (Gehrels et
al.~\cite{geh05}; Hjorth et al.~\cite{hjo05}; Berger et
al.~\cite{ber05}) seem to   
support this hypothesis, because in agreement with expectations
short GRBs are found in galaxies without strong star-formation
activity.

In the most popular scenario, GRBs have in common that they
signal the birth of stellar-mass black holes and the 
associated huge energy release when the growing black hole
accretes matter at 
enormous rates between fractions of a solar mass per second
up to many solar masses per second (Woosley~\cite{woo93a};
Popham et al.~\cite{pop99}; MacFadyen \& Woosley~\cite{mac99}).
The huge amount of gravitational binding energy released during
the accretion process of up to several solar masses of gas
into the black hole could explain the
energetics of even the most distant cosmological gamma-ray bursts
(e.g., GRB981214, see Kulkarni et al.~\cite{kul98}).
Moreover, the compactness of the
stellar-mass black hole could naturally produce the rapid  
variability on time scales of milliseconds observed in many bursts.
For these reasons, massive accretion disks or thick accretion tori
around stellar-mass black holes are considered as favourable model for
the central engines of the cosmological GRBs.

In this scenario the energy of the ultra-relativistically 
expanding fireball or jet, can be provided by
the annihilation of neutrino-antineutrino pairs
(Paczy\'nski~\cite{pac91}; M\'esz\'aros \& Rees~\cite{mes97};
Woosley~\cite{woo93b}; Jaroszy\'nski~\cite{jar93,jar96}; 
Mochkovitch et al.~\cite{moc93,moc95}) or possibly by
magnetohydrodynamical 
processes (Blandford \& Znajek~\cite{bla77}; M\'esz\'aros and
Rees~\cite{mes92}; Popham et al.~\cite{pop99};
Li~\cite{li00}; Lee et al.~\cite{lee00}; see also
Livio et al.~\cite{livio99}; Rosswog \& Ramirez-Ruiz~\cite{ros02};
Rosswog et al.~\cite{ros03a}). In the former
case, the gravitational binding energy of accreted disk matter is
tapped, in the latter case (also) the rotational energy of the central
black hole (BH) could be converted into kinetic energy of the outflow.

In a series of previous papers (Ruffert et
al.~\cite{ruffert1}, \cite{ruffert2}; Ruffert \&
Janka~\cite{ruffert,ruffert3}) it has been shown that the neutrino
emission associated with the {\it dynamical} phase of the merging or
collision of two neutron stars is not sufficiently powerful and 
too short to provide the energy for gamma-ray bursts by
neutrino-antineutrino annihilation.
After a transient phase of gravitational stability, 
the compact massive remnant of the neutron star merger may 
collapse to a black hole and some matter may
remain in a hot toroidal accretion disk around the black hole,
which continues to radiate neutrinos with high luminosities
(for recent discussions and model calculations, see, e.g.,
Popham et al.~\cite{pop99}; Rosswog et al.~\cite{ros99,ros00};
Ruffert \& Janka~\cite{ruffert}; Oechslin et al.~\cite{oe04};
Morrison et al.~\cite{morr04}; Oechslin \& Janka~\cite{oe06}).

\begin{table}
\caption[]{ Parameters of the torus evolution models. The
3D simulations were performed with three levels of nested 
Cartesian grids with the number of grid cells per level
and spatial direction given in column ``zone''. 
$\Delta t_{\rm cal}$ is the time interval over which the model
was evolved. All models started with a black hole mass of
$M_{\rm BH}^{\rm i} = 4.017 M_\odot$. The initial torus masses
are given by $M_{\rm d}^{\rm i}$, $0 \le a \le 1$ is the 
initial value of the dimensionless BH spin parameter, ``pro''
means that torus and BH were chosen to rotate in the same 
direction, ``ret'' means opposite directions of rotation,
and $\alpha$ denotes the dimensionless disk viscosity parameter.}
\begin{center}
\tabcolsep=1.0mm
\begin{tabular}{lcrcccc}
\hline\\[-3mm]
Model & zone & $\Delta t_{\rm cal}$ &
   $M_{\rm d}^{\rm i}$ & BH spin & direction
 & visc\\
 & & ms &
   {\scriptsize $M_\odot$} & $a$
 & & $\alpha$
\\[0.2ex] \hline\\[-3mm]
r00-32  & 32 & 70.0 &  0.0478 & 0.0  &  -  & 0.0    \\  
al1-32  & 32 & 70.0 &  0.0478 & 0.0  &  -  & 0.001  \\  
al2-32  & 32 & 70.0 &  0.0478 & 0.0  &  -  & 0.004  \\  
al3-32  & 32 & 70.0 &  0.0478 & 0.0  &  -  & 0.01   \\  
al4-32  & 32 & 70.0 &  0.0478 & 0.0  &  -  & 0.1    \\  
ro1-32  & 32 & 70.0 &  0.0478 & 0.3  & pro & 0.0    \\  
ro2-32  & 32 & 70.0 &  0.0478 & 0.6  & pro & 0.0    \\  
ro3-32  & 32 & 70.0 &  0.0478 & 0.8  & pro & 0.0    \\  
ro4-32  & 32 & 70.0 &  0.0478 & 0.3  & ret & 0.0    \\  
ro5-32  & 32 & 70.0 &  0.0478 & 0.6  & ret & 0.0    \\  
ir1-32  & 32 & 70.0 &  0.0120 & 0.0  & -   & 0.0    \\  
ir2-32  & 32 & 70.0 &  0.0239 & 0.0  & -   & 0.0    \\  
ir3-32  & 32 & 70.0 &  0.0956 & 0.0  & -   & 0.0    \\  
ir4-32  & 32 & 70.0 &  0.1912 & 0.0  & -   & 0.0        
     \\[0.5ex]
r00-64  & 64 & 40.0 &  0.0478 & 0.0  & -   & 0.0    \\  
al3-64  & 64 & 40.0 &  0.0478 & 0.0  & -   & 0.01   \\  
al4-64  & 64 & 40.0 &  0.0478 & 0.0  & -   & 0.1    \\  
ri4-64  & 64 & 40.0 &  0.1912 & 0.6  & pro & 0.0    \\  
ro2-64  & 64 & 40.0 &  0.0478 & 0.6  & pro & 0.0    \\  
ro5-64  & 64 & 40.0 &  0.0478 & 0.6  & ret & 0.0    \\  
ir1-64  & 64 & 40.0 &  0.0120 & 0.0  & -   & 0.0    \\  
ir4-64  & 64 & 40.0 &  0.1912 & 0.0  & -   & 0.0    \\  
ir5-64  & 64 & 40.0 &  0.1912 & 0.0  & -   & 0.1    \\  
ar1-64  & 64 & 40.0 &  0.0478 & 0.6  & pro & 0.1    \\  
ar2-64  & 64 & 40.0 &  0.1912 & 0.6  & pro & 0.1       
     \\[0.2ex]
\hline
\end{tabular}
\end{center}
\label{tab:models3}
\end{table}

Newtonian hydrodynamic simulations 
including a physical equation of state
(Ruffert et al.~\cite{ruffert1};
Ruffert \& Janka~\cite{ruffert}) indicate that about
$0.1\,M_{\odot}$ of matter might obtain enough angular momentum
during the merging of the neutron stars to avoid
immediate accretion by the black hole. These numbers have recently
been confirmed by simulations including a treatment of
general relativity (Oechslin \& Janka~\cite{oe06}). 
The presence of large angular momentum allows the matter to form and
sustain toroidal disk around the black hole
for a period of time depending on the viscous transport
of angular momentum in the disk.
Therefore there could be enough time for this material to radiate away
in neutrinos a fair fraction of its gravitational binding energy.
A similar situation could result from the merging of a neutron star
with a black hole (Lee \& Klu\'zniak~\cite{lee95,lee99};
Klu\'zniak \& Lee~\cite{klu98}; Janka et al.~\cite{jan99};
Rosswog et al.~\cite{ros04}).

Many questions remain to be explored by quantitative calculations
of the formation, evolution, and energy release of accreting
stellar-mass black holes as possible central 
engines of GRBs. The results of the three-dimensional (3D) hydrodynamic 
simulations presented in this paper aim at studying the time-dependent
evolution of such black holes girded by accretion tori that were formed from
the decompressed matter of a neutron star, which was disrupted in a compact
binary merger. Such tori are generated by a violent dynamical process
in the immediate vicinity of the black hole. 
The assumption of stationary accretion is therefore valid only
approximately and only after a longer phase of adjustment and
relaxation of the matter girding the black hole. This relaxation can
proceed over a fair fraction of the whole period of massive accretion
and requires time-dependent modelling.
Therefore our simulations intend to contribute to answering the
following questions:
How much mass remains in the accretion torus for longer than dynamical
time scales? What is the mass accretion rate of the black hole as
a function of time? What are the properties
of the accretion torus, its density, temperature, neutron-to-proton
ratio, neutrino luminosity as functions of time?
How do they depend on torus mass, gas viscosity, and black hole 
rotation? How efficient is neutrino-antineutrino annihilation in 
depositing energy in the surrounding of the black-hole torus system
in dependence of these parameters?
What are the implications for powering ultrarelativistic outflow 
and gamma-ray bursts by neutrino-antineutrino annihilation?

Our 3D models of the time-dependent, non-stationary, 
evolution of the black-hole torus systems include a detailed
treatment of the equation of state and neutrino physics.
They thus replace approximations of previous work as made in 
the analytic studies of Popham et al.~(\cite{pop99}) and
Di Matteo et al.~(\cite{dim02}) and in the 
numerical 2D models of Lee \& Ramirez-Ruiz~(\cite{lee02}). 
These latter simulations with azimuthal symmetry employed a
simple ideal gas equation state and made the assumption that all the
dissipated energy is radiated away in neutrinos. 
Lee et al.~(\cite{lee04,lee05a,lee05b}), used a more realistic
equation of state and took into account neutrino energy loss and
opacity effects, but with a simplified model for changes lepton number
($Y_e$), considered only Newtonian gravity and nonrotating black
holes, and did not evaluate their 2D hydrodynamic models for the
energy deposition by neutrino-antineutrino annihilation.

The paper is organized as follows. In Sect.~\ref{sec:gabung} we
will introduce the numerical tools used for our simulations, outline
the initial conditions, and give an overview over the different
investigated cases. Sect.~\ref{sec:hasil} will present the results of
our simulations and discuss their implications. The summary
of our main findings will be given in Sect.~\ref{sec:summary}, and
concluding remarks and an outlook will follow in Sect.~\ref{sec:akhir}.

\section{Numerical Models
\label{sec:gabung}}

\subsection{Dynamics
\label{sec:dynamics}}

The three-dimensional computations of black-hole accretion-disk dynamics
in this paper were performed with a Newtonian
hydrodynamics code based on the Piecewise Parabolic Method (PPM) of 
Colella \& Woodward (\cite{col84}) with three levels of nested grids  
(Ruffert~\cite{ruf92}) to ensure both sufficient resolution near the central
black hole and a large computational volume.
A higher grid level has twice the zone size of the level below.
With an equatorial length and width of the total computational volume of
500$\,$km, the smallest zones have a side length of
1.95$\,$km for 64 zones per grid and dimension (``64-resolution'') and
3.90$\,$km for 32 zones on each grid level (``32-resolution'').

\begin{figure}
\psfig{file=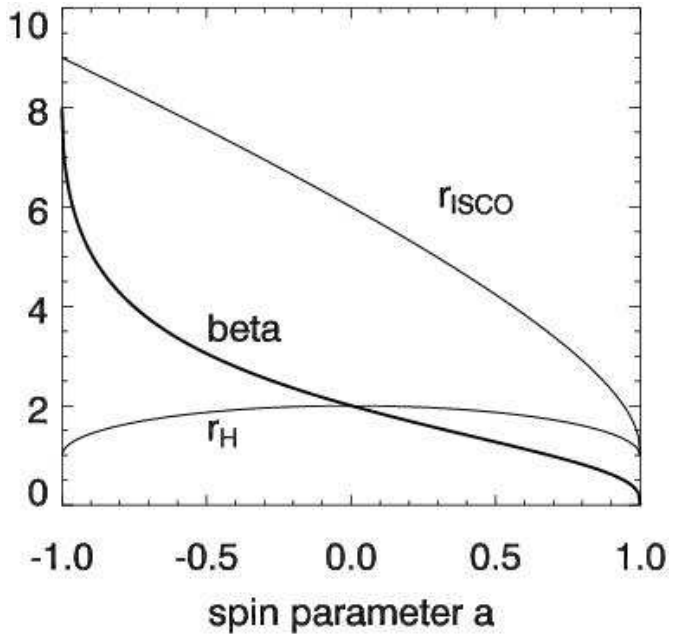,width=8.8cm}
\caption[]{
The black hole event horizon $r_{\rm H}$, the innermost stable
circular orbit $r_{\rm ISCO}$, and $\beta$ as defined in
Eq.~(\ref{eq:beta}), as functions of the black hole spin parameter $a$.
Both $r_{\rm H}$ and $r_{\rm ISCO}$ are given in units of 
$GM_{\rm BH}/c^2 = 0.5R_{\mathrm{s}}$.
\label{fig:Artebeta}
}
\end{figure}

The ``black hole'' with mass $M_{\mathrm{BH}}$ is treated as a 
gravitational centre surrounded by a vacuum sphere on the  
computational grid. Its gravitational potential $\Phi_{\mathrm{BH}}$
as a function of radius $r$ is calculated from 
\begin{equation}
\frac{\displaystyle {\mathrm{d}} \Phi_{\mathrm{BH}}}
{\displaystyle {\mathrm{d}}r} \, = \, - \,
\frac{\displaystyle GM_{\rm BH}}{\displaystyle r^{2-\beta}(r-r_{\rm
    H})^{\beta}} \ ,
\label{eq:phiA}
\end{equation}
where $r_{\rm H}$ is the black hole event horizon, and $\beta$ is a
constant for a given value of the black hole spin parameter $a$.
It is defined by
\begin{equation}
\beta = \frac{\displaystyle r_{\rm ISCO}}{\displaystyle r_{\rm H}} - 1 \ .
\label{eq:beta}
\end{equation}
Here the radius of the innermost stable circular orbit (ISCO), 
$r_{\rm ISCO}$, is the same
as $r_{\rm in}$ in Artemova et al.~(\cite{art96}). Both $r_{\rm H}$
and $r_{\rm ISCO}$ depend on $a$ as shown in Fig.~\ref{fig:Artebeta}.

This potential
reduces to the usual Paczy\'nski-Wiita potential (Paczy\'nski \&
Wiita~\cite{pac80}) when the black hole spin parameter $a$ is zero.
It allows us to mimic important general relativistic effects, in
particular to reproduce the existence of a last stable circular
orbit.

\begin{table*}
\caption[]{ Some results for the torus evolution models. The
dynamical simulations were performed over a time interval of $\Delta
t_{\rm cal}$. All quantities refer to the conditions found at the
end of the simulations. $M_{\rm d}$ is the mass of the gas of the
disk, $T_{\rm max}$ is the maximum gas temperature in energy units,
$L_{\nu_e}$ denotes the electron neutrino luminosity near the end of
the simulation, $L_{\bar{\nu}_e}$ the corresponding electron
antineutrino luminosity, and $L_{\nu_x}$ the luminosity of
heavy-lepton neutrinos (summed for $\nu_{\mu}$,
$\bar\nu_{\mu}$, $\nu_{\tau}$ or $\bar\nu_{\tau}$, which are all
treated equally). The sum of all neutrino luminosities is given by
$L_\nu$, and the mean energies of the different neutrino types by
$\langle\epsilon_{\nu_e}\rangle$,
$\langle\epsilon_{\bar{\nu}_e}\rangle$ and
$\langle\epsilon_{\nu_x}\rangle$. }
\begin{center}
\tabcolsep=1.5mm
\begin{tabular}{llcccccccrr}
\hline\\[-3mm]
Model & $\Delta t_{\rm cal}$ &
   $M_{\rm d}$ & $T_{\rm max}$ &
   $L_{\nu_e}$ & $L_{\bar{\nu}_e}$ & $L_{\nu_x}$ & $L_\nu$ &
   $\langle\epsilon_{\nu_e}\rangle$ &
   $\langle\epsilon_{\bar{\nu}_e}\rangle$ &
   $\langle\epsilon_{\nu_x}\rangle$ \\
 & ms &
   {\scriptsize$10^{-2}M_\odot$} & {\scriptsize MeV} &
   {\scriptsize$10^{50}\frac{\rm erg}{\rm s}$} &
   {\scriptsize$10^{50}\frac{\rm erg}{\rm s}$} &
   {\scriptsize$10^{50}\frac{\rm erg}{\rm s}$} &
   {\scriptsize$10^{50}\frac{\rm erg}{\rm s}$} &
   {\scriptsize MeV} & {\scriptsize MeV} & {\scriptsize MeV}
\\[0.3ex] \hline\\[-3mm]
r00-32  & 70.0 & 1.06 & 2.6
    &  0.08 & 0.22  & 0.002 & 0.3 & 9.7 & 9.3 & 7.8     \\   
al1-32  & 70.0 & 1.14 & 3.3
    &  0.2  & 0.6   & 0.01  & 0.8 &12.5 &12.0 &10.0     \\   
al2-32  & 70.0 & 1.25 & 4.3
    &  1.0  & 2.9   & 0.1   & 4.0 &15.4 &15.2 &12.8     \\   
al3-32  & 70.0 & 1.43 & 5.0
    &  1.4  & 3.9   & 0.2   & 5.5 &17.5 &17.2 &14.7     \\   
al4-32  & 70.0 & 0.84 & 5.9
    &  4.0  & 9.0   & 2.0   &15.0 &17.5 &19.2 &16.9     \\   
ro1-32  & 70.0 & 1.54 & 3.8
    &  0.3  & 0.8  & 0.008 & 1.1 & 11.3 & 11.0 & 9.4    \\   
ro2-32  & 70.0 & 2.10 & 4.5
    &  0.7  & 1.8  & 0.03 & 2.5 & 11.4 & 11.2 & 9.7    \\   
ro3-32  & 70.0 & 2.02 & 6.4
    &  9.0  & 14.0  & 0.4 &23.0 & 18.2 & 19.1 & 18.3  \\   
ro4-32  & 70.0 & 0.67 & 2.1
    & 0.009 & 0.02 & 0.0006 & 0.03 & 8.5 & 8.3 & 6.9  \\   
ro5-32  & 70.0 & 0.56 & 1.8
    & $<$0.005 & $<$0.01 & $<$0.0002 & $<$0.015 & 5.8 & 6.1 & 5.4      \\   
ir1-32  & 70.0 & 0.46 & 1.4
    & $<$0.0005  & $<$0.0005  & $<$0.0001 & $<$0.001 & 5.1 & 5.6 & 4.7  \\   
ir2-32  & 70.0 & 0.57 & 1.9
    & $<$0.0025  & $<$0.0025  & $<$0.0005 & $<$0.005 & 6.8 & 6.7 & 5.7  \\   
ir3-32  & 70.0 & 2.35 & 3.1
    &  1.7 & 2.4  &0.035 &  4.1 & 11.9 & 11.4 & 9.6                     \\   
ir4-32  & 70.0 & 4.31 & 7.1
    &  34.0 & 50.0 &2.0 & 86.0 & 17.3 & 17.3 & 17.0                        
     \\[0.7ex]
r00-64 & 40.0 & 2.22 & 4.2
    &  0.6 &  1.2  & 0.01& 1.8 & 12.0 & 11.6 & 10.2 \\  
al3-64 & 40.0 & 2.44 & 5.6
    &  3.4 &  7.7 & 0.3   &11.4 & 16.4 & 17.8 & 15.4 \\  
al4-64 & 40.0 & 1.76 & 8.7
    & 18.0 & 55.0 & 7.0   &80.0 & 21.3 & 22.8 & 20.1 \\  
ri4-64 & 40.0 &12.22 & 7.0
    & 26.0 & 43.0 & 1.0 &70.0 & 14.5 & 16.5 & 17.9 \\  
ro2-64 & 40.0 & 3.10 & 4.6
    &  1.6 & 1.8  & 0.02 & 3.4 & 12.9 & 11.7 & 11.7  \\  
ro5-64 & 40.0 & 1.13 & 3.9
    &  0.3 & 0.7  & 0.01 & 1.0 & 12.2 & 11.8 &  9.8  \\  
ir1-64 & 40.0 & 0.71 & 2.4
    & 0.07 & 0.03 & 0.001 & 0.1 &  9.4 &  8.4 & 7.0  \\  
ir4-64 & 40.0 & 7.87 & 6.2
    &  7.2 & 15.0 & 0.2 &22.4 & 13.8 & 12.8 & 14.9 \\  
ir5-64 & 40.0 & 8.77 & 10.4
    &150.0 &350.0 &110.0 &610.0 & 20.7 & 20.9 & 26.7 \\  
ar1-64 & 40.0 & 2.34 & 6.6
    & 25.0 &70.0  &15.0  &110.0 & 21.5 & 22.7 & 20.3 \\  
ar2-64 & 40.0 &10.44 & 9.5
    &120.0 &240.0 & 80.0 &440.0 & 19.4 & 19.7 & 25.2     
     \\[0.7ex]
\hline
\end{tabular}
\end{center}
\label{tab:models}
\end{table*}

The local Keplerian velocity for the Artemova-Bj\"ornsson-Novikov
potential is
\begin{equation}
v_{\rm Kepler}(r) = \sqrt{\frac{GM(r)r^{\beta-1}}{(r-r_{\rm H})^{\beta}}}\ ,
\label{eq:vkepA}
\end{equation}
with $M(r)$ being
the mass enclosed by the sphere of radius $r$ and $\beta$ being given by
Eq.~(\ref{eq:beta}).
This becomes the Keplerian velocity of the Paczy\'nski-Wiita potential
when the black hole spin parameter $a$ is zero and thus $\beta = 2$.

The radius of the vacuum sphere mimicking the black hole is chosen to
be the arithmetic mean of event horizon and innermost stable
circular orbit. The outflow boundary condition allows mass and angular
momentum (and energy) to leave the grid. The mass and the spin of
the black hole are accordingly updated during the simulations.
The corresponding changes of the black hole spin are, however,
minor (cf.\ Setiawan et al.~\cite{setia04}) because the accreted mass
is small compared to the initial mass of the black hole in all
cases.

So the only GR effect being modeled is the reduction in radius of the
ISCO. Explicit frame dragging is omitted, as well as the gravitational
lensing effect which might capture some of the neutrinos emitted in
this region. A detailed discussion of the lensing effect can be found
in Sect.~7.2 of Ruffert \& Janka~\cite{ruffert}), where we find a
moderate decrease of the annihilation energy by 10\%-30\%.

\subsection{Viscosity
\label{sec:simulasi}}

In our simulations we compare the evolution of the disk with
and without shear viscosity.
The viscosity source terms for energy, momentum components, and
entropy equations can be expressed, respectively, as follows
\begin{equation}
V_{\mathrm {E}}\,=\, {\eta_p\over 2}\,
\left(\frac{\partial v^i}{ \partial x^j} + \frac{\partial
  v^j}{\partial x^i} - {2\over 3} \delta_{ij} \frac{\partial
  v^k}{\partial x^k} \right)^{\! 2} \, \, ,
\label{eq:1}
\end{equation}
\begin{equation}
V_{\mathrm {mom}}^i\,=\, \frac{\displaystyle \partial}{\displaystyle
  \partial x^j} \Biggl[ {\eta_p}\, \left(\frac{\partial v^i}{ \partial
    x^j} + \frac{\partial v^j}{\partial x^i} - {2\over 3} \delta_{ij}
  \frac{\partial v^k}{\partial x^k} \right) \Biggr]  \, \, ,
\label{eq:2}
\end{equation}
\begin{equation}
V_{\mathrm {ent}}\,=\, {1\over k_{\mathrm B}T}\,\Biggl[ {\eta_p\over 2}\,
\left(\frac{\partial v^i}{ \partial x^j} + \frac{\partial
  v^j}{\partial x^i} - {2\over 3} \delta_{ij} \frac{\partial
  v^k}{\partial x^k} \right)^{\! 2}
\Biggr] \, \, ,
\label{eq:3}
\end{equation}
where $\eta_p$ is the dynamical shear viscosity coefficient
parameterized by the simple expression:

%
%

%
\begin{equation}
\eta_p = \alpha \rho c_s^2 / \Omega_K \, .
\end{equation}
Here $\alpha$ is the standard dimensionless disk-viscosity parameter 
introduced by Shakura and Sunyaev~(\cite{sha73}), $\Omega_K$ is the 
Keplerian angular velocity,
and $c_s = (\Gamma P/\rho)^{1/2}$ is the adiabatic
sound speed with $\Gamma$ being the adiabatic index.
The expressions of Eqs.~\ref{eq:1} and~\ref{eq:2} were added as source
terms to the energy and momentum evolution equations of our PPM scheme.

\begin{figure*}[htp!]
\begin{tabular}{cc}
  \psfig{file=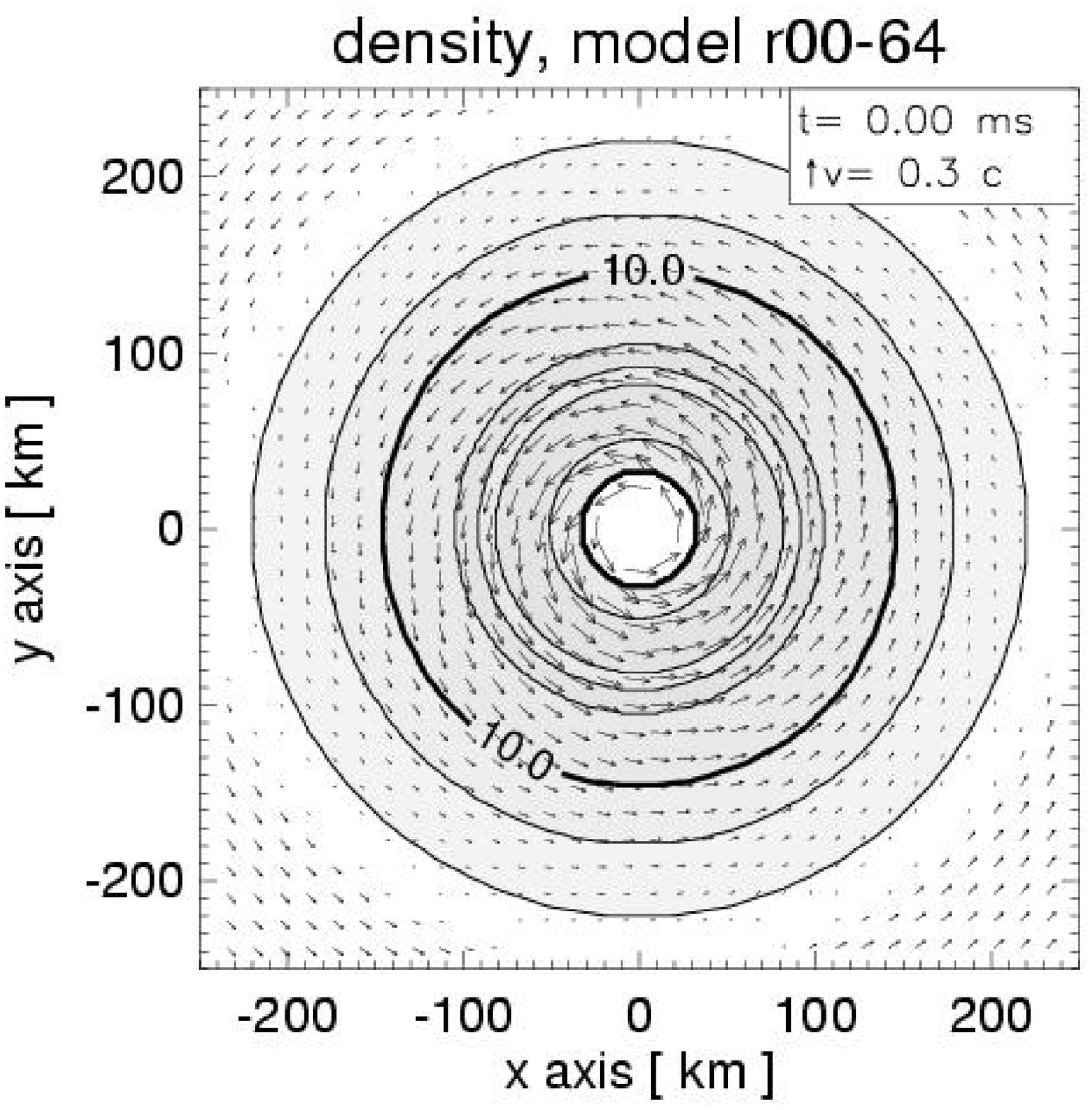,width=8.2cm} &
  \psfig{file=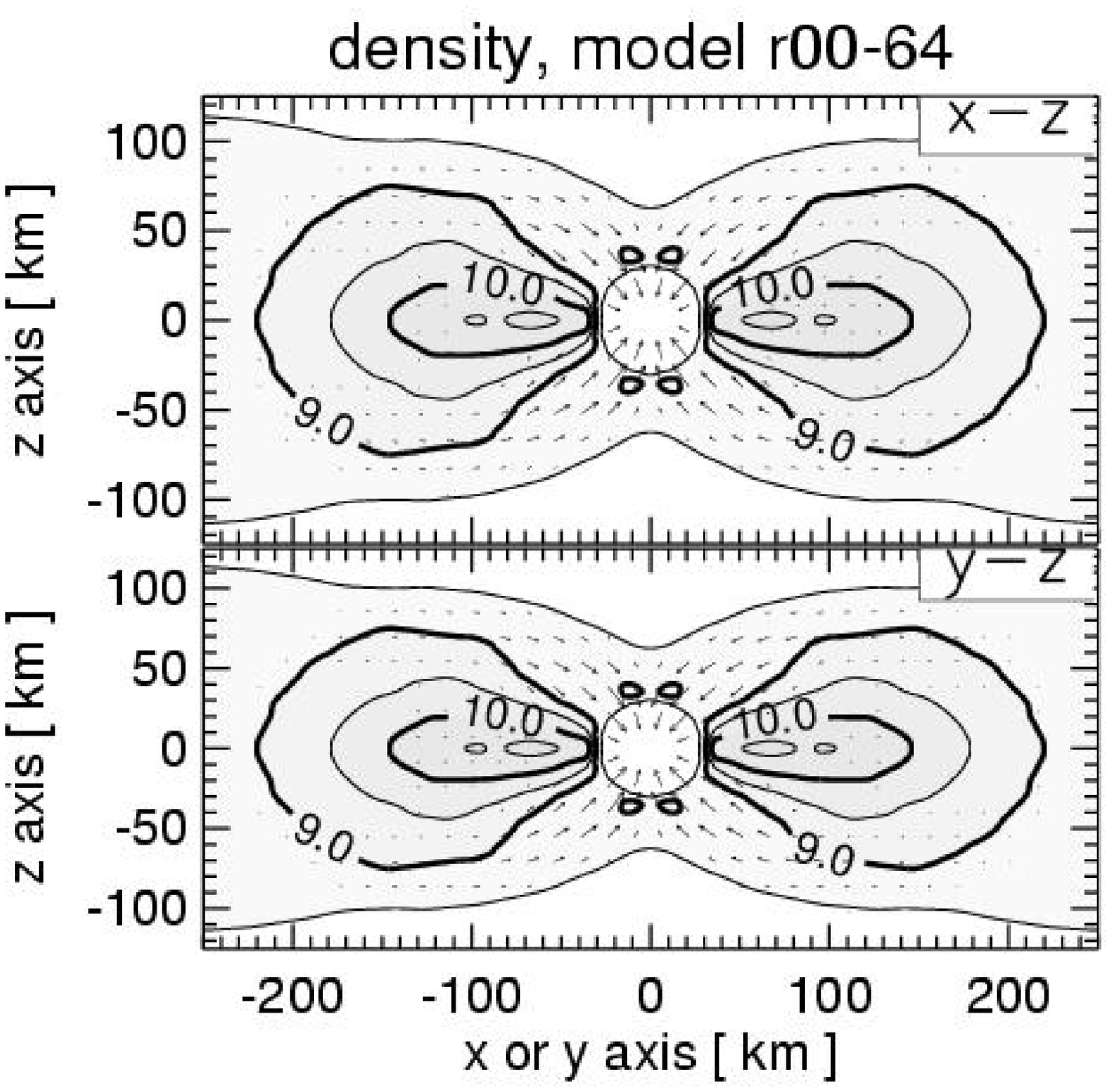,width=8.2cm} \\[-2ex]
  \psfig{file=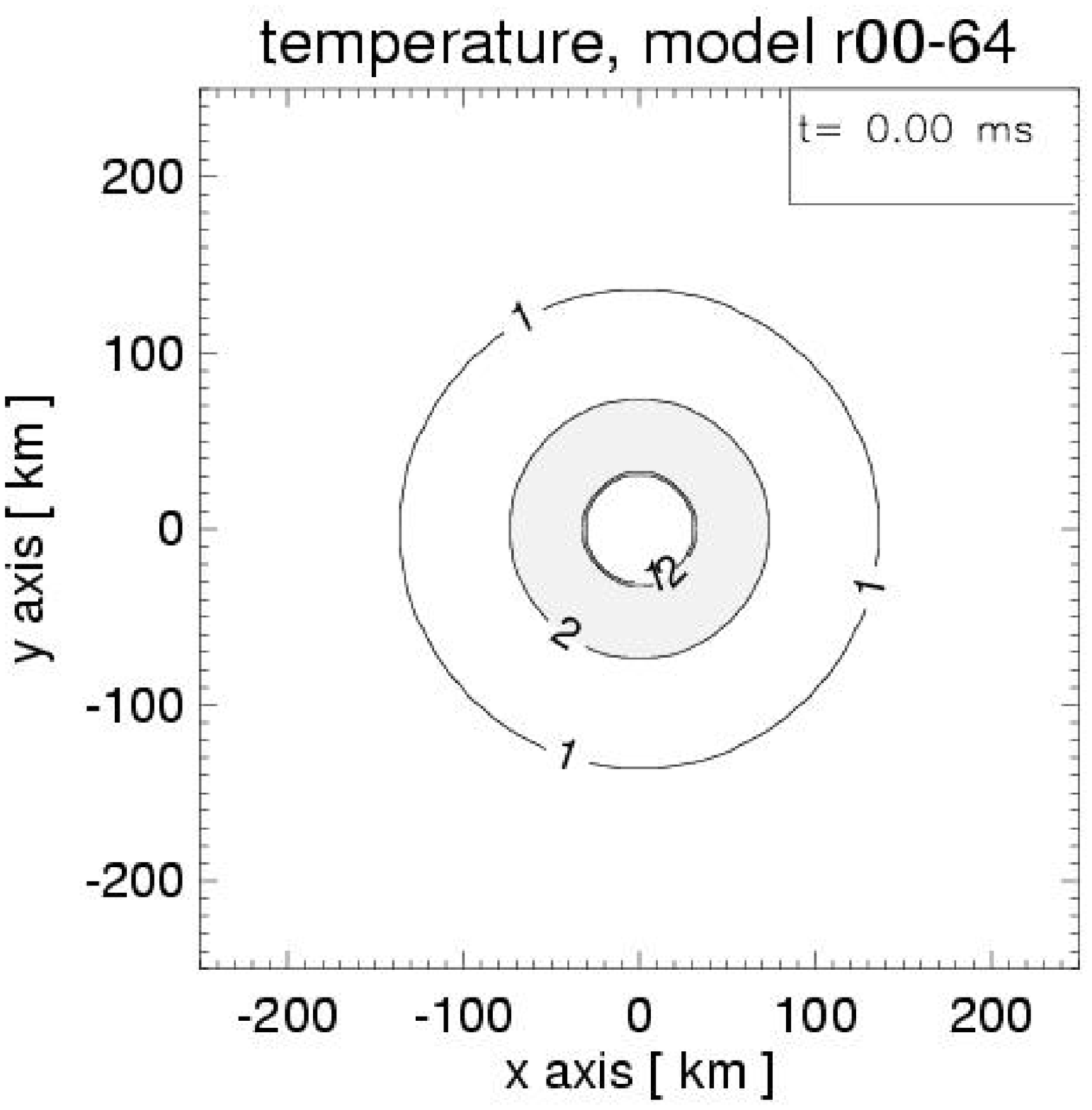,width=8.2cm} &
  \psfig{file=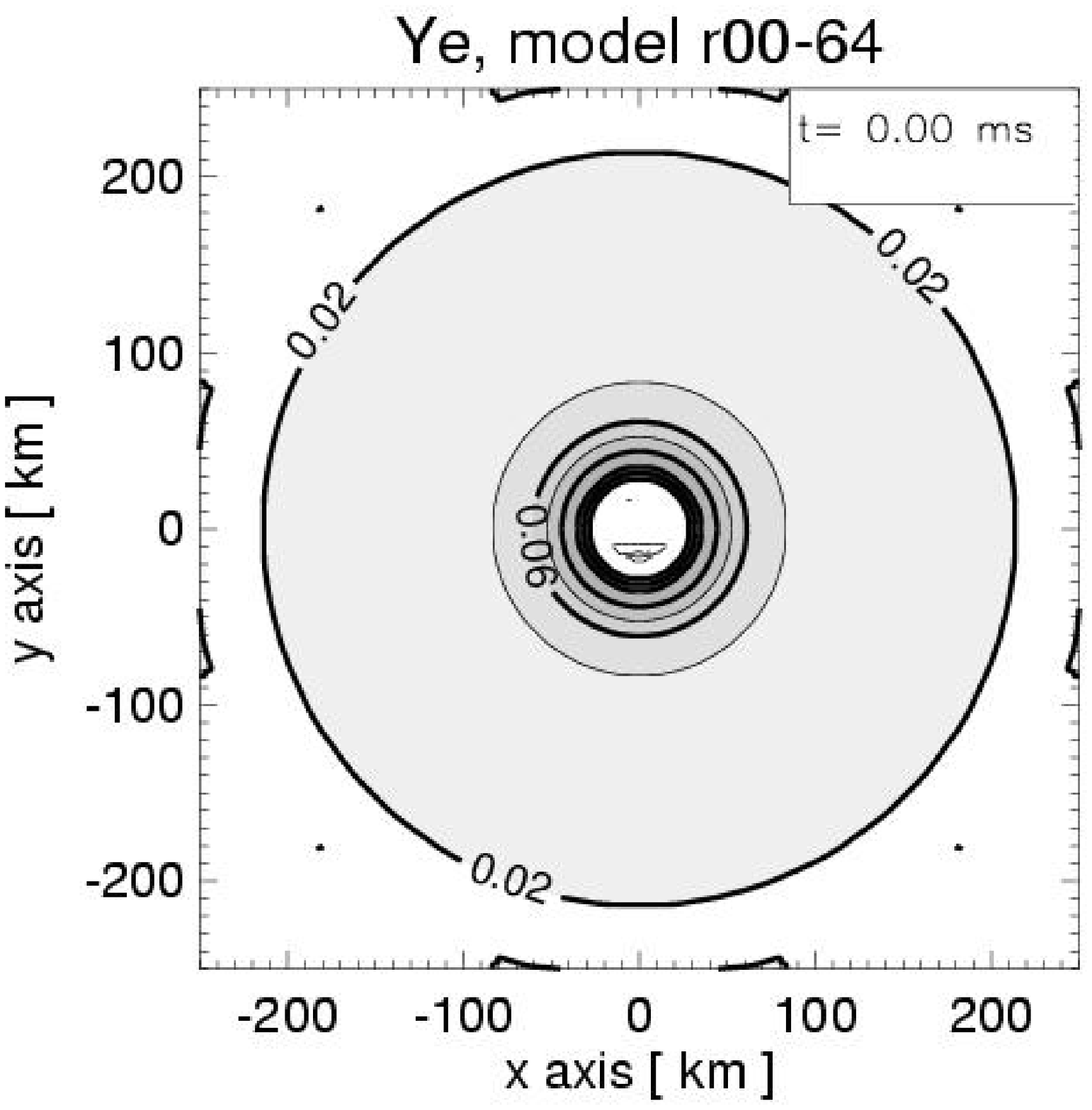,width=8.2cm}  \\[-2ex]
\end{tabular}
\caption[]{
Initial configuration at the start of the simulations for Model~r00-64.
The density is displayed in the orbital plane (left) and in two orthogonal
planes perpendicular to the equator (right).
It is given in g$\,$cm$^{-3}$
with contours spaced logarithmically in steps of 0.5~dex. 
The arrows in the density plots indicate the velocity field. 
The temperature and electron fraction ($Y_e$) are displayed in the
orbital plane. 
The temperature is measured in MeV, its contours are labelled with the
corresponding values. The contours of electron fraction are spaced
linearly in steps of 0.02. 
\label{fig:1}
}
\end{figure*}

We use the subscript $p$ for the shear viscosity that is assumed to
result from physical processes in the disk (e.g.~turbulence or
magnetohydrodynamic processes, see Balbus \& Hawley~\cite{bal98},
Duez et al.~\cite{duez04}, Thompson et al.~\cite{tho05})
to distinguish it from the numerical viscosity which was discussed
in Ruffert \& Janka~(\cite{ruffert3}) and Ruffert \& Janka~(\cite{ruffert}). 
The approach used in this paper
is similar to the disk formalism developed by Chen et
al.~(\cite{chen95}), Abramowicz et al.~(\cite{abram95}), Lee \&
Ramirez-Ruiz~(\cite{lee02}), and McKinney \& Gammie~(\cite{mckinney});
see also Hawley \& Krolik~(\cite{hawleykrolik}), 
Narayan et al.~(\cite{nar01}), and Kohri \& Mineshige~(\cite{kohri02}).

The resolution dependent viscous dissipation due to numerical
viscosity can be estimated to correspond to an $\alpha$ value of
approximately $0.01$ for our code and the chosen grid zoning (see
Ruffert \& Janka~\cite{ruffert3}, and Janka et al.~\cite{ruffert}).

\begin{figure*}[htp!]
\begin{tabular}{cc}
  \psfig{file=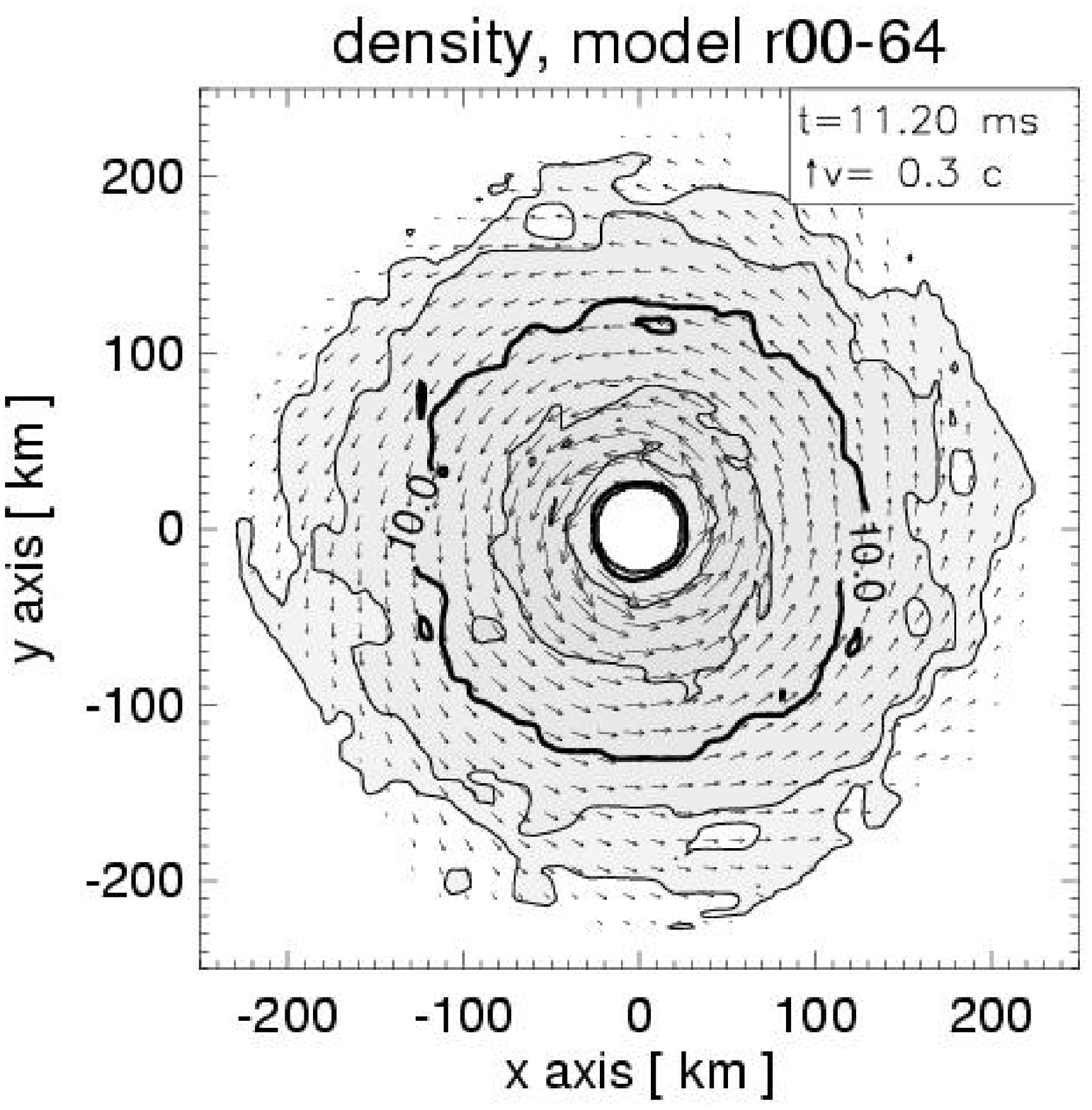,width=8.2cm} &
  \psfig{file=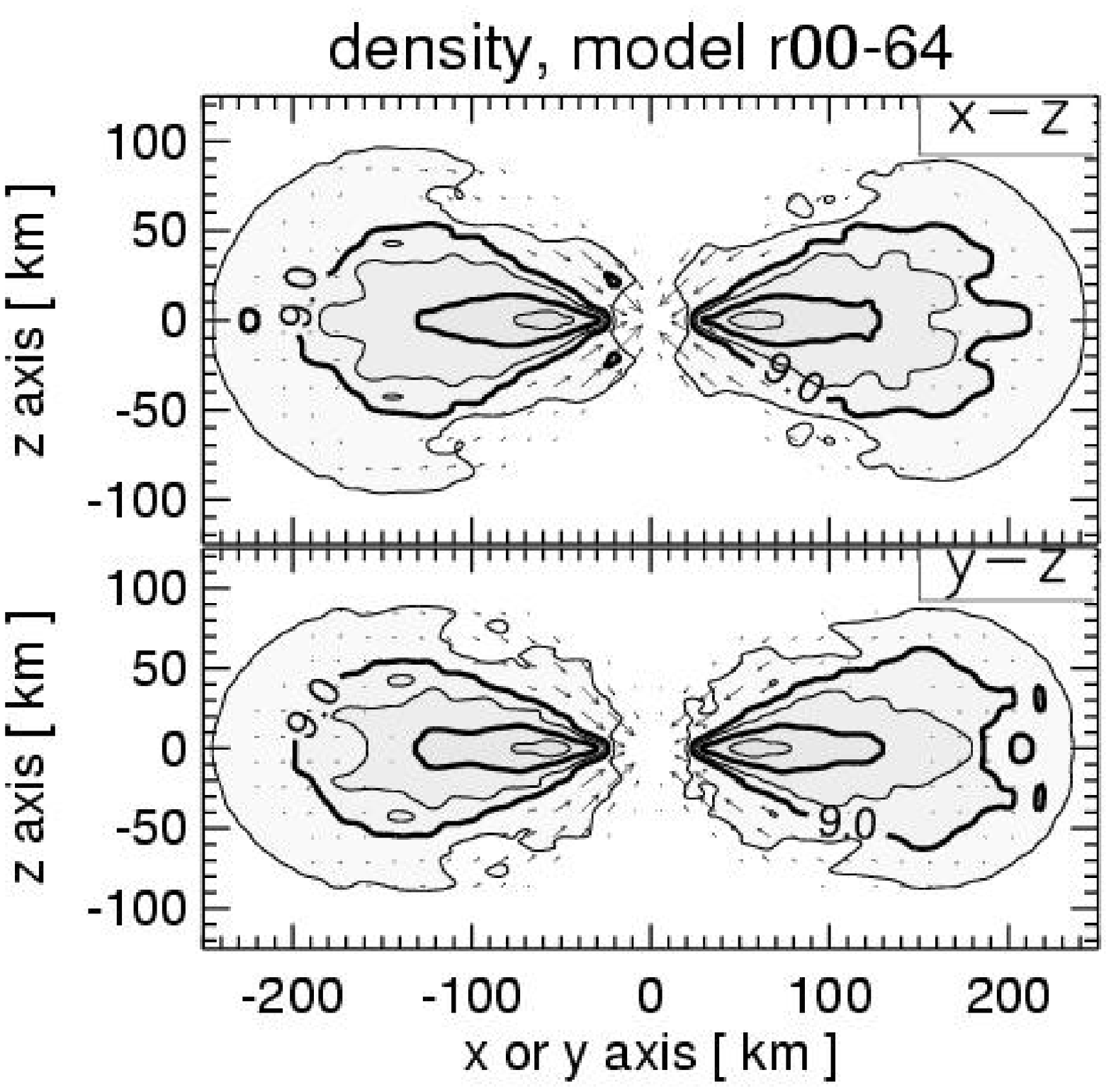,width=8.2cm} \\[-2ex]
  \psfig{file=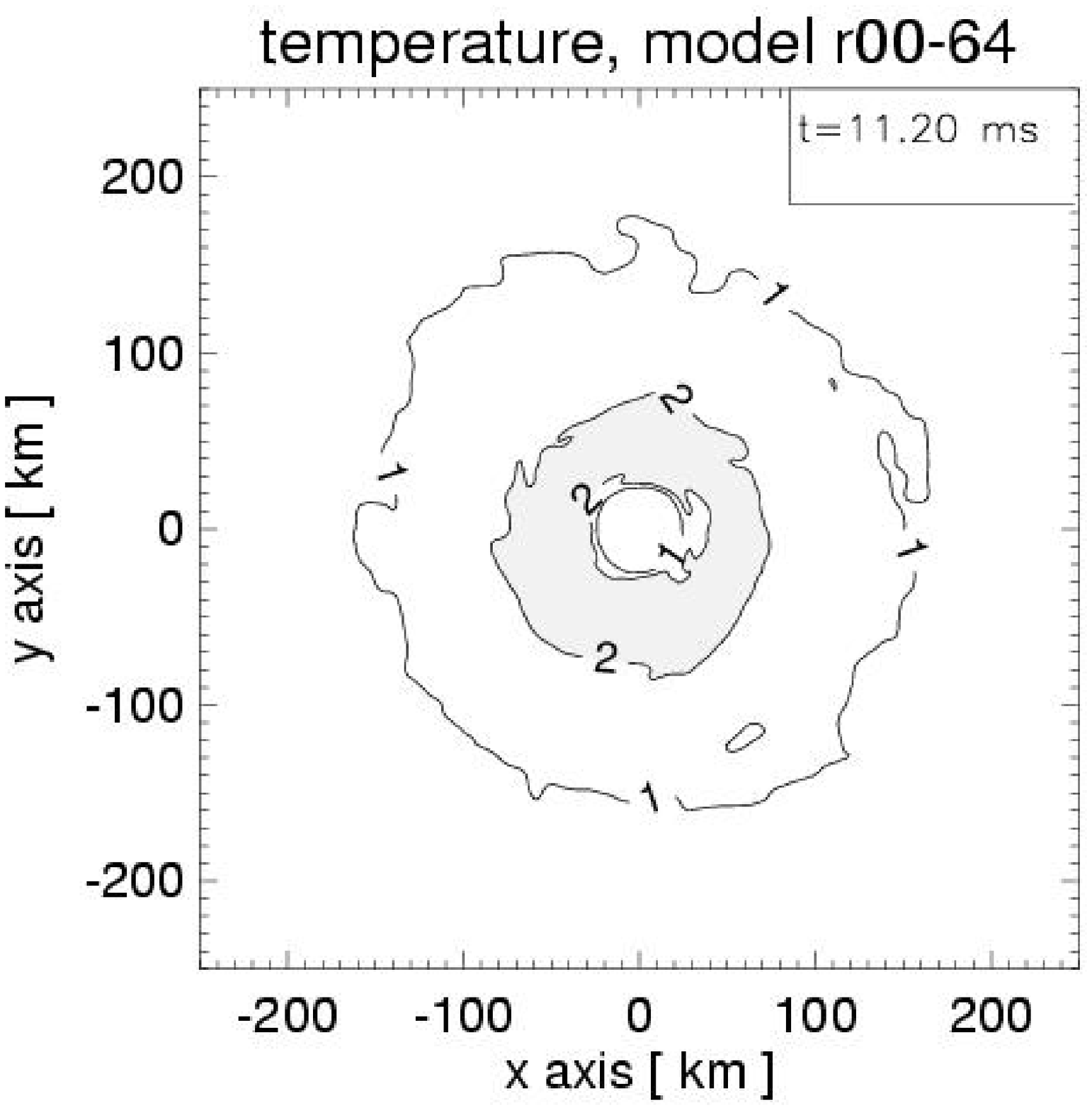,width=8.2cm} &
  \psfig{file=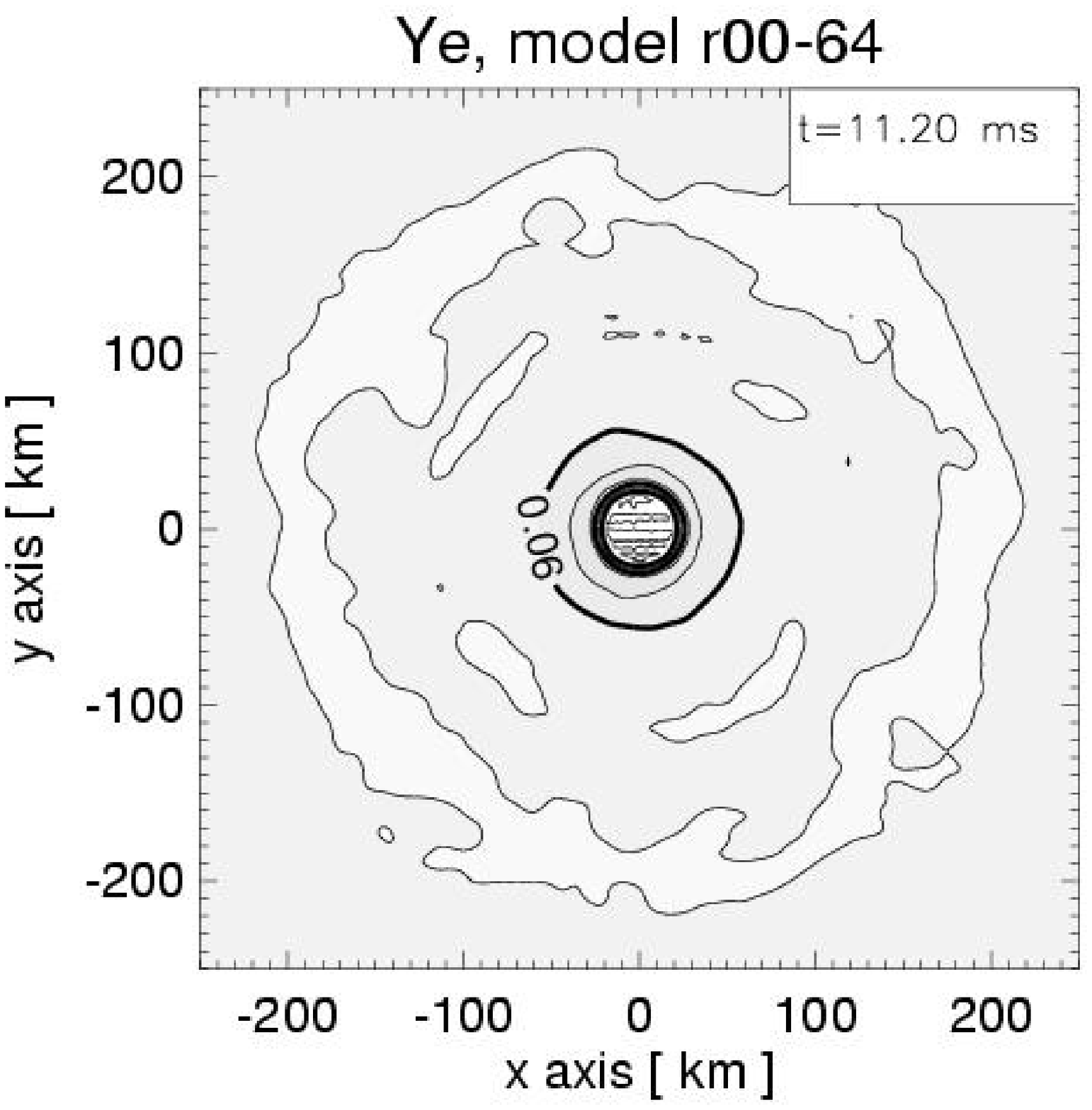,width=8.2cm}  \\[-2ex]
\end{tabular}
\caption[]{
Density distribution for Models~r00-64 in the orbital plane (left) and
perpendicular to it (right) at about 11$\,$ms after the start of the
simulations.
It is given in g$\,$cm$^{-3}$
with contours spaced logarithmically in steps of 0.5~dex. 
The arrows in the density plots indicate the velocity field. 
The temperature and electron fraction ($Y_e$) are displayed in the
orbital plane. 
The temperature is measured in MeV, its contours are labelled with the
corresponding values. The contours of electron fraction are spaced
linearly in steps of 0.02. 

\label{fig:2}
}
\end{figure*}

\subsection{Thermodynamics}

The thermodynamics of the gas is described by using the equation of
state (EoS) of Lattimer \& Swesty~(\cite{lat91}). This allows us to
follow the neutrino production and escape by a leakage and
trapping scheme as
detailed in previous papers (Ruffert et al.~\cite{ruffert1},
\cite{ruffert2}; Ruffert \& Janka~\cite{ruffert,ruffert3}) in order
to calculate the energy and lepton number changes by neutrino
losses. This approach is used in the present paper, because our main
aim is the investigation of the relevance for gamma-ray burst
scenarios, in particular for those where the neutrino emission has
been suggested to provide the energy for the relativistic gamma-ray
burst fireball via neutrino-antineutrino annihilation. Also the
amount of mass ejection during the dynamical interaction and the
properties of the ejected matter depend on the EoS.

The density is bounded from below in our simulations by the 
assumption of an environmental medium of about
$10^8\,{\rm g\,cm}^{-3}$ (determined by the minimum density of
the employed equation of state table) and is more than
two orders of magnitude below the average densities inside 
our simulated tori.

In addition to the standard equations of continuity and conservation
of momentum and energy, we also evolve an entropy equation (more
information can be found in Ruffert \& Janka~(\cite{ruffert3}).
The terms due to viscosity are given in Sect.~{\ref{sec:simulasi}}.
This allows us to obtain the temperature from both the
energy and the entropy equation, allowing a cross-check.
For the simulations of the torus evolution shown in this paper,
the two temperatures are in fairly good overall agreement, 
so this double-check of the temperature is redundant in the context
for the simulations presented in this paper.
Local differences between the two temperatures remain local and
are therefore of no major importance for the overall hydrodynamical
results.

\begin{figure*}[htp!]
\begin{tabular}{cc}
  \psfig{file=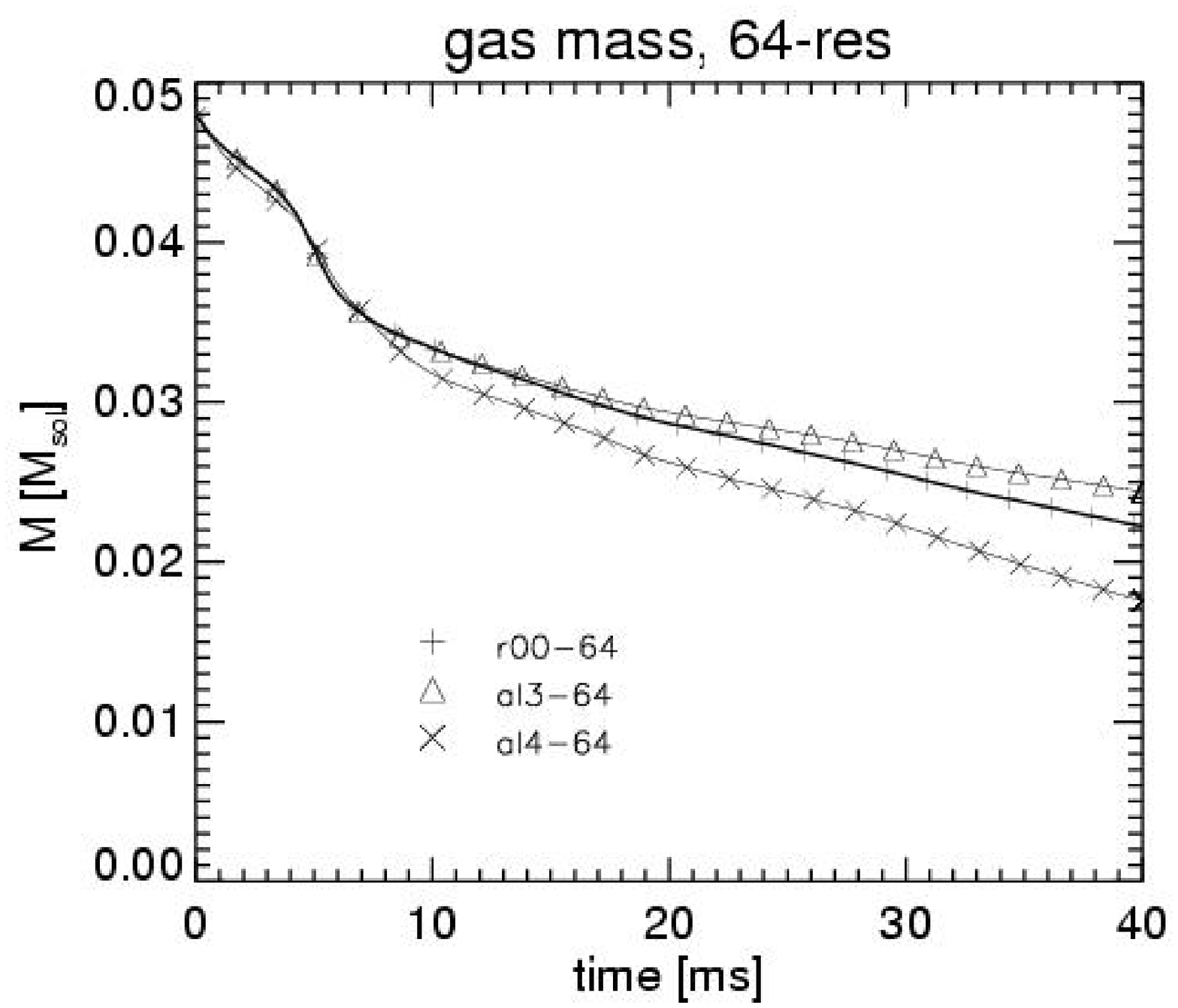,width=8.8cm} &  \\[-30ex]
  \psfig{file=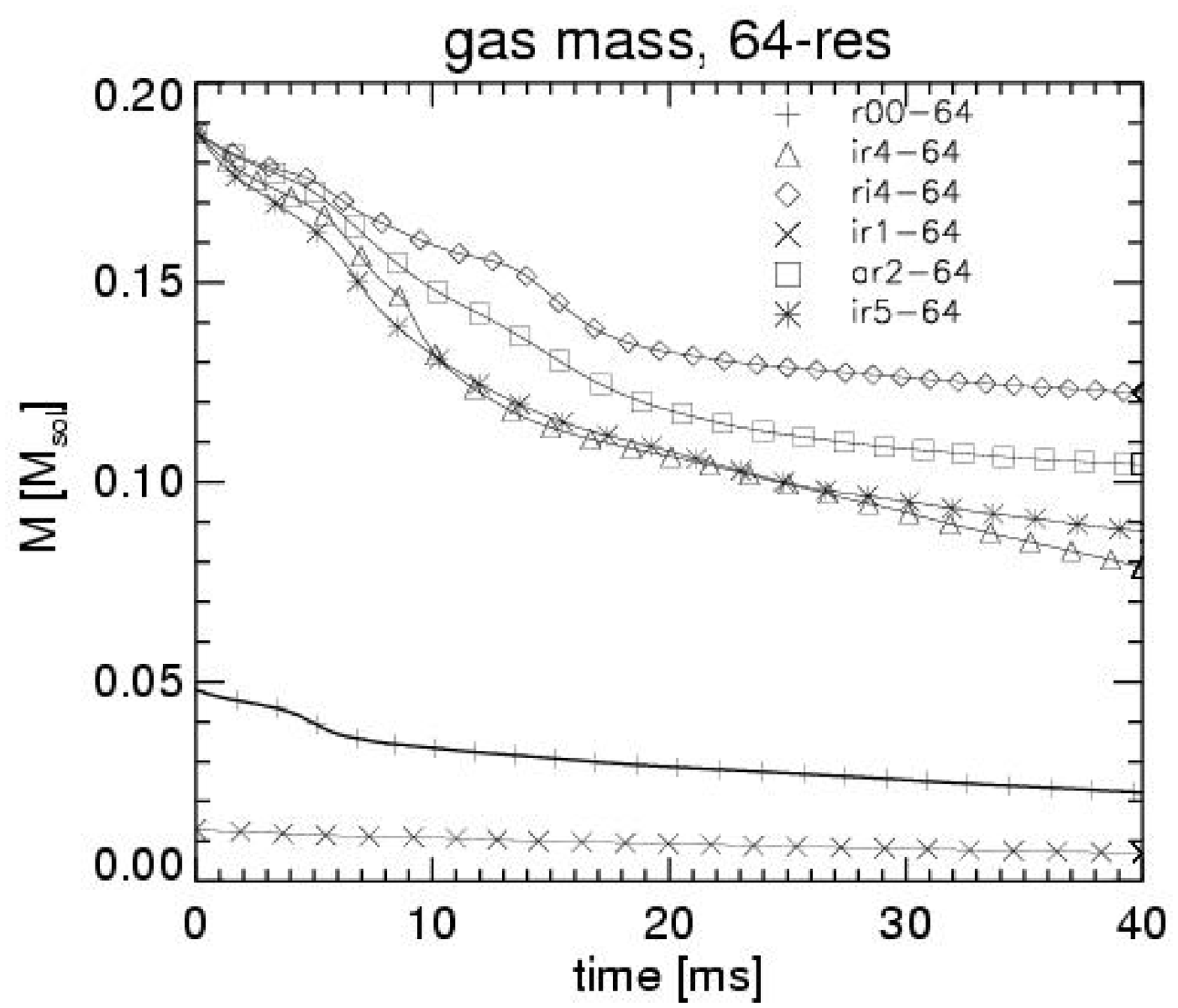,width=8.8cm}  &
  \psfig{file=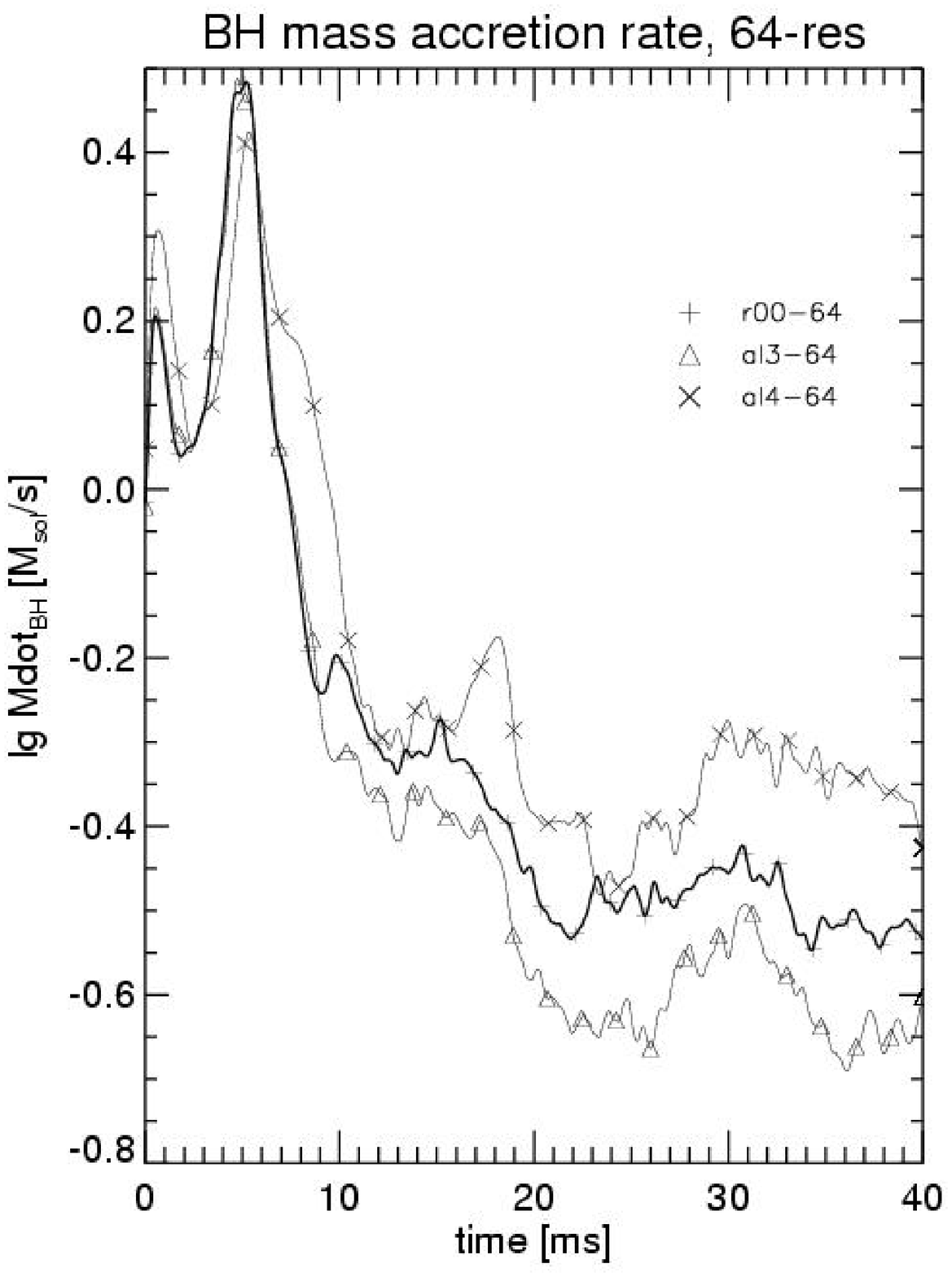,width=8.8cm} \\[-2ex]
\end{tabular}
  \caption[]{Mass accretion rate of the black hole and and gas mass 
on the grid as functions of time for Models~r00-64, al3-64 and al4-64
with increasing disk viscosity.
Gas mass on the grid as functions of time for the
    reference Model~r00-64, the low-mass torus Model~ir1-64, and the
    high-mass torus Models~ir4-64, ir5-64, ri4-64, and ar2-64.}
\label{fig:3a}
\end{figure*}

\begin{figure*}[htp!]
\begin{tabular}{cc}
  \psfig{file=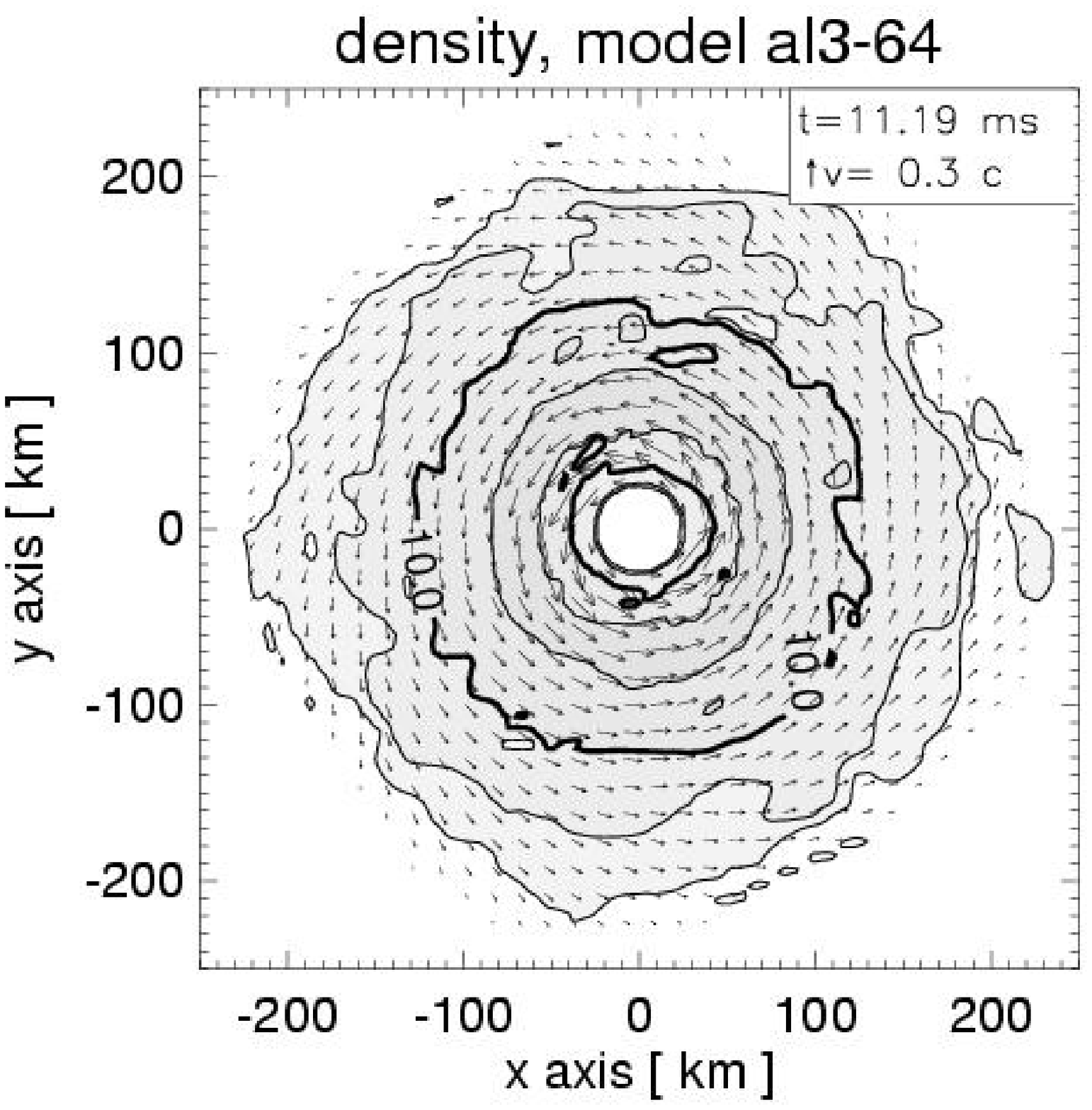,width=8.2cm} &
  \psfig{file=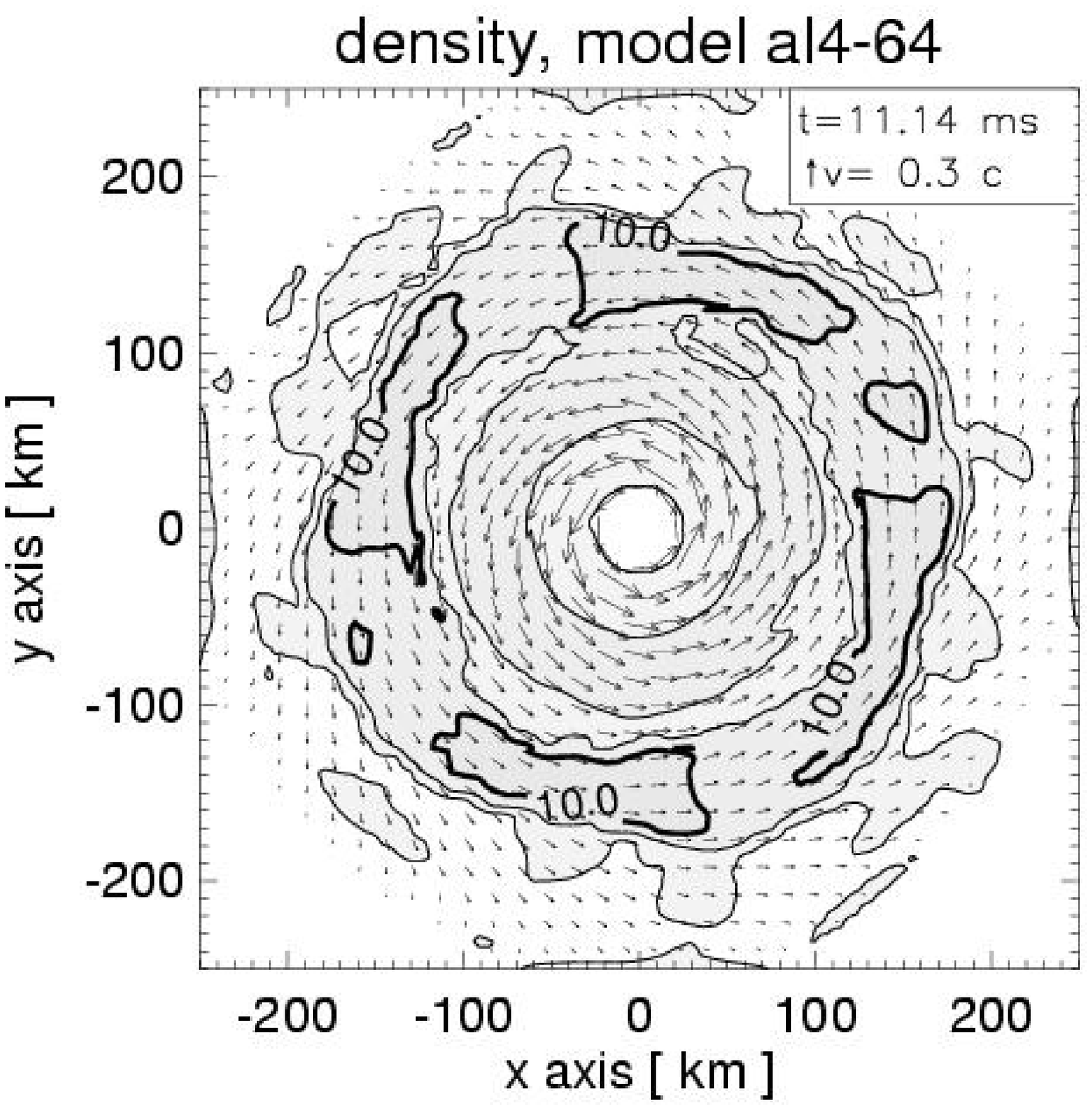,width=8.2cm} \\
  \psfig{file=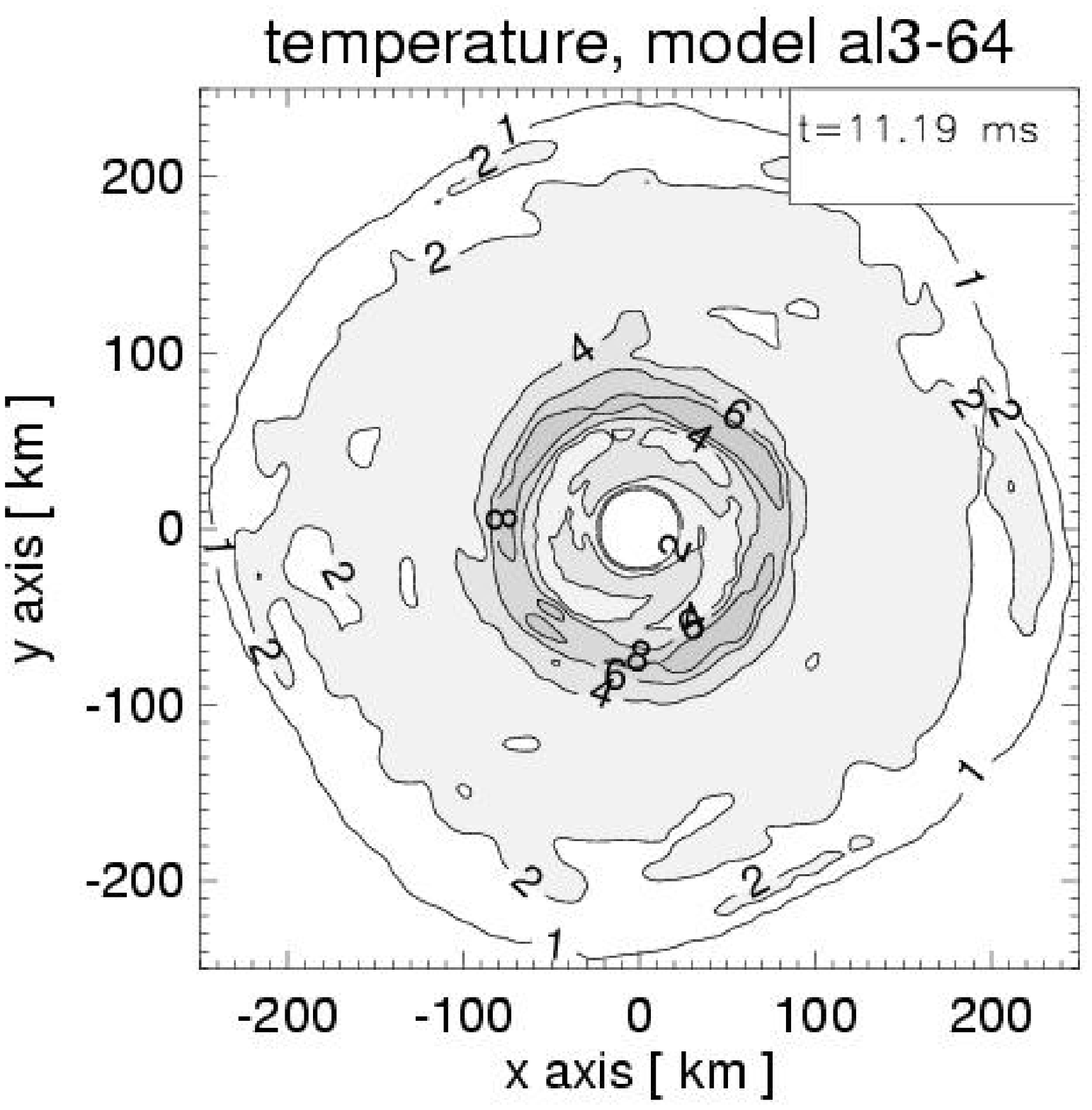,width=8.2cm} &
  \psfig{file=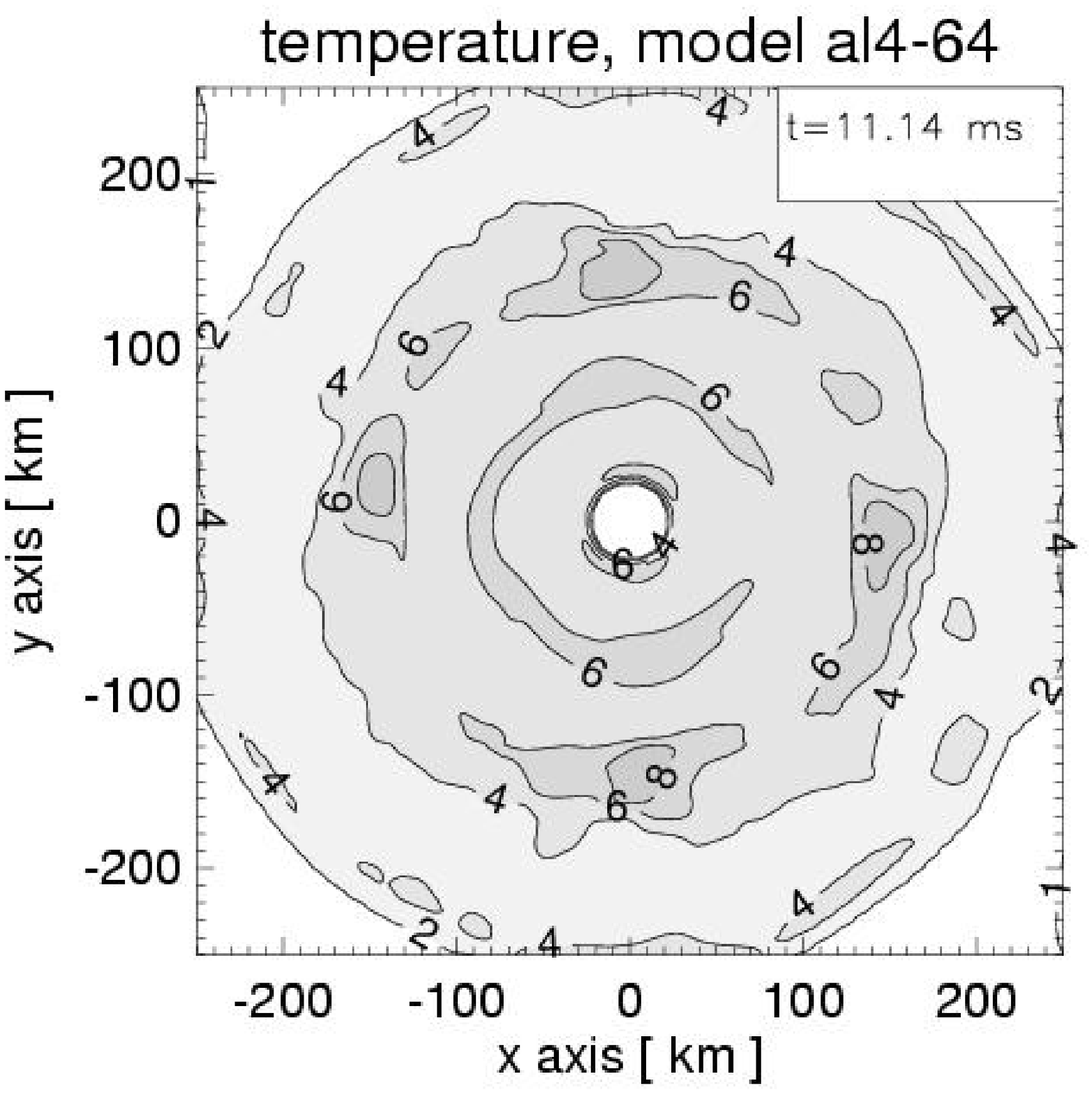,width=8.2cm} \\[-2ex]
\end{tabular}
\caption[]{
Density and temperature distribution for Models~al3-64 (left)
and al4-64 (right) with disk viscosities of $\alpha = 0.01$
and $\alpha = 0.1$, respectively.
The density is given in g$\,$cm$^{-3}$
with contours spaced logarithmically in steps of 0.5~dex. 
The arrows in the density plots indicate the velocity field. 
The temperature is measured in MeV, its contours are labelled with
the corresponding values.
\label{fig:2a}
}
\end{figure*}

\subsection{Models and initial conditions}

Table~\ref{tab:models3} contains a list of computed models with
their specific parameters, including the values of $\alpha$. We
follow the non-stationary evolution of the toroidal disk around the
central black hole on much longer time scale than the dynamical time
scale of the merger of black hole/neutron star systems, considering
the viscous transport of angular momentum which depends on the
assumed shear viscosity of the disk which we parametrise in the
models. Model ``r00'' defines a reference case without black hole
rotation and without disk viscosity. The ``al'' models include
disk viscosity but no black hole spin, models ``ro'' and ``ri''
consider rotating black holes but no disk viscosity, and models
``ar'' take into account non-zero values for the disk viscosity
and BH spin. 
The names of the models also carry information about the
number of zones of each grid level (32 or 64). 

The data for our initial setups, which consist of a black hole and
an axisymmetric, toroidal 
accretion disk with chosen masses, were taken from one of the final
configurations obtained in the simulations described by Janka et
al.~(\cite{jan99}; see also Ruffert et al.~(\cite{ruffert1,ruffert2}) 
and Ruffert \& Janka~(\cite{ruffert3}) for more information).  
Density, temperature, and neutron-to-proton ratio of the matter,
which is assumed to be in nuclear statistical equilibrium,
were azimuthally averaged.
Because we varied the disk mass by simply scaling the density, and
moreover switched from the Newtonian potential of the merger runs to
the pseudo-Newtonian gravity in this work, the tori are initially
slightly out of rotational equilibrium.
Despite the artificial adjustment which
happens during the first milliseconds of our simulations, we do not
consider this as a major problem because the compact object merger as
well as a subsequent collapse of the merger remnant to a black hole
are violent processes, followed naturally by a relaxation phase of 
the mass left in a torus.

The simulations of the present work started with a black hole 
mass of 4.017 $M_\odot$. This is the mass produced by the final
configurations of the models described above.
We will use the model with a torus mass of 0.0478$\,M_\odot$, 
as ``reference model''. The mass was then also changed to values
between 0.0120$\,M_\odot$ and 0.1912$\,M_\odot$ (see
Table~\ref{tab:models3}) by simply scaling the density distribution of
the reference configuration with an appropriate factor. The initial
conditions of temperature, density, and electron fraction (i.e.,
proton-to-baryon ratio) for the reference case are shown in 
Fig.~\ref{fig:1}.

The initial black hole spin parameter was varied between values of 
0, 0.3, 0.6 or 0.8 for
32-resolution models and between 0 or 0.6 for 64-resolution models,
respectively. We did not choose $a$ closer to unity because our
code is not fully relativistic. When $a$ is near unity, the disk
moves in so close to the event horizon that general relativistic
effects are extremely strong and the fluid velocities approach
the speed of light, thus making the approximative description
of relativistic gravity and the nonrelativistic fluid dynamics 
of our code questionable.


\section{Results \label{sec:hasil}}

In this section we will discuss only the models calculated with a
resolution of 64 zones on each level of the nested grid, so the 
specification ``-64'' in the model name,
e.g.~r00-64, will be mostly omitted. Comparing runs with 32 zones
and 64 zones at the same time, one finds only relatively small
differences of the model evolution and global 
disk properties, of which some are given in Table~\ref{tab:models}.
The 3D simulations with 32 zones, however, could be 
performed for a time interval of 70$\,$ms, whereas computer
resources did not allow us to follow the evolution of the 
better resolved models for more than 40$\,$ms. 
In Table~\ref{tab:models}
we also list the corresponding results of the 32-zone runs, which 
include cases that we studied with high resolution, too.

\begin{figure*}[htp!]
\begin{tabular}{cc}
  \psfig{file=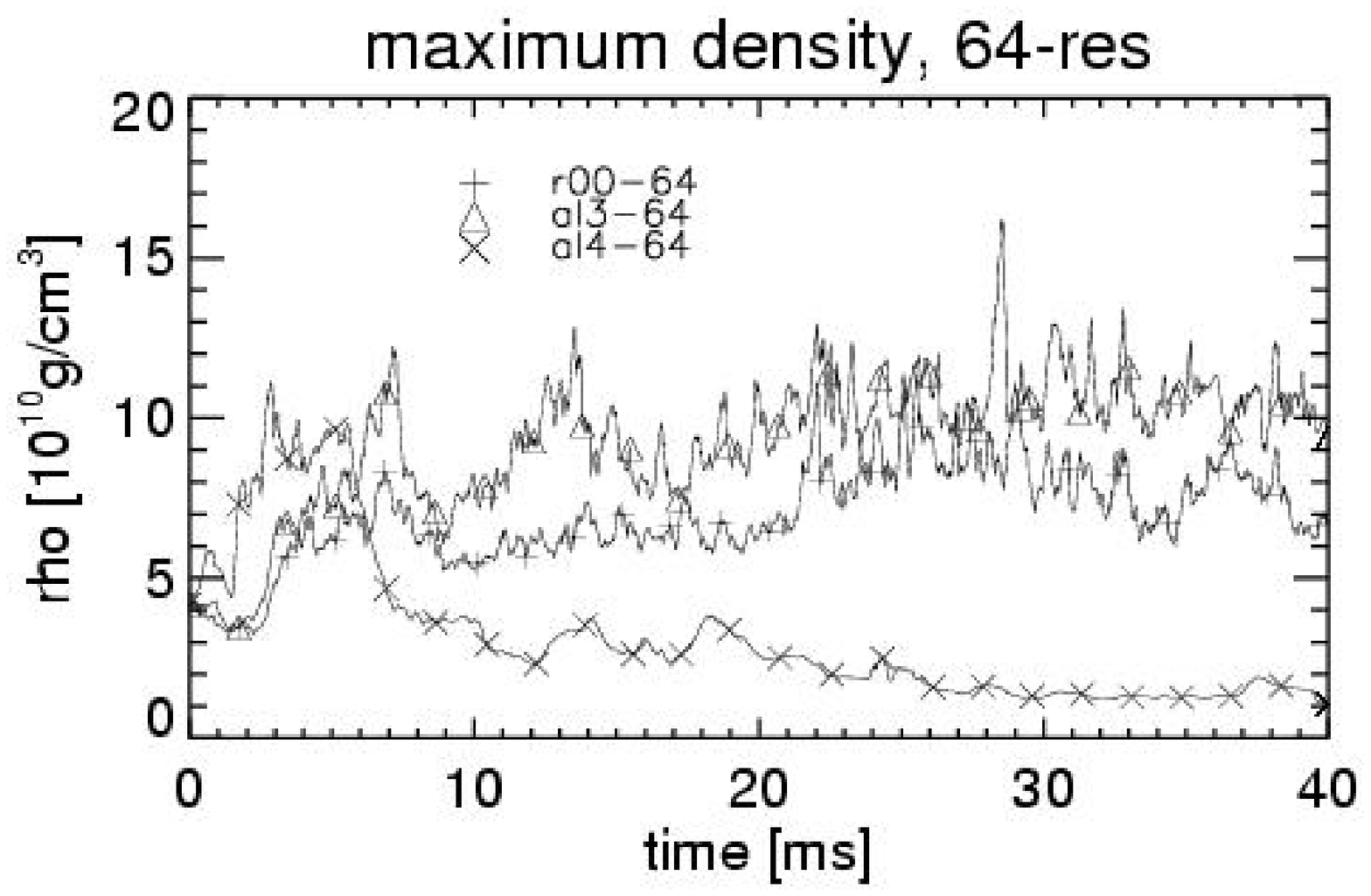,width=8.8cm} &
  \psfig{file=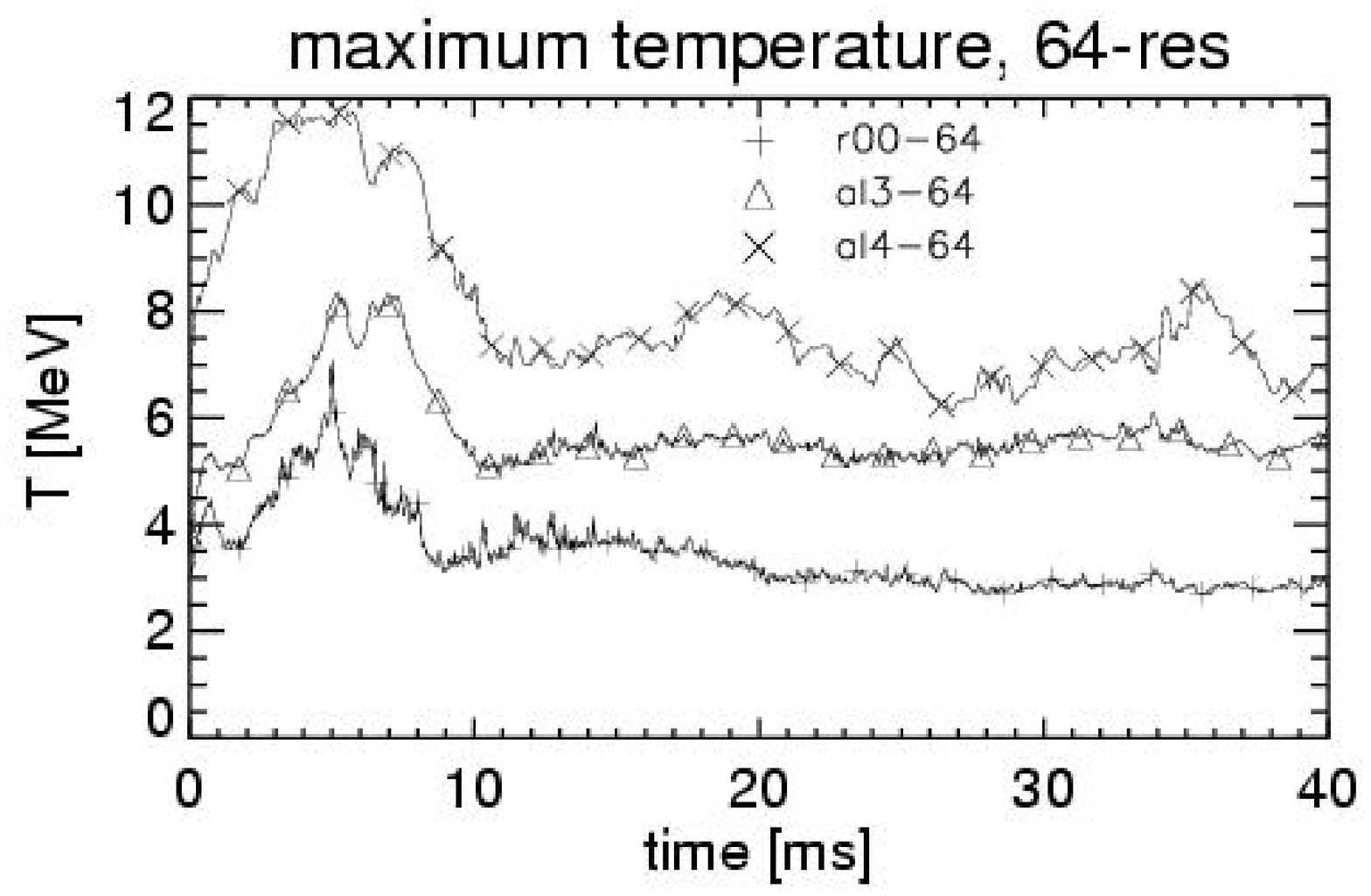,width=8.8cm} \\[-2ex]
\end{tabular}
  \caption[]
{Maximum values of gas density (left) and temperature (right)
on the grid as functions of time for Models~r00-64, al3-64 and al4-64
with increasing disk viscosity.}
\label{fig:8bnew}
\end{figure*}

\subsection{Dynamical evolution of reference Model~r00}

We begin with discussing the evolution of the disk in Model~r00,
which will serve as reference model in this paper.
Figure~\ref{fig:1} shows the initial distributions of density,
velocity, temperature, and electron fraction for this model.
The obvious axisymmetric construction of these distributions
was described above in the section on initial conditions.
Figure~\ref{fig:2} shows these distributions after $11.2\,{\rm ms}$. 
At this time the disk has reached a quasi-equilibrium state, and its
properties in Model~r00 are the following: Within 100$\,$km of the 
black hole the densities are a few times 
$10^{10}\,{\rm g}\,{\rm cm}^{-3}$, the temperatures are 
around 2$\,$MeV, and the electron fraction is 0.04--0.06.

Given our nested Cartesian grid, one cannot expect a precisely
axisymmetric distribution to retain its symmetry when evolved with an
explicit code. Some degree of numerical noise will always creep into
the simulation and break the symmetry.
This is expected to be on the scale of a couple of zones, so on a small
scale as compared to the radius of the disk.
On the other hand, the nested grid structure, specifically the
boundaries of the grids and the transition to a lower resolution
outside of a central cubic volume, will imprint a distinct 4-fold
symmetry, which mostly remains small in amplitude.
So, although it is known that disks also suffer physical
instabilities, e.g.~Papaloizou \& Pringle~(\cite{papa84}), Kleiber \&
Glatzel~(\cite{klei99}), our simulations do not permit to connect
the observed destruction of the initial axial symmetry with
such physical instabilities. The numerical noise must be expected
to drown all but the fastest growing high-order and low-order modes.

Figure~\ref{fig:3a} shows the temporal evolution of the black hole
accretion rate and the integrated mass of the accretion torus.
A phase of high mass accretion rate is visible from the outset until
about $t\approx 10\,{\rm ms}$.
This is a transient effect linked to the initial construction of the
disk. 
This phase can also easily be spotted in the plot for the evolution of
the maximum temperature in Fig~\ref{fig:8bnew}: The temperature
increases transiently. 
The disk settles into a quasi-steady state characterised by a nearly
constant black hole accretion rate at a lower value and by only slowly
changing values of the average density and temperature in the torus.
This relaxation is more pronounced in the mass accretion rate of the 
black hole (Fig.~\ref{fig:3a}), a fact which might indicate that 
it is the aftermath of the settling in the outer parts of the
torus, which has now made its way to the centre.

\begin{figure*}[htp!]
\begin{tabular}{cc}
  \psfig{file=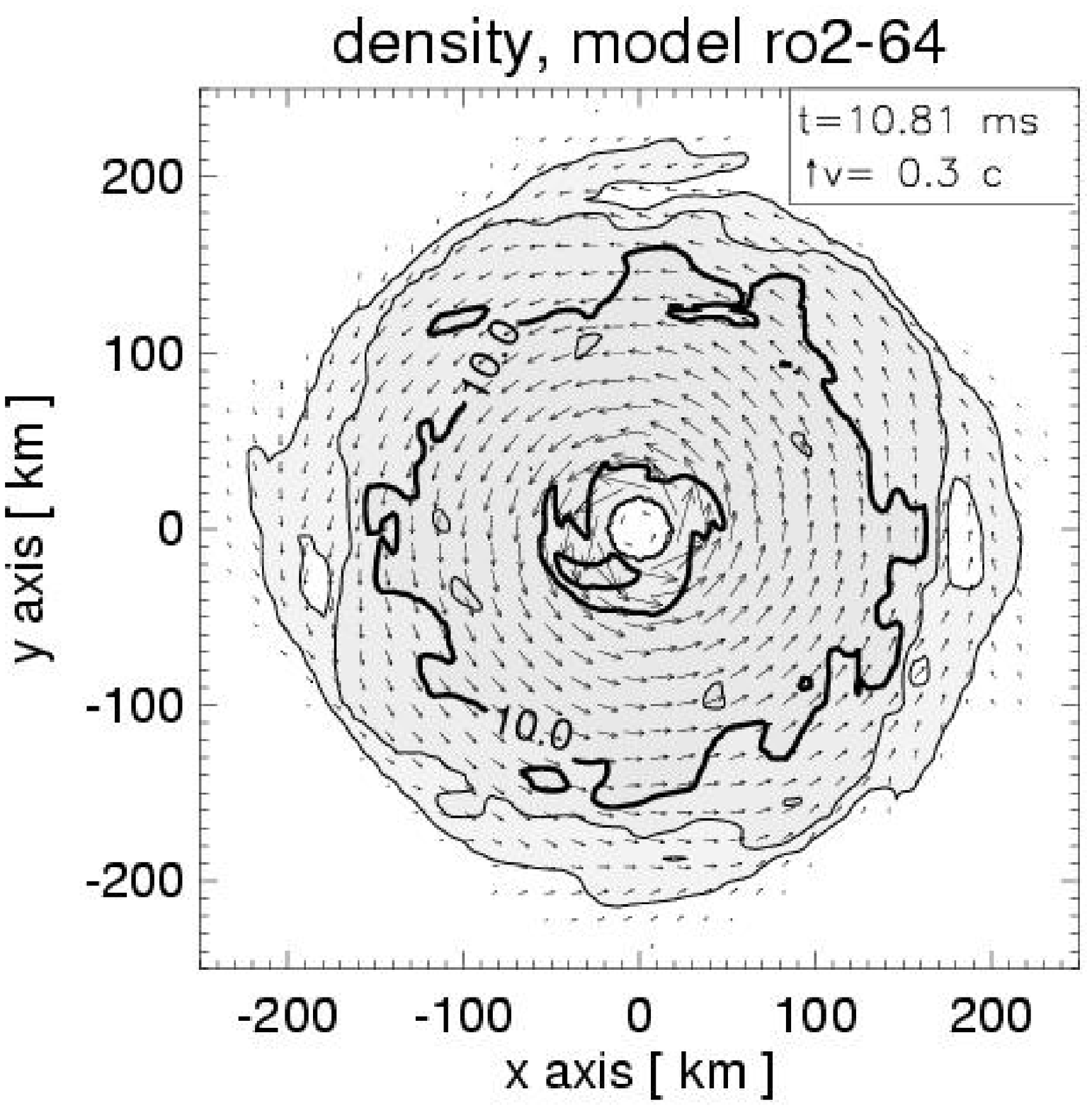,width=8.2cm} &
  \psfig{file=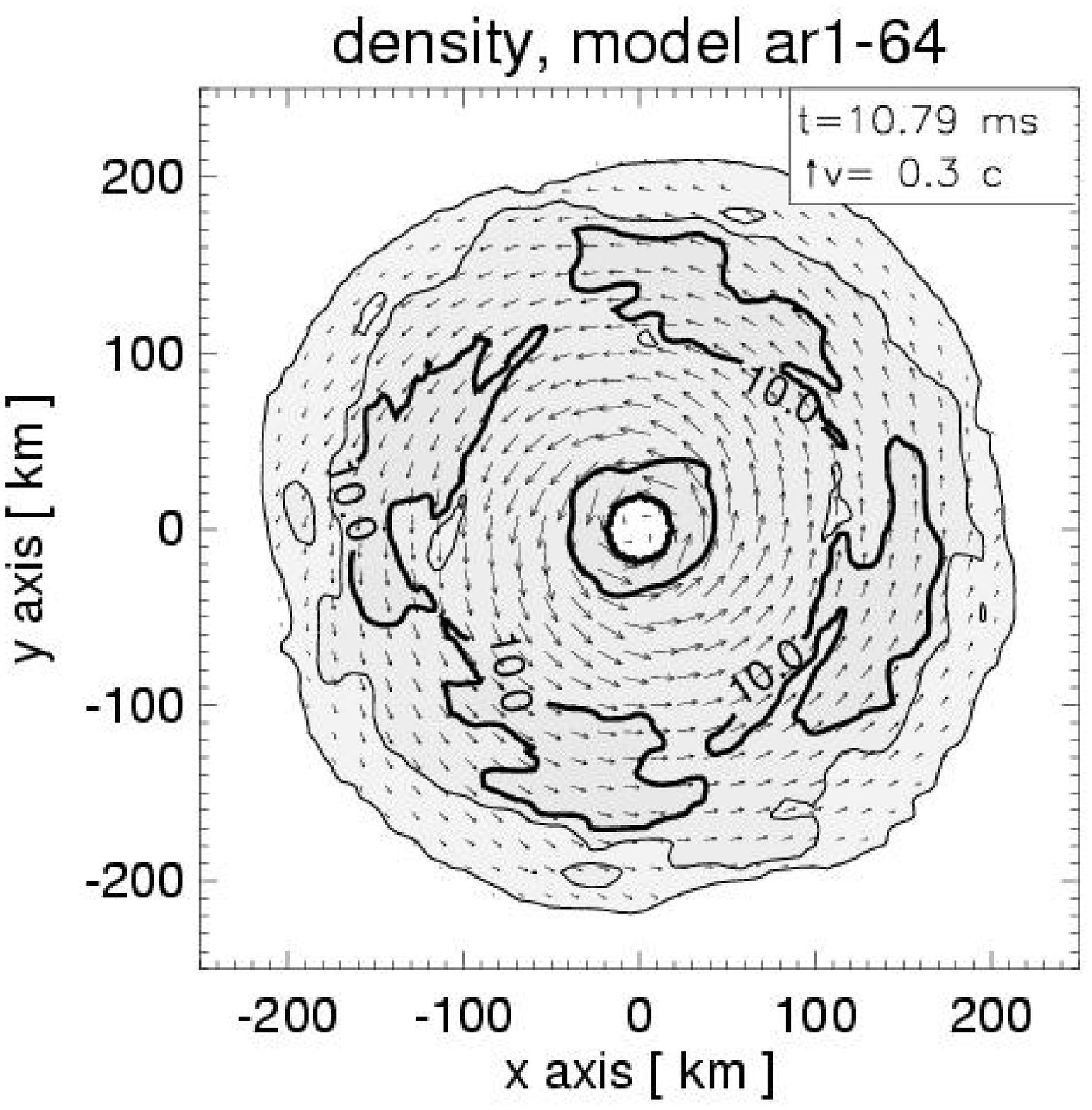,width=8.2cm} \\[-2ex]
  \psfig{file=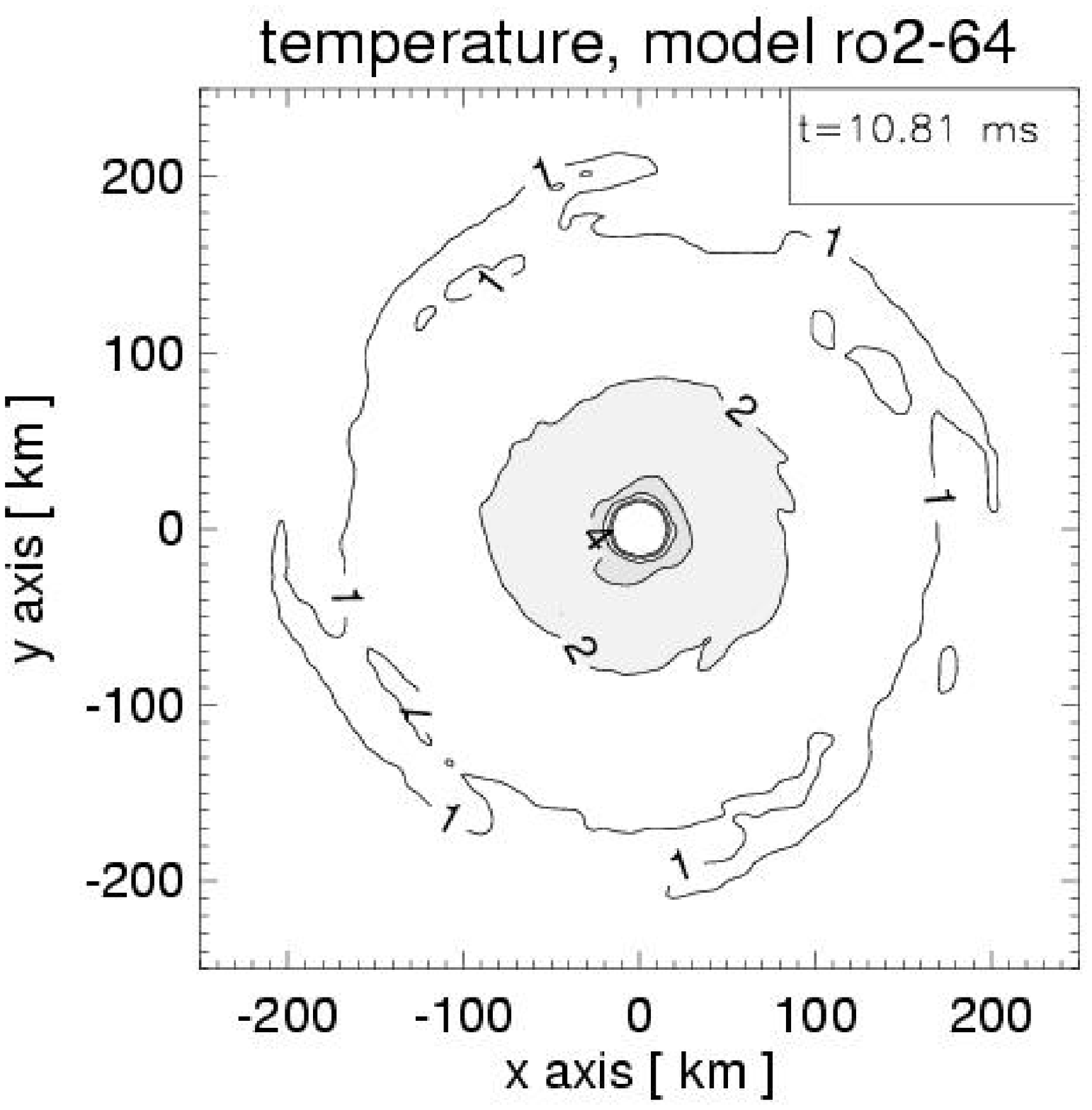,width=8.2cm} &
  \psfig{file=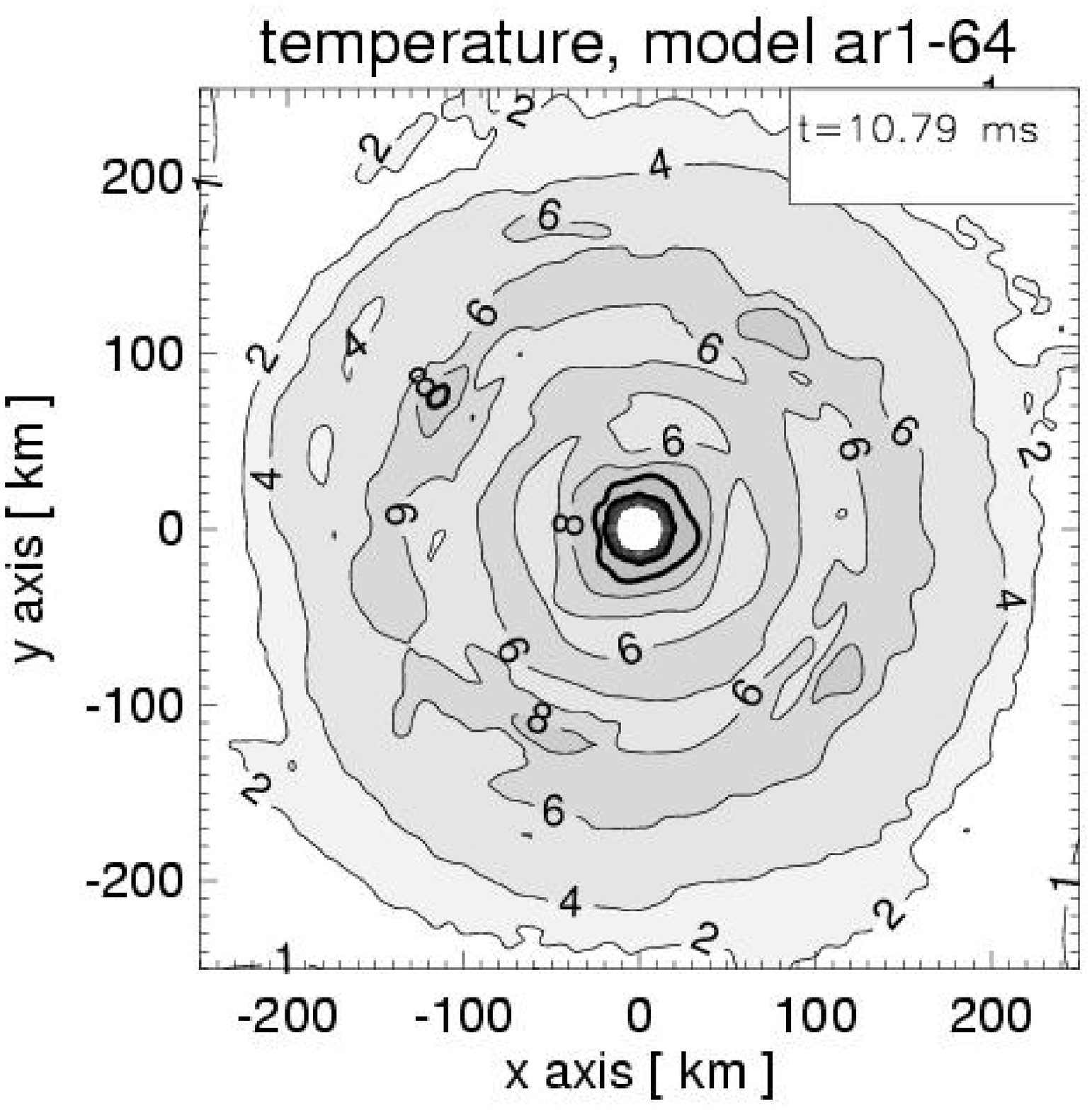,width=8.2cm}  \\[-2ex]
\end{tabular}
\caption[]{
Density and temperature distribution for Models~ro2-64 (left)
and ar1-64 (right) for a corotating black hole, with disk viscosities
of $\alpha = 0.0$, and $\alpha = 0.1$, respectively. 
The density is given in g$\,$cm$^{-3}$
with contours spaced logarithmically in steps of 0.5~dex. 
The arrows in the density plots indicate the velocity field. 
The temperature is measured in MeV, its contours are labelled with
the corresponding values.
\label{fig:II2}
}
\end{figure*}

\begin{figure*}[htp!]
\begin{tabular}{cc}
  \psfig{file=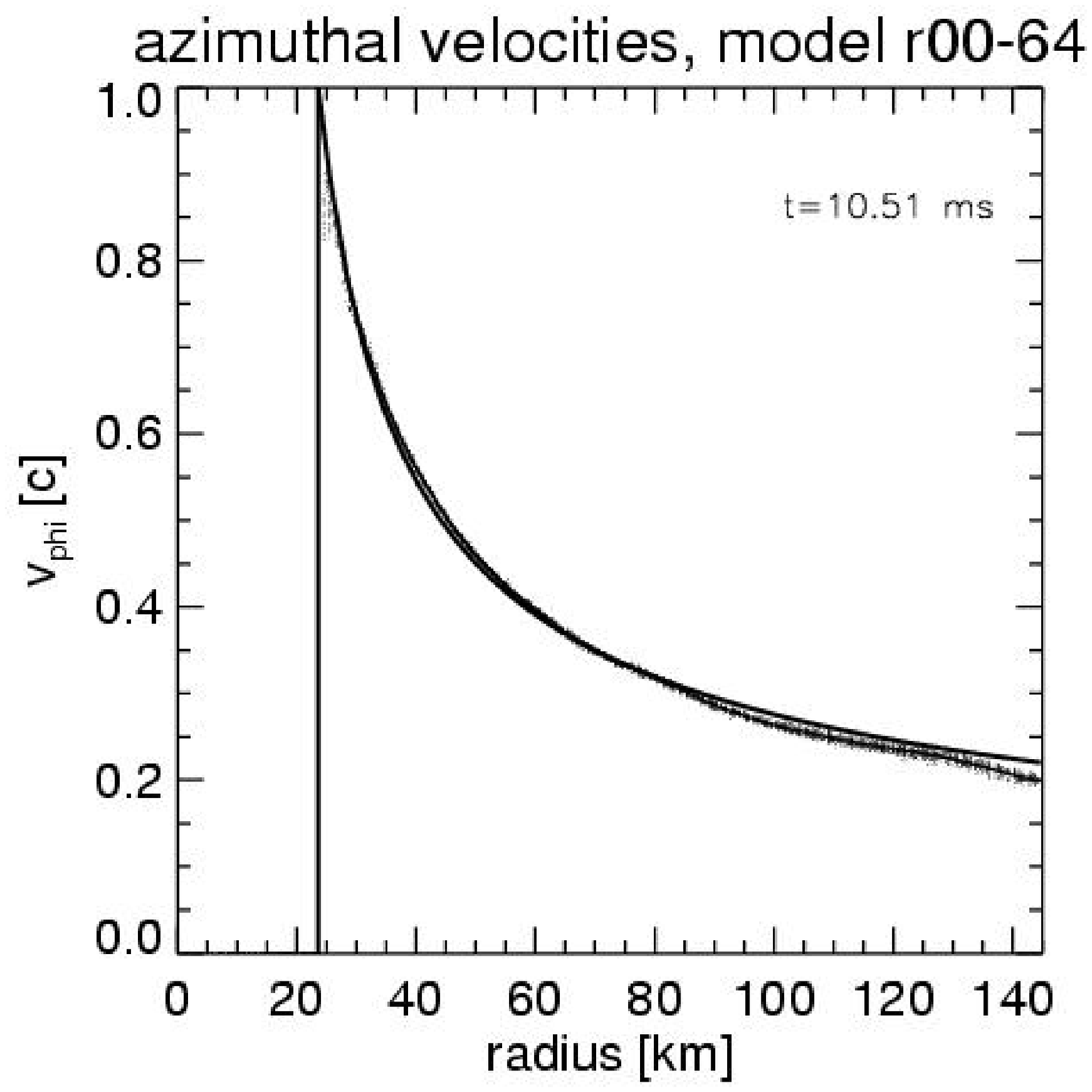,width=8.0cm} &
  \psfig{file=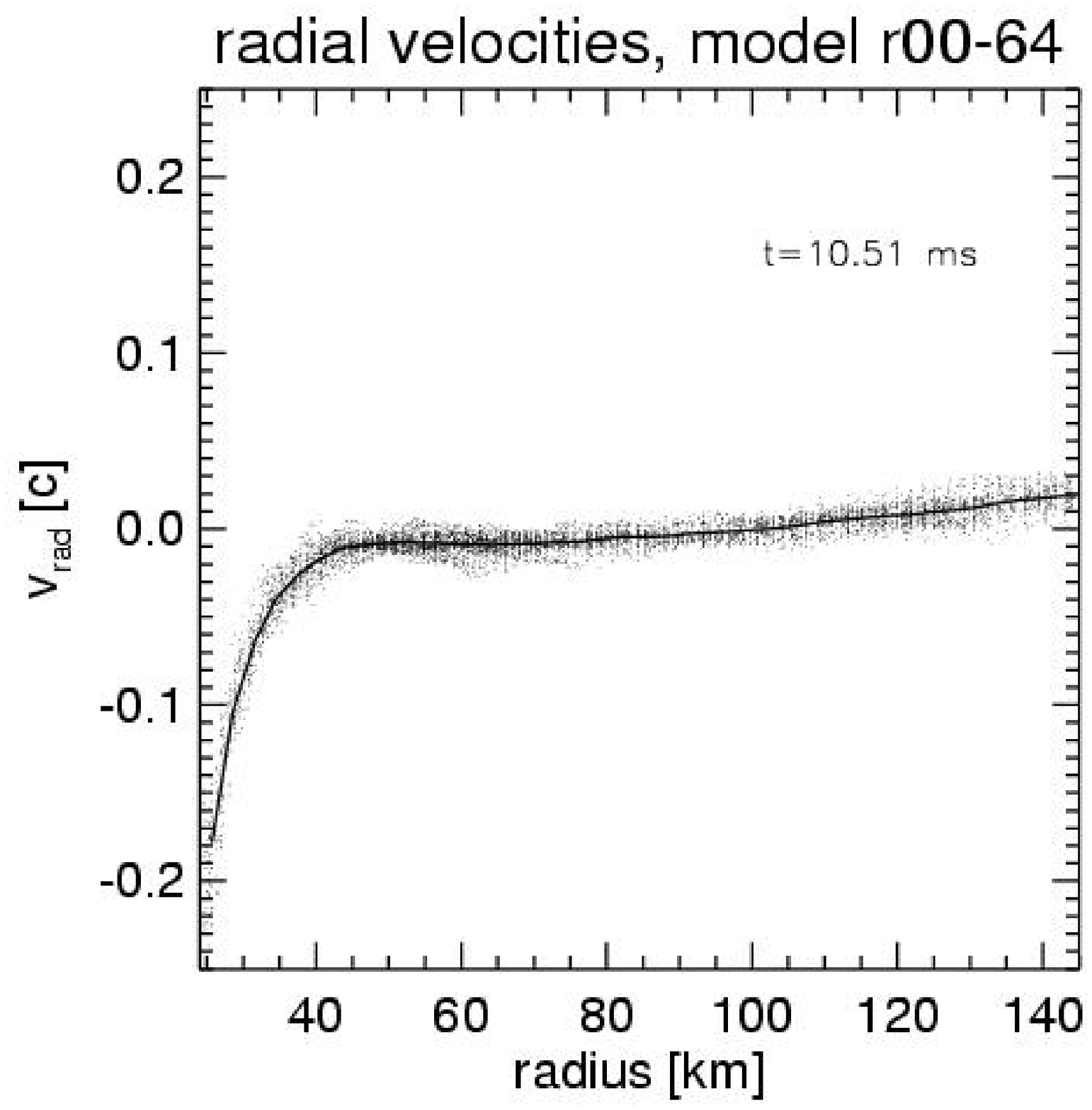,width=8.0cm} \\[-2ex]
  \psfig{file=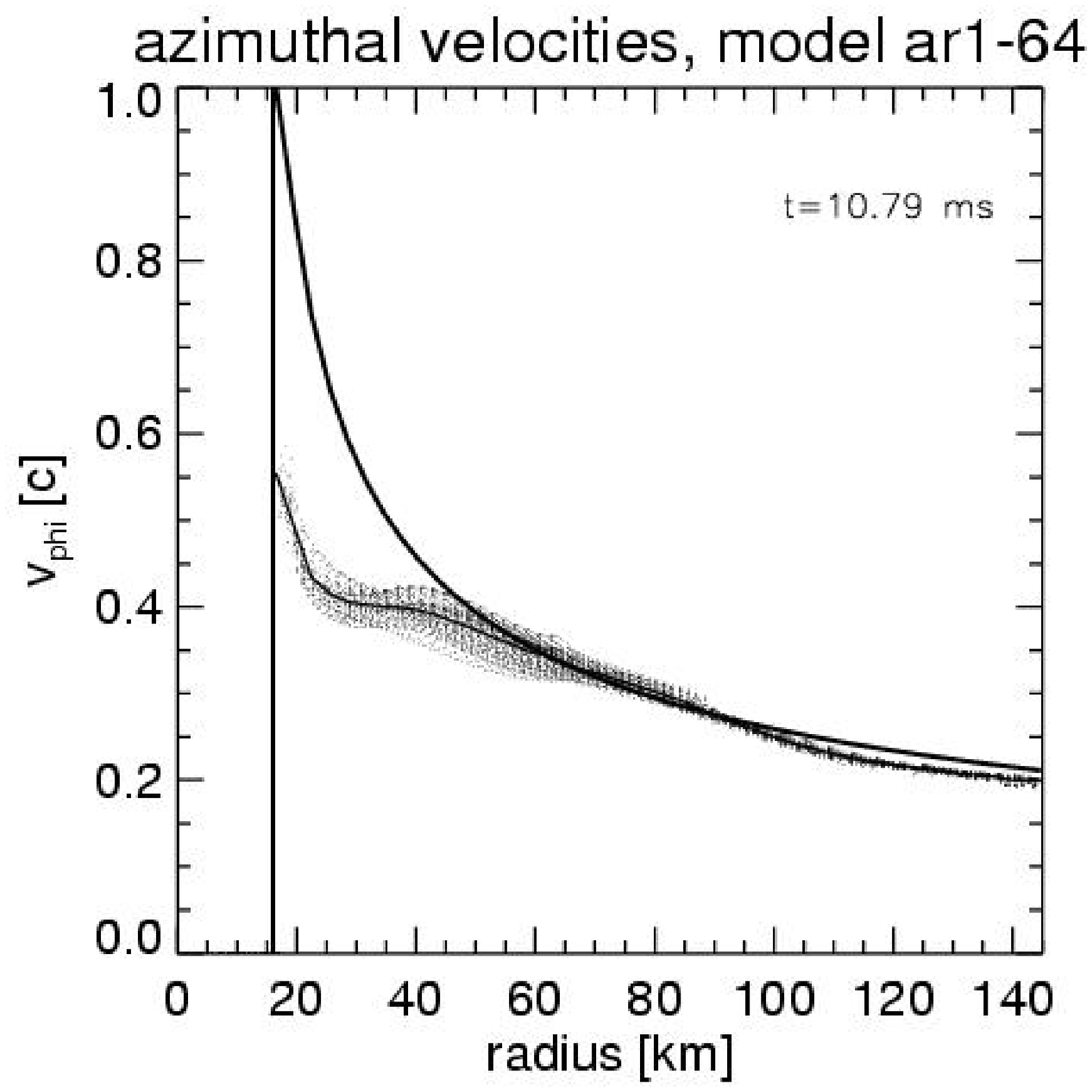,width=8.0cm} &
  \psfig{file=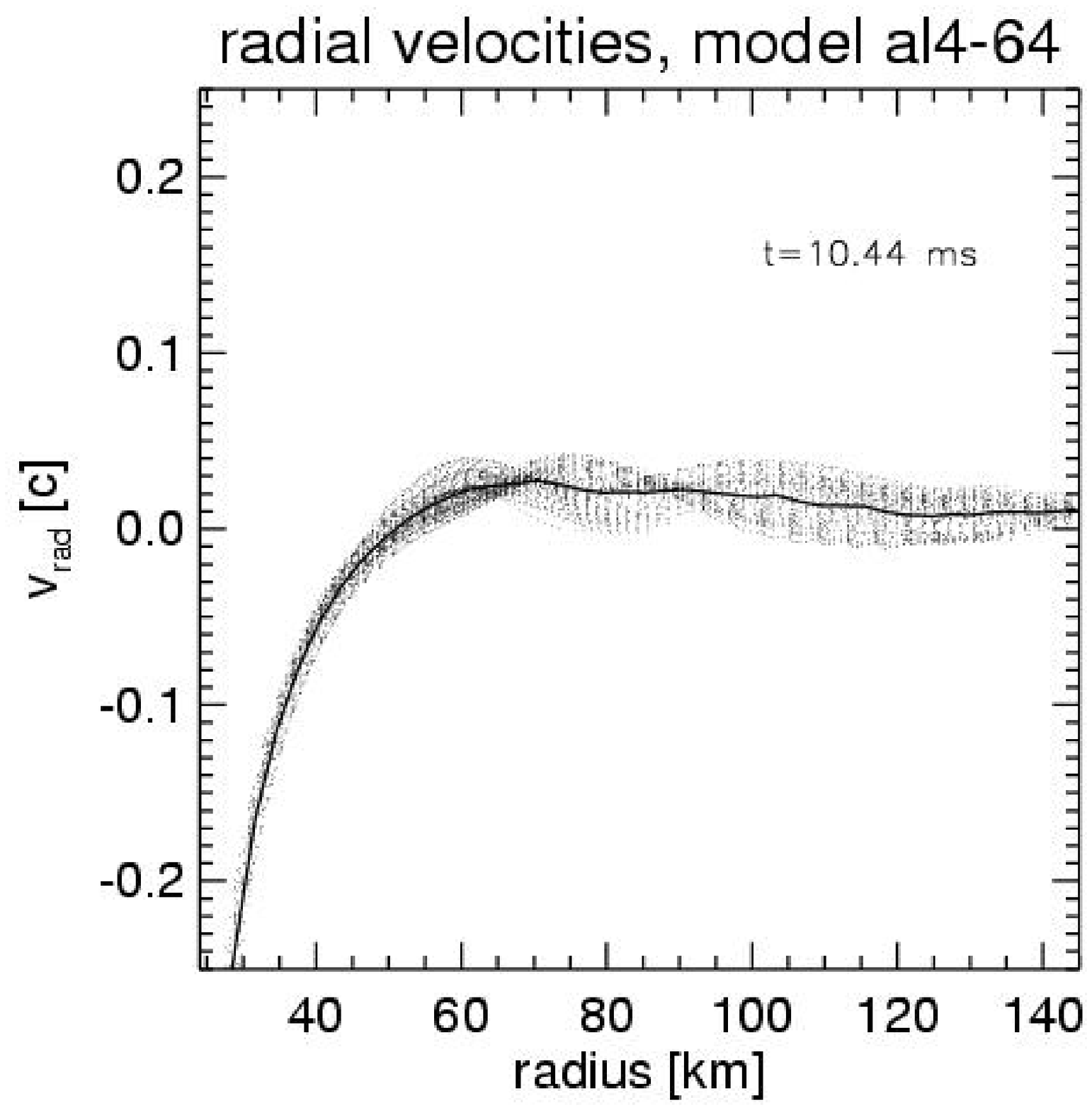,width=8.0cm} \\[-2ex]
\end{tabular}
  \caption[]{
    {\em Left panels:}
     The dots represent the azimuthal
    velocities $v_\varphi(d)$ (normalised to the speed of light)
    of all zones in the equatorial plane of the reference
    Model~r00-64, and the high disk-viscosity Model~ar1-64 with
    corotating black hole, at the times after the start of the
    simulation indicated in the plots.
    The thin solid line gives the average value of the
    azimuthal velocities and the bold solid line the local Keplerian
    velocity $v_{\rm Kepler}(d)$ of the Artemova et al.~(\cite{art96})
    potential as a function of the equatorial  
    radius $d$ (Eq.~\ref{eq:vkepA}).
    The inner vacuum boundary of the computational grid,
    is located at the arithmetic mean of the event horizon and
    innermost stable circular orbit (ISCO) of the central black hole.
    {\em Right panels:} The dots give the radial velocities $v_r(d)$
    (normalised to the speed of light) as a function of the equatorial
    radius $d$ for all grid zones in the equatorial plane
    of the reference Model~r00-64 and the high disk-viscosity
    Model~al4-64, at the same times.
    The solid lines represent the mean values of all zones within
    binning intervals of 3$\,$km. }
\label{fig:rphi}
\end{figure*}

\begin{figure*}[htp!]
\begin{tabular}{cc}
  \psfig{file=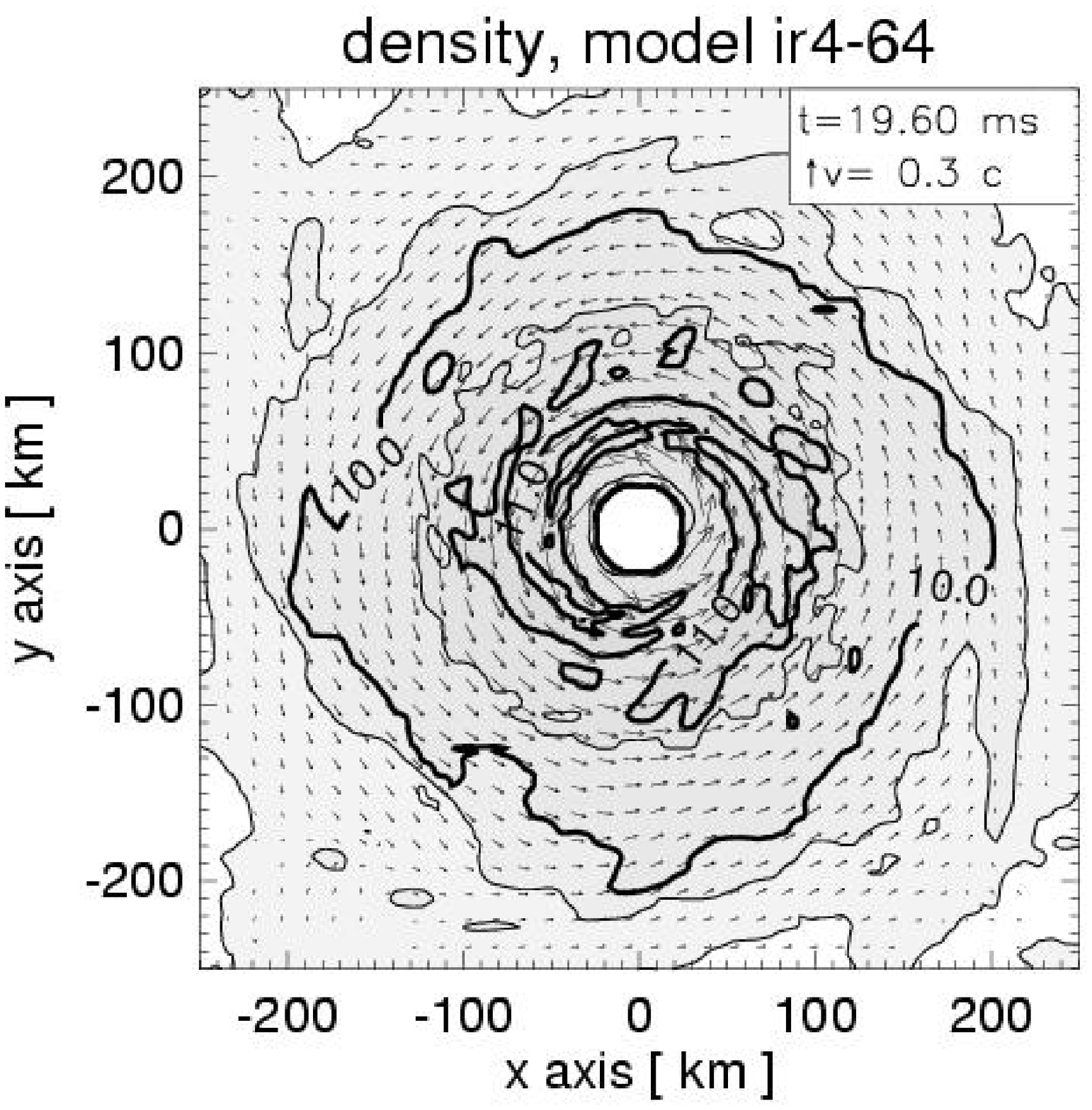,width=8.2cm} &
  \psfig{file=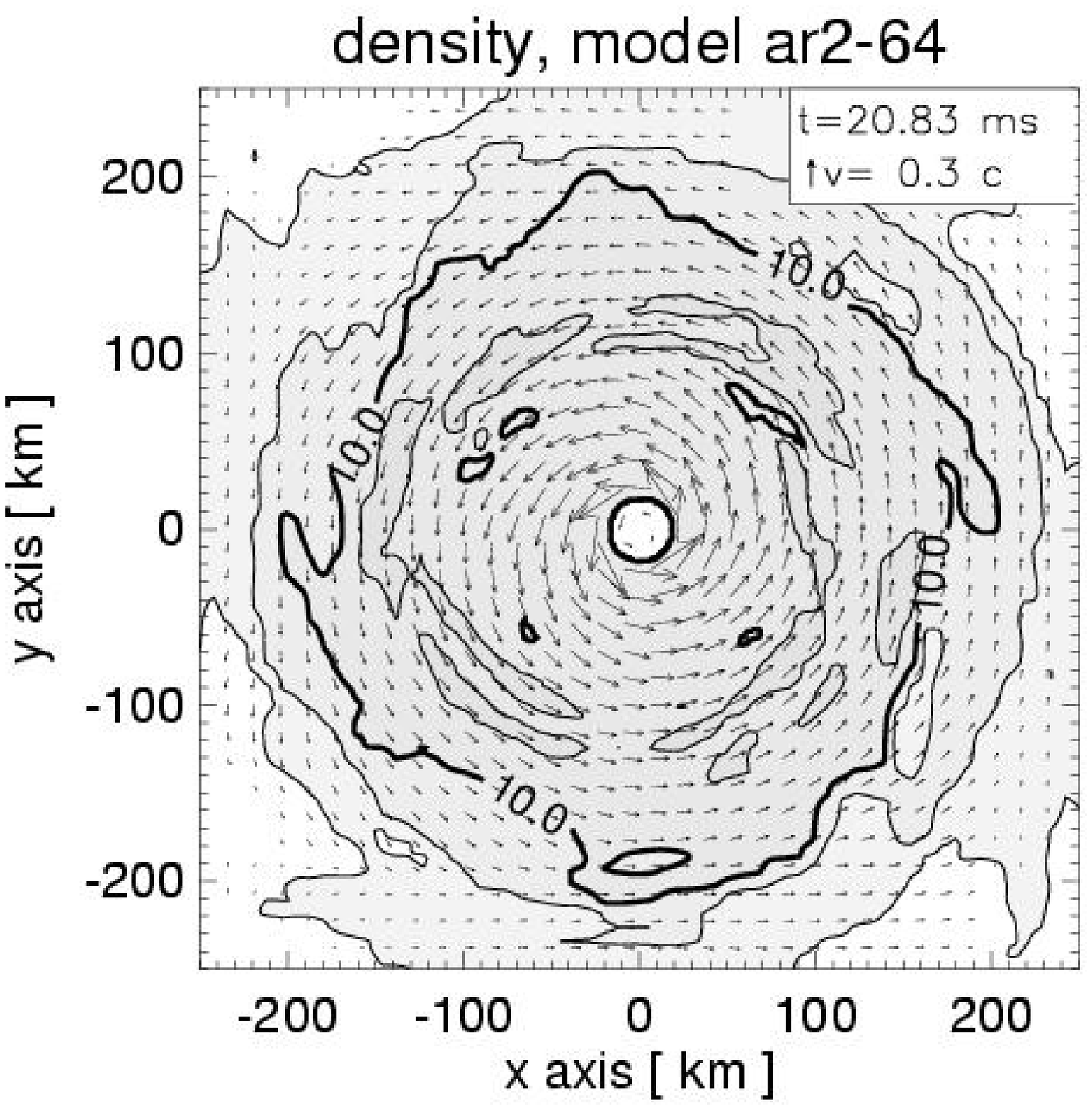,width=8.2cm} \\[-2ex]
  \psfig{file=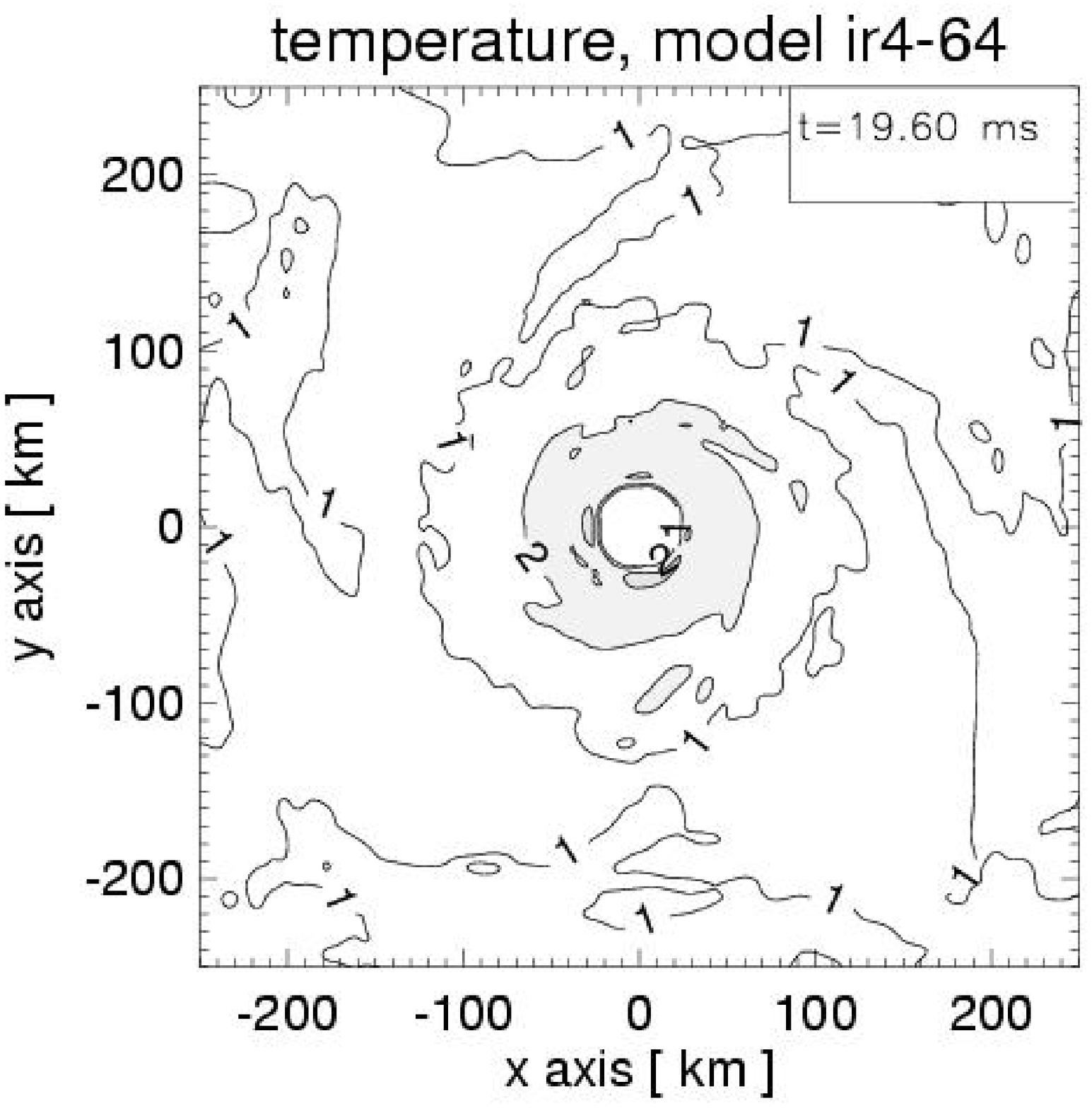,width=8.2cm} &
  \psfig{file=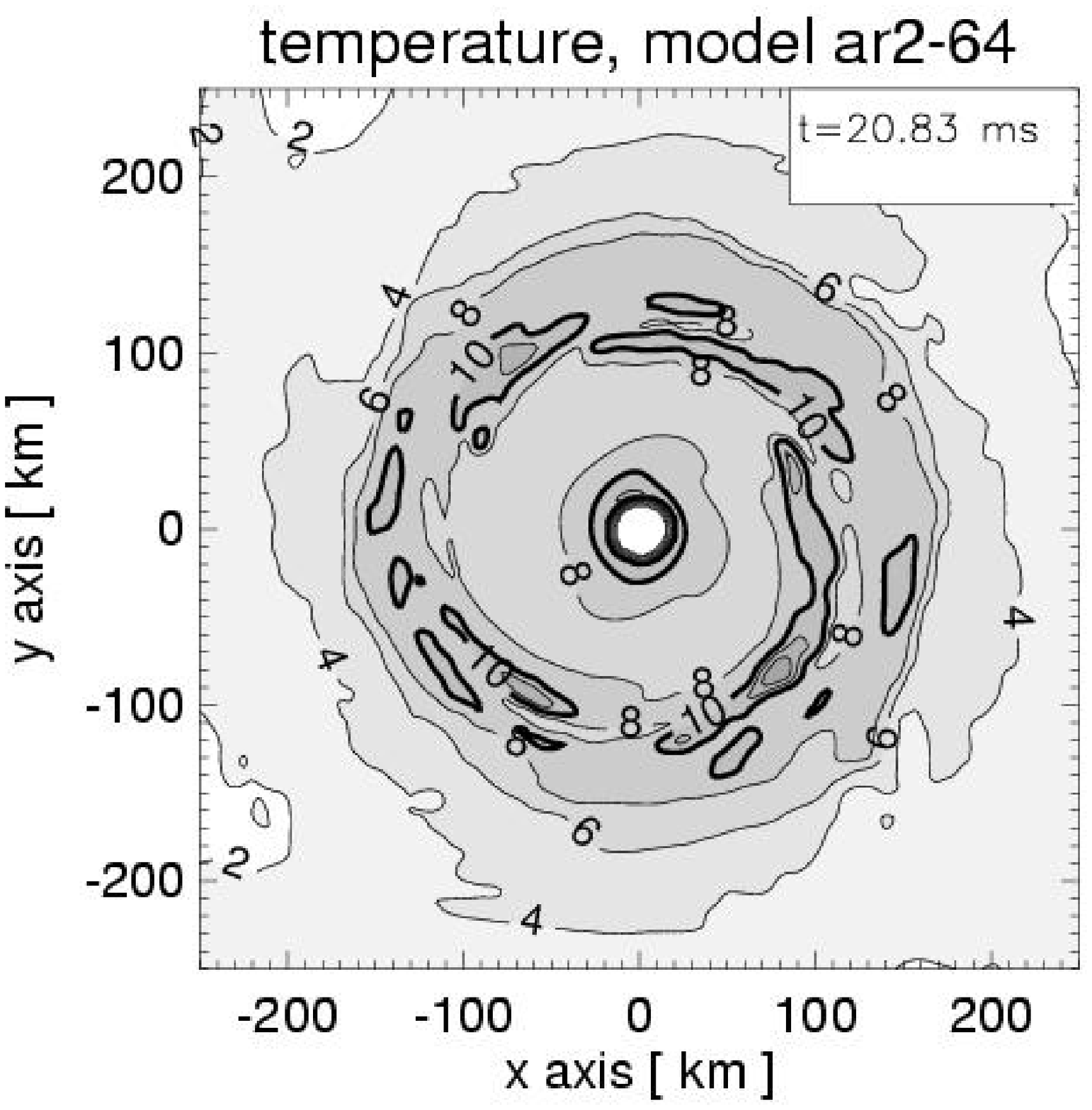,width=8.2cm} \\[-2ex]
\end{tabular}
\caption[]{
Density and temperature distribution for high-mass torus models with
nonrotating black hole and without disk viscosity (ir4-64; left) and
with corotating black hole and disk viscosity $\alpha = 0.1$ 
(ar2-64; right) at about 20$\,$ms after the start of the
simulations.
The arrows indicate the velocity field. The density
is given in g$\,$cm$^{-3}$ with
contours spaced logarithmically in steps of 0.5~dex.
The temperature is measured in MeV, its contours are labelled with
the corresponding values.
\label{fig:III2}
}
\end{figure*}

\subsection{Influence of viscosity and varied disk mass}

The distributions of density, velocity, and temperature
for Models al3 and al4 are shown in Fig.~\ref{fig:2a}. 
These have the same parameters as
Model r00 except for an increased viscosity (see Table~\ref{tab:models3}).
While both models show, in Fig.~\ref{fig:8bnew}, a clear increase in
disk temperature with increasing viscosity, the spatial distribution is
different: Model al3 with the small value of the viscosity shows high
temperatures between 4$\,$MeV and 8$\,$MeV only within a central
region of radius 100$\,$km (Fig.~\ref{fig:2a}), whereas Model al4 
has 4$\,$MeV and higher in the whole extended disk.
Since the pressure increases with temperature, this heating by the
increased viscosity expands the disk and reduces its density
(Fig.~\ref{fig:2a}).
Assuming otherwise equal velocities, a reduced density implies a
reduced mass accretion rate.
However, the large-scale effect of viscosity is to
redistribute angular momentum from the centre of the disk out to the
periphery, thus allowing mass to drift inward and be accreted by the
black hole.
The interplay of these two effects, the increase of the accretion
due to viscosity on large scales and the decrease of the accretion rate
due to a local decrease of the density in the innermost parts of the 
disk, can be observed in
Fig.~\ref{fig:3a} for Models~al3 and al4.
In Model al3 the modest viscosity affects only the inner, most rapidly
orbiting regions of the disk where shear effects are largest.
The associated local central heating raises the thermal pressure and
thus conspires to reduce the mass accretion rate.
When the viscosity is further increased as done in Model~al4, the global
viscosity effect dominates, and the mass accretion rate increases
beyond the values of Model~r00 where no viscosity was added. As a
consequence, the disk mass drops more rapidly than in the other models
(Fig.~\ref{fig:3a}), leading to a clearly lower maximum density in the
disk (Fig.~\ref{fig:8bnew}). The maximum disk temperature, on the other
hand, is clearly largest in Model~al4 (Fig.~\ref{fig:8bnew}). 

The model pair ir4 and ir5, which also only differs in the amount of
added viscosity, shows the same relationship with respect to
the mass accretion rate as Models~r00 and al3:
In Fig.~\ref{fig:3a} 
one can see a smaller mass accretion for the
model with higher viscosity (ir5).
The reasoning is similar to what has been described above: The
temperatures are higher for
Model~ir5, but the densities distinctly lower.
It seems that for tori with higher mass, which is the case for Models ir4 
and ir5
relative to Models~r00, al3, and al4, the onset of the full effect of
the increase of the mass accretion rate due to global angular momentum
redistribution is offset to even higher values of the viscosity
parameter.
Increasing the disk mass by a factor of 4 does not produce any
other major changes in flow dynamics.
However the larger densities do have a large effect on the emission of
neutrinos as will be described further below.

\subsection{Influence of black hole rotation and of a more compact
gravitational potential}

We mimic some effects of relativistic gravity by using a
Paczy\'nski-Wiita~(\cite{pac80}) potential and its extension to
rotating black holes by Artemova et al.~(\cite{art96}).
Varying the spin parameter between negative (counter-rotating black hole)
and positive (black hole corotating with disk) values, the radius of 
the horizon changes and the ISCO is shifted to smaller values compared
to the case of a black hole without rotation (Fig.~\ref{fig:Artebeta}).
Looking at the central circle
in Fig.~\ref{fig:II2} as compared to,
e.g., Model r00, one sees this slight shrinking
of the inner grid boundary (at the arithmetic mean of $r_{\mathrm{H}}$
and $r_{\mathrm{ISCO}}$) when the black hole rotates in the
prograde direction. Models~ro2 and ro5 can be directly compared to
r00, the only difference being the presence of rotation. The same
applies to the pair al4 and ar1.

No major difference can be seen in the density or temperature
distributions, except at the very centre close to the black hole. The
temperatures for the rotating black holes are about 2$\,$MeV higher than
in the non-rotating cases
The counter-rotating Model~ro5 retains the low temperatures of the
reference Model~r00, i.e.~around 2$\,$MeV. In case of the viscous tori with 
corotating black hole the higher temperatures in the close vicinity of the
black hole correlate with a distinct narrowing of the accretion torus in the
vertical ($z$) direction.

The combined increase of central temperatures brought about by
viscosity and rotation has a clear impact on the velocities in the
direct vicinity of the black hole.
Fig.~\ref{fig:rphi} shows separately the azimuthal and radial
velocities in the equatorial plane versus the distance from the
centre of the black hole.
Each zone is represented by a dot, and the spread of the points at a
given radius reflects the deviations from rotational symmetry.
For an axially symmetric configuration all dots at a specific radius
would cluster on top of each other.

Comparing the radial velocities of Models~al4 and r00 (top and bottom
right panels in Fig.~\ref{fig:rphi}), one sees
that the large viscosity produces the inward facing accretion flow,
i.e.~negative velocities, at larger radii, inside of 60$\,$km for
Model~al4 as opposed to 40$\,$km for Model~r00.
This reflects the canonical viscosity effect of globally shifting
angular momentum outward allowing mass to fall inward.
The added effect in Model~ar1 of narrowing the potential well with a
rotating black hole does not change this situation significantly.
Note that the outer parts of the tori show slow expansion at the
displayed times, indicating ongoing relaxation as well as the
consequence of outward transport of angular momentum. 

The azimuthal velocities in Model~r00 closely follow the Keplerian
values for the Paczy\'nski-Wiita potential, see the top left panel in
Fig.~\ref{fig:rphi}.
Although the added viscosity in Model~al4 (bottom panel) increases the
temperature in the disk, the pressure change due to this increase is
not sufficient to appreciably change the velocity distribution.
Only when black hole rotation is added, with its additional temperature
increase and corresponding pressure increase in the vicinity of the 
black hole, 
does the matter in the disk become more strongly pressure supported within
a radius of 50$\,$km. The azimuthal velocities are significantly
below the Keplerian values in this region, as given by
Eq.~(\ref{eq:vkepA}) (with the radius $r$ replaced by the equatorial
distance $d$) for the Artemova-Bj\"ornsson-Novikov potential.

In all four pairs of models, r00/ro2, ri4/ir4, al4/ar1, and ir5/ar2 
(see Figs.~\ref{fig:III2} for snapshots of
the most massive torus Models~ir4 and ar2 at $t\approx 20\,$ms)
the first member of the pair is the non-rotating case, while the second
contains a corotating black hole. All other parameters are kept equal.
Examining Table~\ref{tab:models} one notices that for each case
the rotating model has a higher disk mass than the
corresponding non-rotating model at the end of the simulated evolution
(cf.\ also Fig.~\ref{fig:3a}). This is a strong indication that a
corotating black hole accretes matter from the initial tori less
quickly, because the angular momentum at the ISCO is lower and a rotating
black hole thus allows matter with lower angular momentum to remain
on orbits. Moreover, the higher temperatures close to the black hole
provide pressure support and additionally stabilise the torus.
A confirmation is found by looking at the counter-rotating case ro5,
which can be compared to Models~r00 and ro2. Indeed, the disk of 
Model~ro5 has an even smaller mass than that of Model~r00.

\begin{figure*}[htp!]
\begin{tabular}{cc}
  \psfig{file=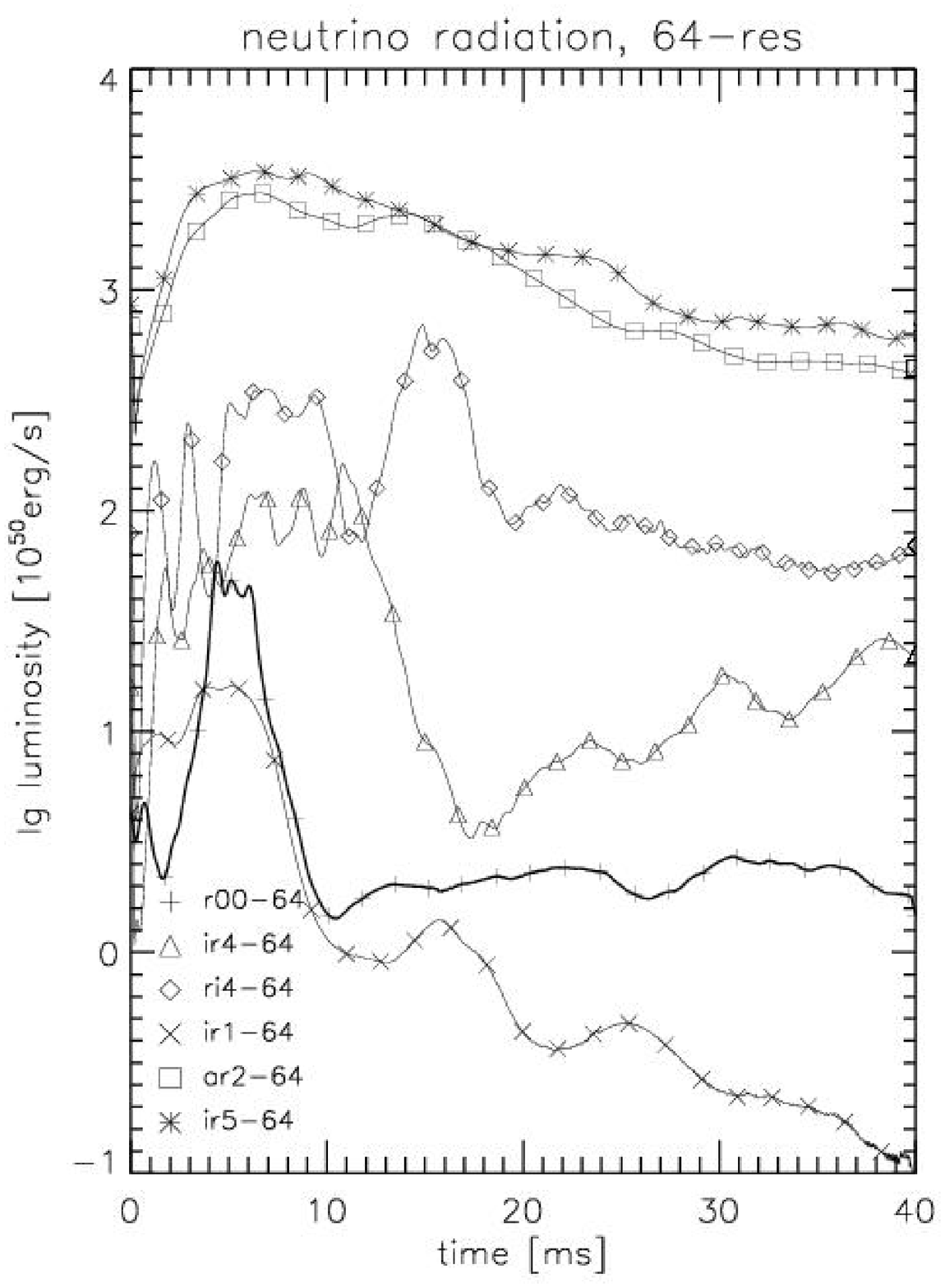,width=8.8cm} &
  \psfig{file=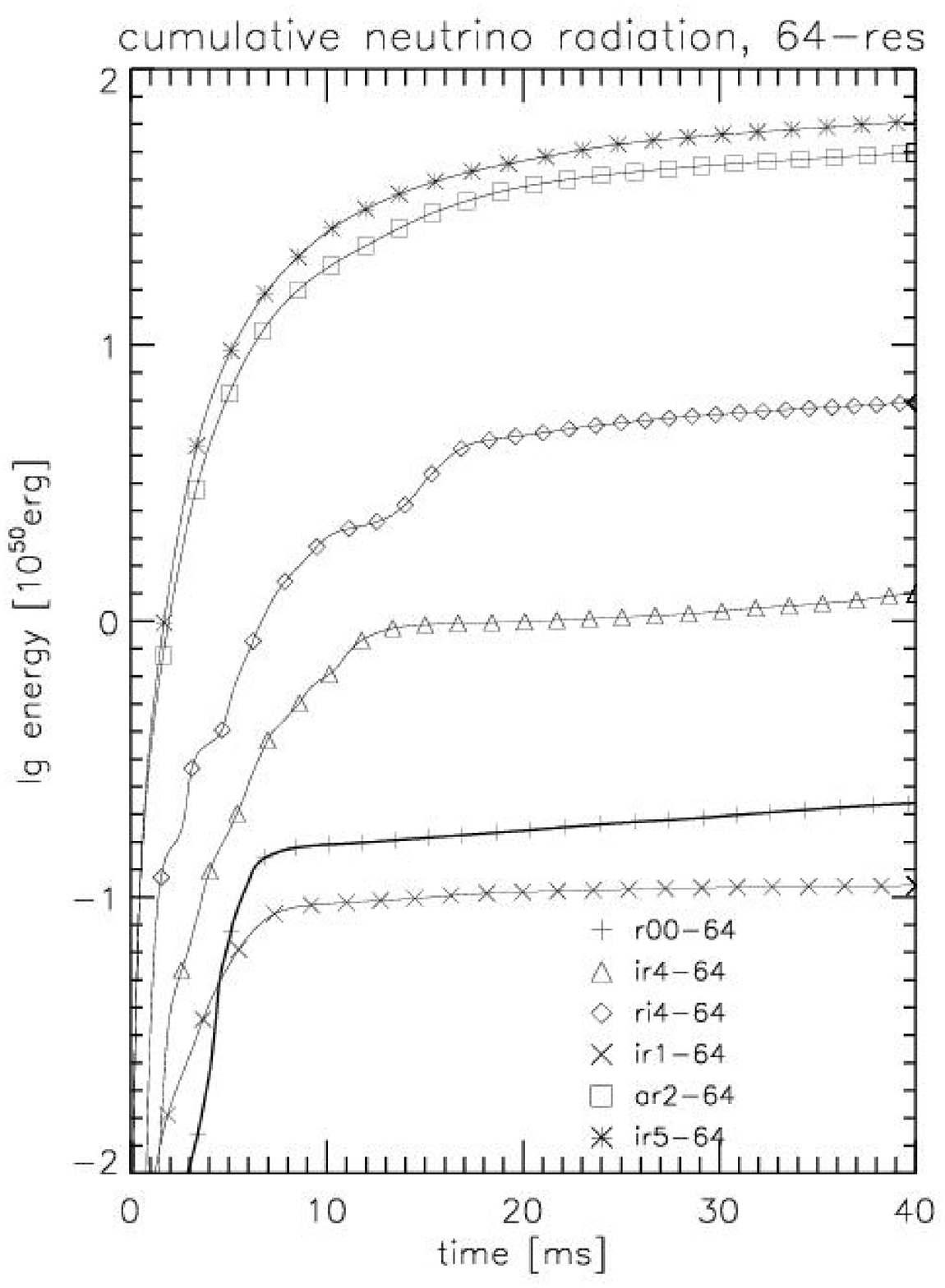,width=8.8cm} \\[-2ex]
\end{tabular}
  \caption[]{Total neutrino luminosities (left) and cumulative energy
   radiated in neutrinos (right) as functions of time for the 
   reference Model~r00-64, the low-mass torus Model~ir1-64, and the
   high-mass torus Models ir4-64, ir5-64, ri4-64, and ar2-64.}
\label{fig:III45a}
\end{figure*}

\subsection{Neutrino emission}

The neutrino luminosities and cumulative energy emitted in neutrinos as
functions of time for a sample of 64-resolution models are shown in
Fig.~\ref{fig:III45a}.
Note that electron neutrinos, $\nu_e$, and electron antineutrinos, 
$\bar\nu_e$, dominate the emission by a large factor, because at 
the density and temperature conditions in the accretion tori 
their production rate by charged-current 
absorption of electrons and positrons on free protons and neutrons,
respectively, 
\begin{eqnarray}
e^- + p  &\longrightarrow& n + \nu_e\ ,
\label{eq:reac1} \\
e^+ + n  &\longrightarrow& p + \bar\nu_e \ ,
\label{eq:reac2}
\end{eqnarray}
is much higher than the production rate of 
neutrino-antineutrino pairs of all flavors by thermal neutrino
processes and nucleon-nucleon bremsstrahlung (see also 
Ruffert \& Janka~\cite{ruffert}, Rosswog \&
Liebend\"{o}rfer~\cite{ros03}). 

\begin{figure*}[htp!]
\begin{tabular}{cc}
  \psfig{file=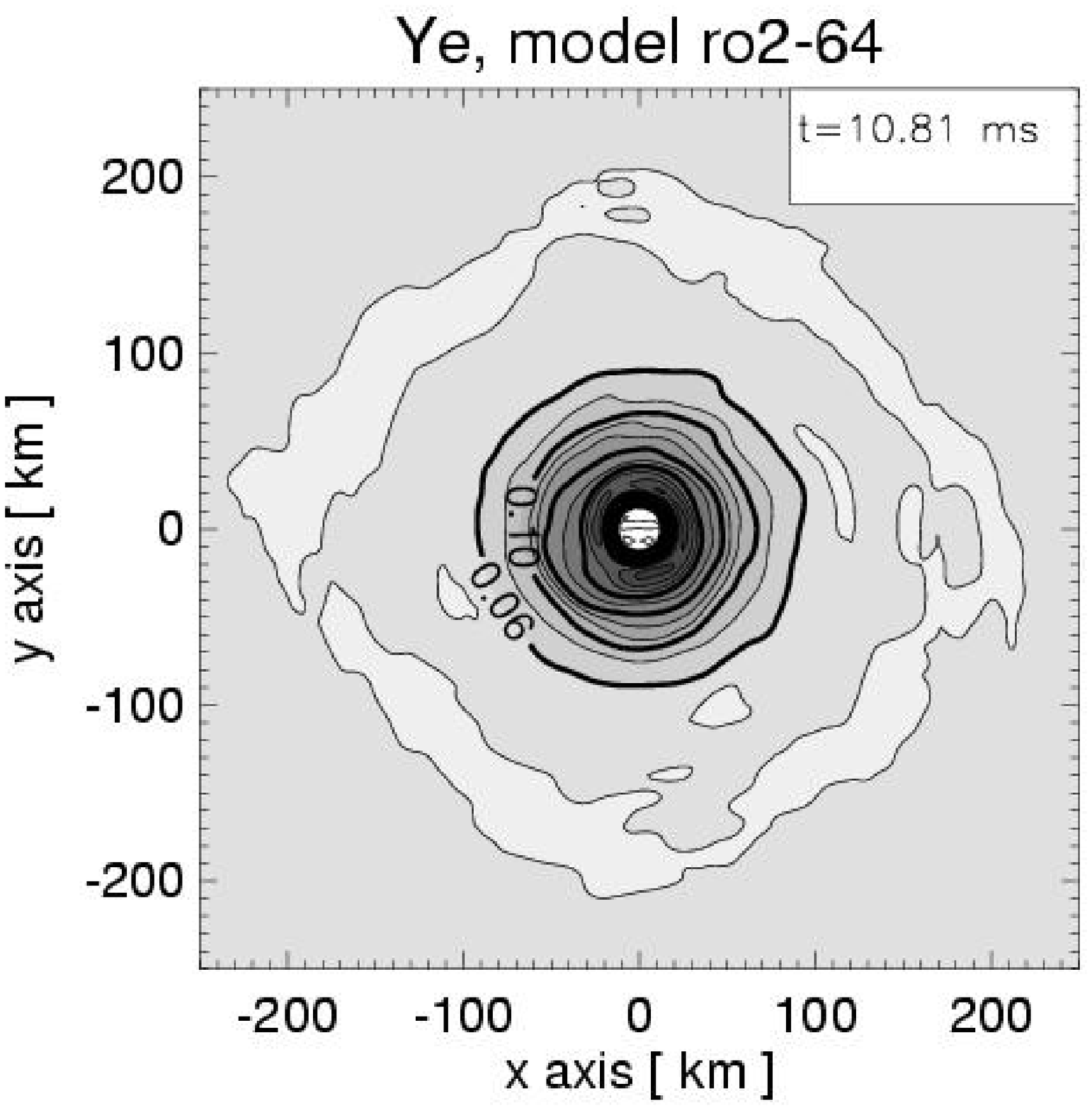,width=8.2cm} &
  \psfig{file=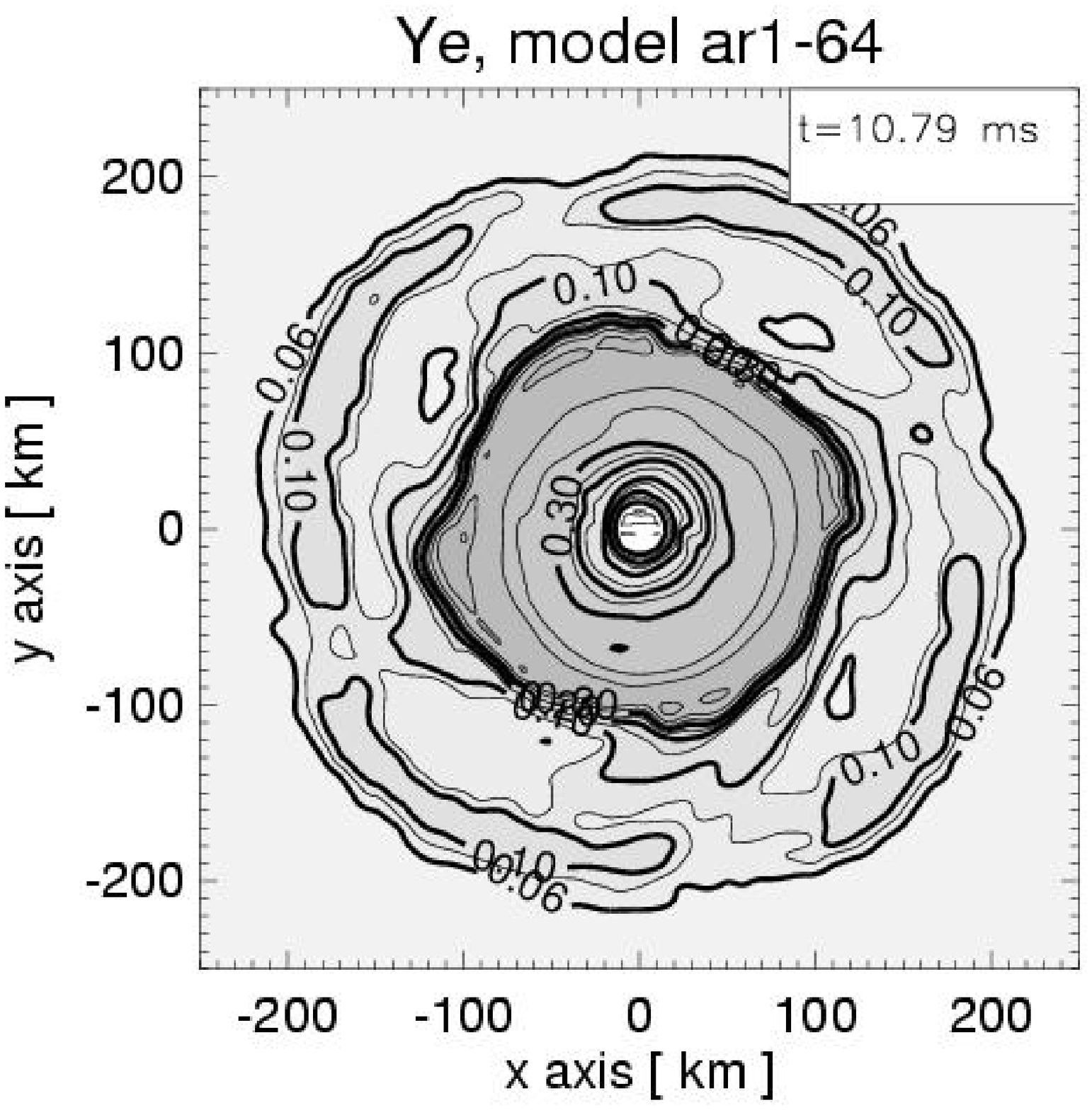,width=8.2cm} \\[-2ex]
  \psfig{file=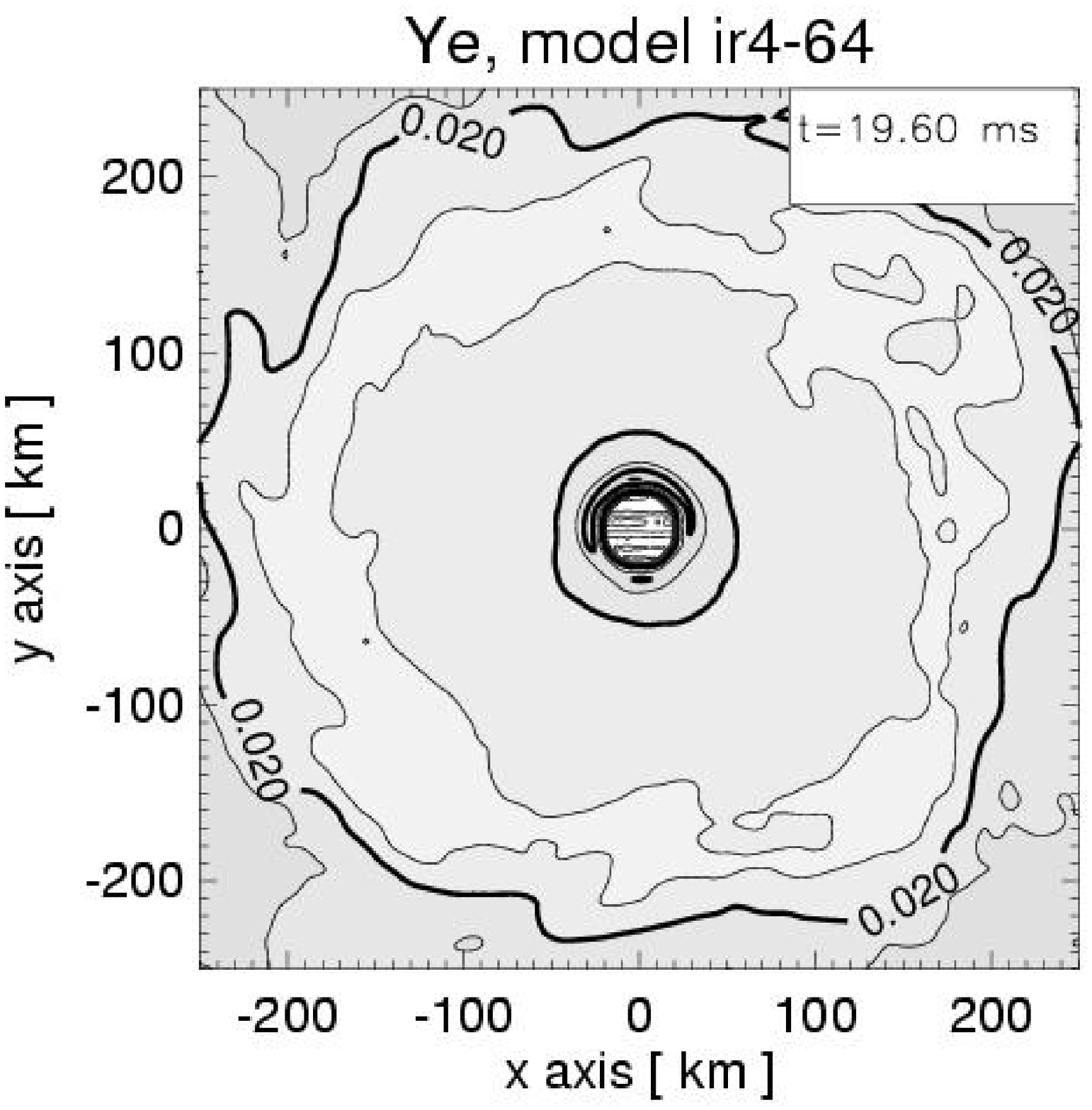,width=8.2cm} &
  \psfig{file=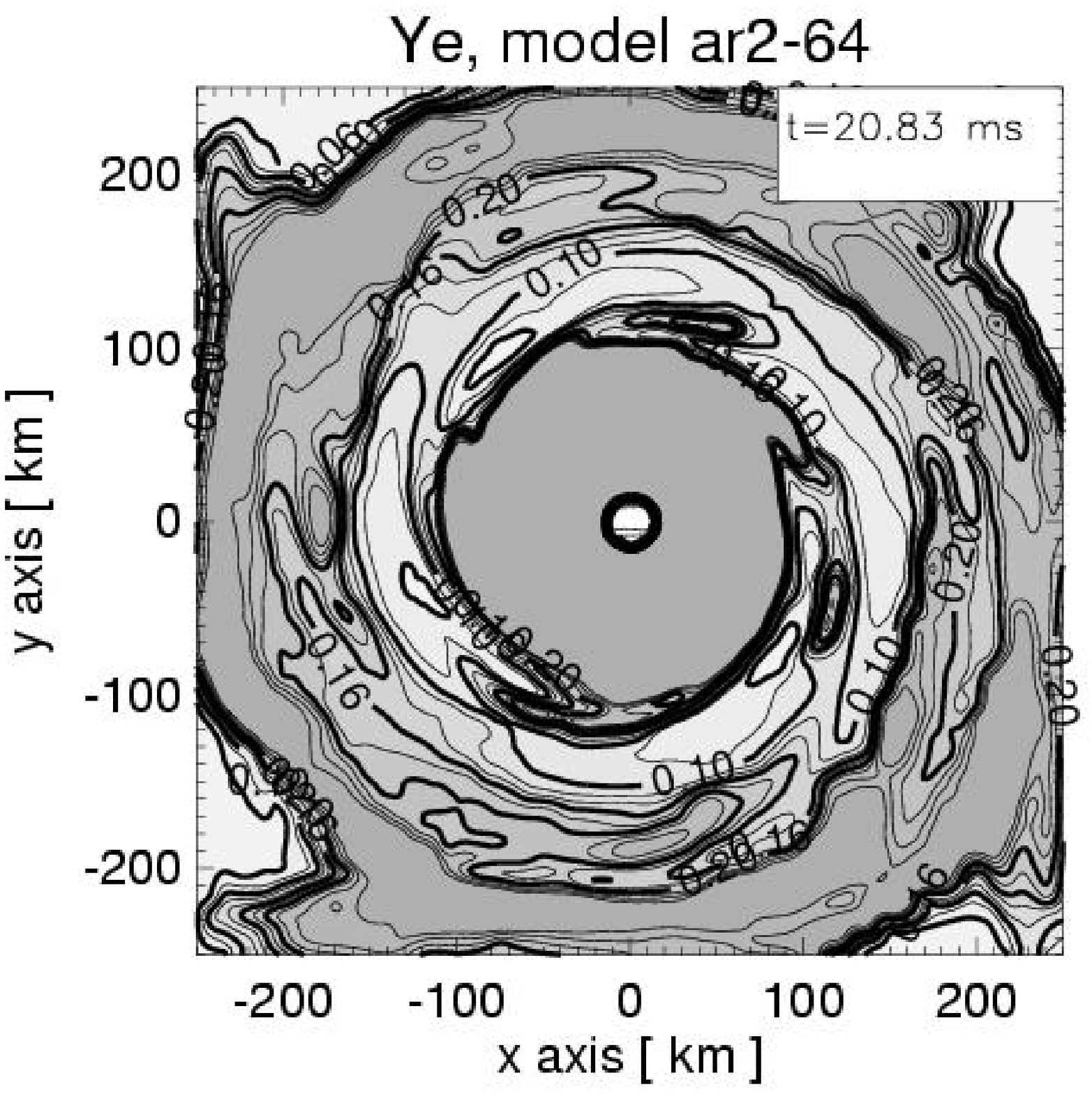,width=8.2cm} \\[-2ex]
\end{tabular}
\caption[]{
Electron fraction in the orbital plane for Models~ro2-64, ar1-64, 
ir4-64 and ar2-64, in the orbital plane 
at about 11$\,$ms after the start of the simulations. 
The contours are spaced linearly in steps of 0.02.
\label{fig:II1a}
}
\end{figure*}

\begin{figure*}[htp!]
\begin{tabular}{cc}
  \psfig{file=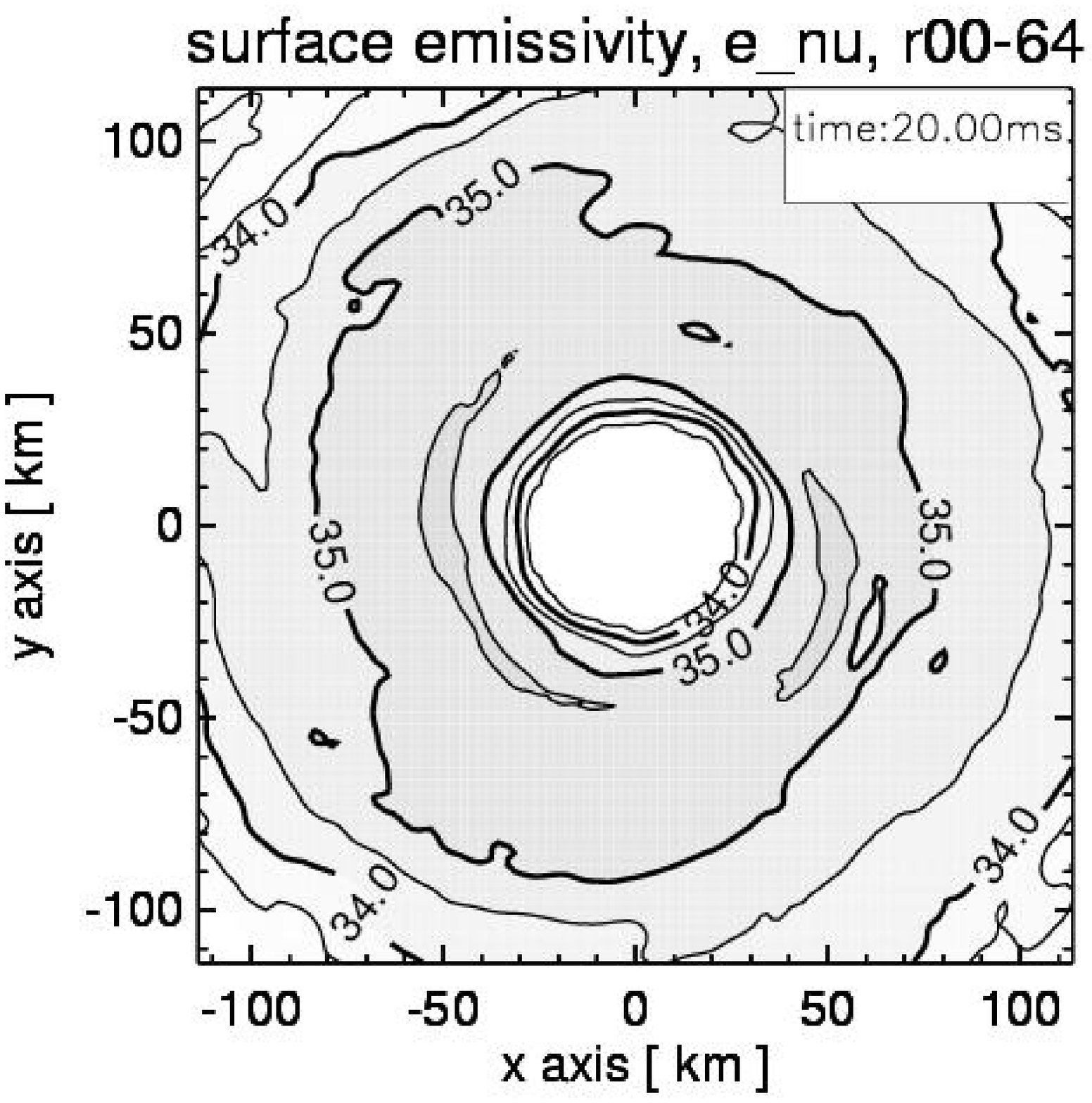,width=8.2cm} &
  \psfig{file=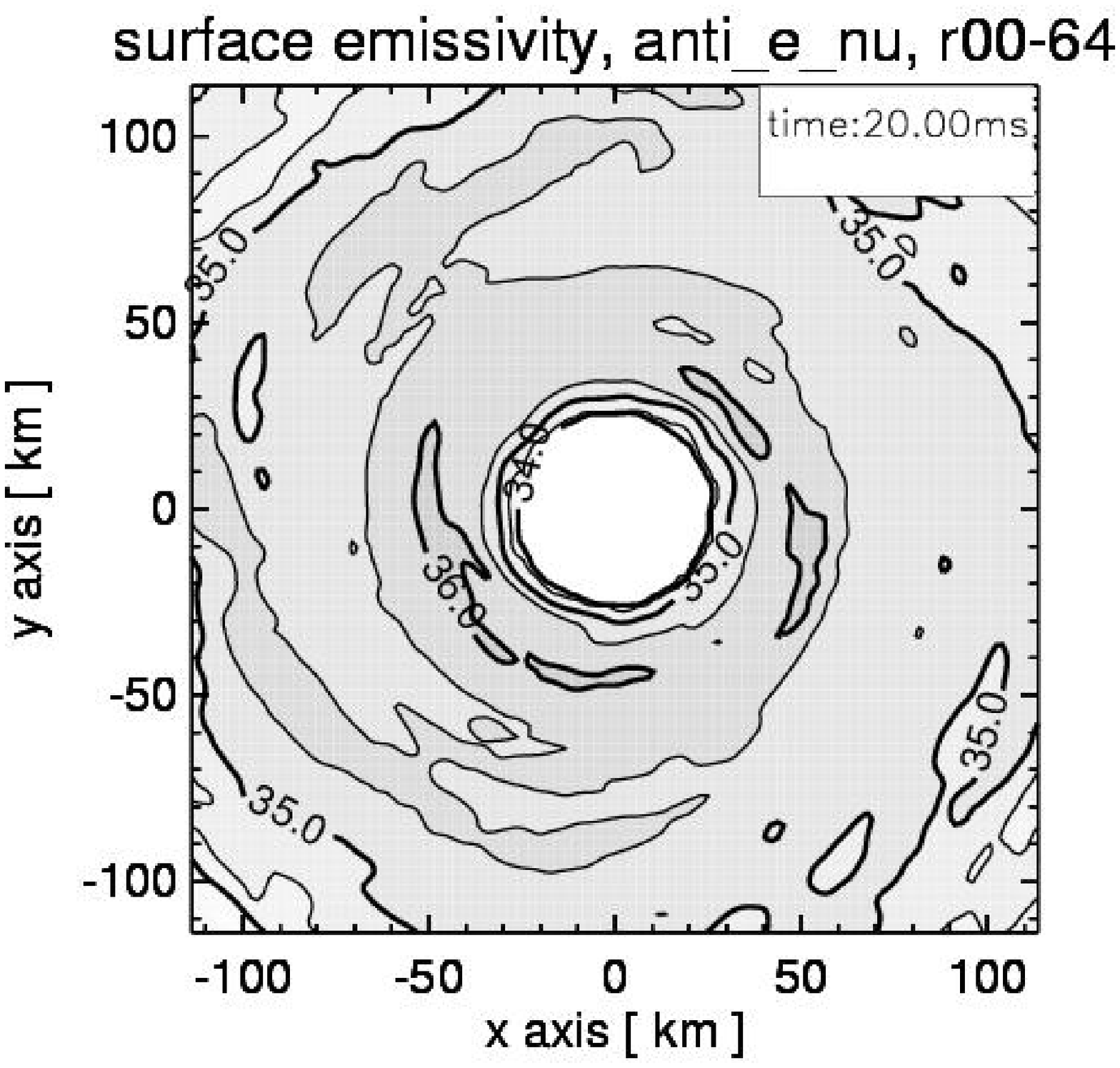,width=8.2cm} \\[-2ex]
  \psfig{file=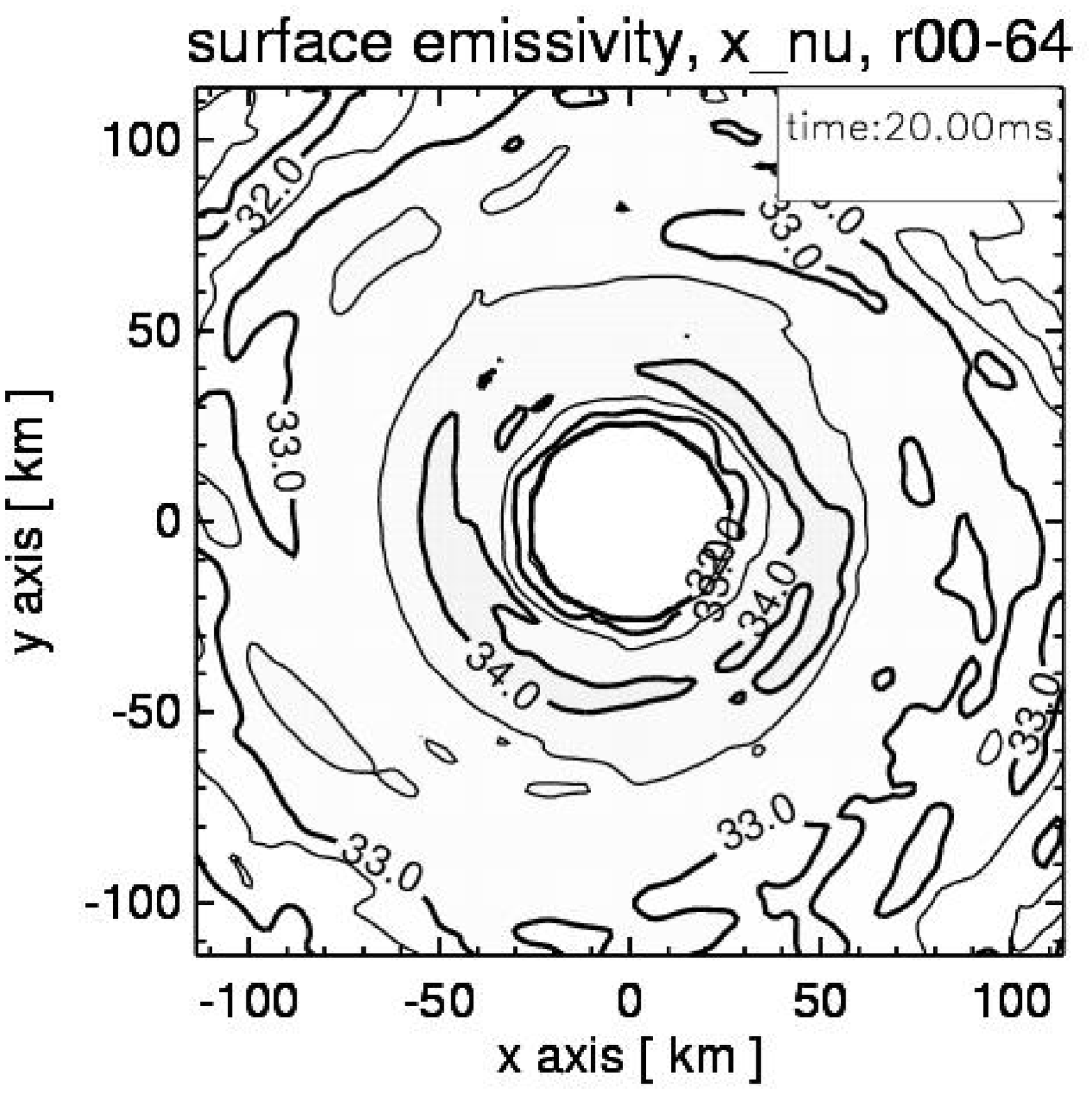,width=8.2cm} &
  \psfig{file=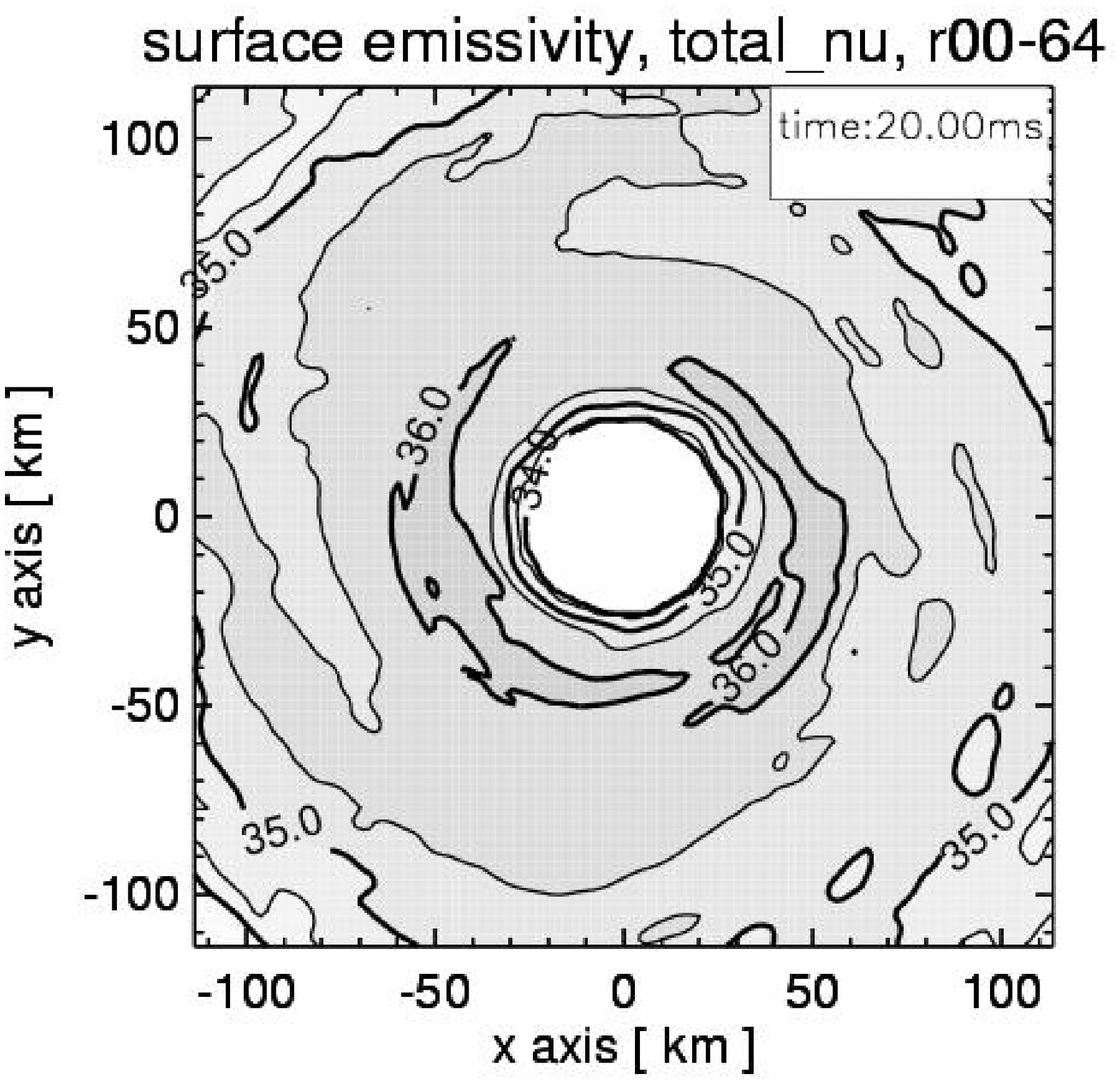,width=8.2cm} \\[-2ex]
\end{tabular}
\caption[]{
Neutrino energy loss rates per unit area in the orbital plane for
Model~r00-64 at 20$\,$ms after the start of the simulations.
The plotted values show the logarithm of the rates in 
erg$\,$cm$^{-2}$s$^{-1}$, obtained
by integration of the local energy loss rates per unit of volume 
from $z = 0$ to infinity. The top left panel gives the results for
electron neutrinos, the top right panel for electron antineutrinos,
the lower left panel for the sum of muon and tau neutrinos and 
antineutrinos, and the lower right panel the total values for
neutrinos and antineutrinos of all flavors. The contours are
spaced in steps of 0.5~dex, bold lines are labelled with
their corresponding values.
The grey shading emphasises the emission levels, dark grey
representing the largest energy loss by neutrino emission.
\label{fig:16n1}
}
\end{figure*}

The $\bar\nu_e$ luminosities are initially much larger than the
$\nu_e$ luminosities and at the end of the calculated evolution 
still typically 2--3 times higher (see Table~\ref{tab:models}).
The tori consist of decompressed neutron star matter, which 
inside of the neutron star was initially in a very neutron-rich
($Y_e\la 0.1$) state of neutrinoless beta equilibrium 
(the initial setup chosen for our simulations was adapted to neutron
star merger results and therefore still  
reflects this origin of the torus matter, cf.\ Fig.~\ref{fig:1}).
Chemical or kinetic equilibrium at densities much below 
nuclear matter corresponds
to larger values of $Y_e$ and a higher proton abundance.
At decompression, matter tries to approach this new equilibrium
state, in which the production of $\nu_e$ and $\bar\nu_e$  
are in balance. During this ``protonization'' 
electron antineutrinos are emitted in larger numbers than $\nu_e$.
The consequences of this process are visible in two ways. First,
comparing the $Y_e$ distribution in Fig.~\ref{fig:II1a} after 11$\,$ms
and 20$\,$ms of evolution with the initial state shown in
Fig.~\ref{fig:1} one sees 
significantly higher values of $Y_e$, i.e.~more protons.
Secondly, in Fig.~\ref{fig:16n1}, the comparison of the local energy
emission rates per unit area or volume, of $\nu_e$, $\bar\nu_e$, and
heavy-lepton  neutrinos plus antineutrinos (``$\nu_x$'' for the sum of
muon and tau neutrinos, $\nu_{\mu}$, $\bar\nu_{\mu}$, $\nu_{\tau}$, and 
$\bar\nu_{\tau}$), respectively, shows that the $\bar\nu_e$ emission
is significantly higher than that of $\nu_e$.
The protonization is faster and leads to higher 
values of $Y_e$ at the end of our simulations in case of hotter and
less dense tori; this trend is very similar for larger viscosity
and black hole rotation (Models~ro2, ar1, ir4 and~ar2 in
Fig.~\ref{fig:II1a}). 
All tori
are at most marginally opaque to neutrinos in their denser parts and
the diffusion times short in most regions. The $Y_e$ evolution is
therefore governed by the capture rates directly. Since 
positron captures in reaction~(\ref{eq:reac2}) increase
extremely steeply with rising temperature, the highest values of
$Y_e$ develop in regions of highest temperatures.

The mean energies of the radiated neutrinos ($\left\langle\epsilon_{\nu_e}
\right\rangle$, $\left\langle\epsilon_{\bar\nu_e}\right\rangle$, and
$\left\langle\epsilon_{\nu_x}\right\rangle$) are defined as the ratio of
integral energy loss (rate) to total number loss (rate) in neutrinos of a
given kind (see Ruffert et al.~\cite{ruffert1,ruffert2}). 
They correlate with the 
average and maximum torus temperatures as visible from the parallel trends of
$T_{\mathrm{max}}$ and the mean neutrino energies in Table~\ref{tab:models}.
Electron neutrinos and antineutrinos are radiated away with mean energies
of 10--20$\,$MeV and are similarly energetic, because
electron degeneracy plays a minor role and the cross sections of both
production reactions scale equally with particle energies;
moreover, reabsorption of neutrinos is of minor importance because
the tori are essentially transparent to neutrinos.
Heavy-lepton neutrinos, on the other hand, tend to have a slightly
lower mean energy except in cases with the largest torus masses
(Models~ri4, ir4, ir5, and ar2), which also tend to be among the
hottest models. Only in the two cases with the highest torus temperatures
and shear viscosity parameter $\alpha = 0.1$ (Models~ir5 and ar2) 
do muon and tau neutrinos yield a non-negligible
contribution to the total energy loss in neutrinos.

The results summarised in Tables~\ref{tab:models} and 
\ref{tab:efficiencies} and Fig.~\ref{fig:III45a} show that
the neutrino luminosities rise steeply with torus mass. In
simulations without shear viscosity ($\alpha = 0$) we find a
roughly quadratic increase, $L_{\nu}\propto M^2$ (compare
Model~r00-64 with ir4-64 and ro2-64 with ri4-64),
whereas with shear viscosity $\alpha = 0.1$ 
a certain saturation can be observed and the luminosities increase
only slightly steeper than linearly, $L_{\nu}\propto M^{1+\xi}$
for $\xi > 0$ (Models al4-64 vs.\ ir5-64 and ar1-64 vs.\ ar2-64).
Shear viscosity has the most dramatic influence on the neutrino
emission: the luminosities grow by a factor of 30--40 when 
$\alpha = 0.1$ instead of $\alpha = 0$ (Models~r00-64 vs.\ al4-64; 
ir4-64 vs.\ ir5-64). In contrast, a black
hole with a spin parameter up to $a\sim 0.6$ in corotation with the
torus has a much weaker influence on the neutrino emission. 
While for calculations
without shear viscosity ($\alpha = 0$) the neutrino luminosities
increase at most by a factor of 2--3 (compare Model~r00-64 with ro2-64
and Model~ri4-64 with ir4-64 in Table~\ref{tab:models} and
Fig.~\ref{fig:III45a}), an additional increase of the neutrino 
emission is hardly visible or absent for cases with $\alpha = 0.1$ 
(compare Models~al4-64 with ar1-64 and Models~ir5-64 with ar2-64).
Black hole rotation with more extreme spin parameter, however, 
leads to a large boost of the neutrino emission as can be seen from
Models~ro2-32 with $a = 0.6$ and ro3-32 with $a = 0.8$ in 
Table~\ref{tab:models}. Despite of nearly equal torus masses, the
latter model has a 9 times higher neutrino luminosity.

The series of models with reduced resolution exhibits the tendency
of slightly lower torus temperatures compared to the corresponding
simulations with finer grid zoning at the same time of the evolution.
The neutrino luminosities and mean energies, however, show no clear
trend and can be higher or lower, depending mostly on whether the 
torus has a larger or smaller mass than in the better resolved
calculations. This sensitivity together with the fact that the
less massive tori have lost most of their mass to the black hole 
after an evolution of $70\,$ms (Table~\ref{tab:models}), disfavours
the late-time models with lower resolution for a quantitative 
comparison of the
influence of viscosity and black hole rotation. The neutrino 
emission during these late stages has dropped to a low level in 
most cases and is therefore mostly dependent on the
torus mass which is left around the black hole 
(see Table~\ref{tab:models}). The general trends associated with
viscosity, torus mass, and black hole rotation, however, can also
be seen in the 32-resolution models.
 
The upper left panels of Fig.~\ref{fig:IIIeff64} and right panel of 
Fig.~\ref{fig:eff64} provide information about the instantaneous
conversion efficiency of rest-mass energy to neutrino-antineutrino
energy, 
\begin{equation}
q_{\nu}\ \equiv\ {L_{\nu}\over \dot M\,c^2}\ , 
\label{eq:acceff}
\end{equation}
for a selection of 64-resolution models as function of time.
It is obvious that the values are extremely low ($q_{\nu}\ll 1$\%)
in case of zero physical shear viscosity ($\alpha = 0$) and 
become of order 1\% only for large torus masses ($M_{\mathrm{d}}
> 0.1\,M_\odot$ as in case of Model~ri4-64, cf.\ 
Table~\ref{tab:efficiencies}). For $\alpha = 0.1$ the efficiencies
reach several percent (e.g.~Model~ar1) and 
Models~ar2 and ir5 in Fig.~\ref{fig:IIIeff64}). It is interesting
that $q_{\nu}$ for the low-mass torus Model~ar1 reaches its
maximum at early times and then decreases towards the end of our
simulations, whereas the values are stable or even increase over
the simulated period of evolution in Models~ir5 and ar2.

\begin{table}
\caption[]{
Torus mass $M_{\rm d}$, black hole mass accretion rate $\dot M_{\rm d}$,
estimated accretion time scale of the torus,
$t_{\rm acc}\equiv M_{\rm d}/\dot M_{\rm d}$,
total neutrino luminosity $L_\nu$, integral rate of energy
deposition by neutrino-antineutrino annihilation around the accretion
torus, $\dot{E}_{\nu\bar{\nu}}$, and
total energy deposition by $\nu\bar\nu$ annihilation in the time interval
$t_{\rm fin}$, $E_{\nu\bar{\nu}}$, for all 64-resolution models.
All quantities are given at time $t_{\rm fin}$ after the start of
the simulation.
}
\begin{flushleft}
\tabcolsep=1.0mm
\begin{tabular}{lcccccccc}
\hline\\[-3mm]
Model & $t_{\rm fin}$ & $M_{\rm d}$ & $\dot M_{\rm d}$
      & $t_{\rm acc}$ & $L_\nu$       & $\dot{E}_{\nu\bar{\nu}}$
                      & $E_{\nu\bar{\nu}}$ \\
 & ms &
   {\scriptsize$10^{-2}M_\odot$} & {\scriptsize$M_\odot\,{\rm s}^{-1}$} &
   ms &
   {\scriptsize$10^{50}\frac{\rm erg}{\rm s}$} &
   {\scriptsize$10^{50}\frac{\rm erg}{\rm s}$} &
   {\scriptsize$10^{50}{\rm erg}$}
\\[0.3ex] \hline\\[-3mm]
r00-64   & 40.0 & 2.22 & 0.29 & 77. &  1.8 & 1.8$\cdot 10^{-4}$  & 4.4$\cdot 10^{-4}$  \\
al3-64   & 40.0 & 2.44 & 0.25 & 98. & 11.4 & 8$\cdot 10^{-3}$    & 3.8$\cdot 10^{-3}$  \\
al4-64   & 40.0 & 1.76 & 0.38 & 46. & 80.0 & 0.4                 & 0.26                \\
ri4-64   & 40.0 &12.22 & 0.41 &298. & 70.0 & 0.35                & 0.13                \\
ro2-64   & 40.0 & 3.10 & 0.16 &194. &  3.4 & 7.2$\cdot 10^{-4}$  & 3.3$\cdot 10^{-3}$  \\
ro5-64   & 40.0 & 1.13 & 0.25 & 45. &  1.0 & 3$\cdot 10^{-5}$    & 1.8$\cdot 10^{-4}$  \\
ir1-64   & 40.0 & 0.71 & 0.11 & 65. &  0.1 & 5$\cdot 10^{-7}$    & 8$\cdot 10^{-5}$    \\
ir4-64   & 40.0 & 7.87 & 1.28 & 61. & 22.4 & 2.8$\cdot 10^{-2}$  & 1.8$\cdot 10^{-3}$  \\
ir5-64   & 40.0 & 8.77 & 0.58 &151. &610.0 & 21.0                & 2.40                \\
ar1-64   & 40.0 & 2.34 & 0.35 & 67. &110.0 & 0.7                 & 0.29                \\
ar2-64   & 40.0 &10.44 & 0.28 &373. &440.0 & 11.0                & 1.90
    \\[0.7ex]
\hline
\end{tabular}
\end{flushleft}
\label{tab:efficiencies}
\end{table}

\subsection{Neutrino-antineutrino annihilation}

Energy deposition in the vicinity of the black-hole torus system
by the annihilation of neutrinos ($\nu$) and
antineutrinos ($\bar\nu$) of all flavors, which are radiated 
away from the accretion torus, to electron-positron pairs
in the process
\begin{equation}
\nu + \bar\nu\ \longrightarrow\ e^+ e^- 
\end{equation}
is considered as a way to produce a highly relativistic
pair-plasma, if the density of baryonic matter in
the deposition region is sufficiently low. Provided the 
entrainment of additional baryons during the subsequent 
expansion can be prevented, the pressure by radiation and
electron-positron pairs can then accelerate the plasma to
ultrarelativistic velocities. Recent general relativistic 
hydrodynamic simulations have demonstrated that this indeed
happens and that the ultrarelativistic outflow becomes
collimated to semi-opening angles between $\sim\,$5$^{\mathrm{o}}$
and $\sim\,$12$^{\mathrm{o}}$ (Aloy et al.~\cite{aloy05}).

\begin{figure*}[htp!]
\begin{tabular}{cc}
  \psfig{file=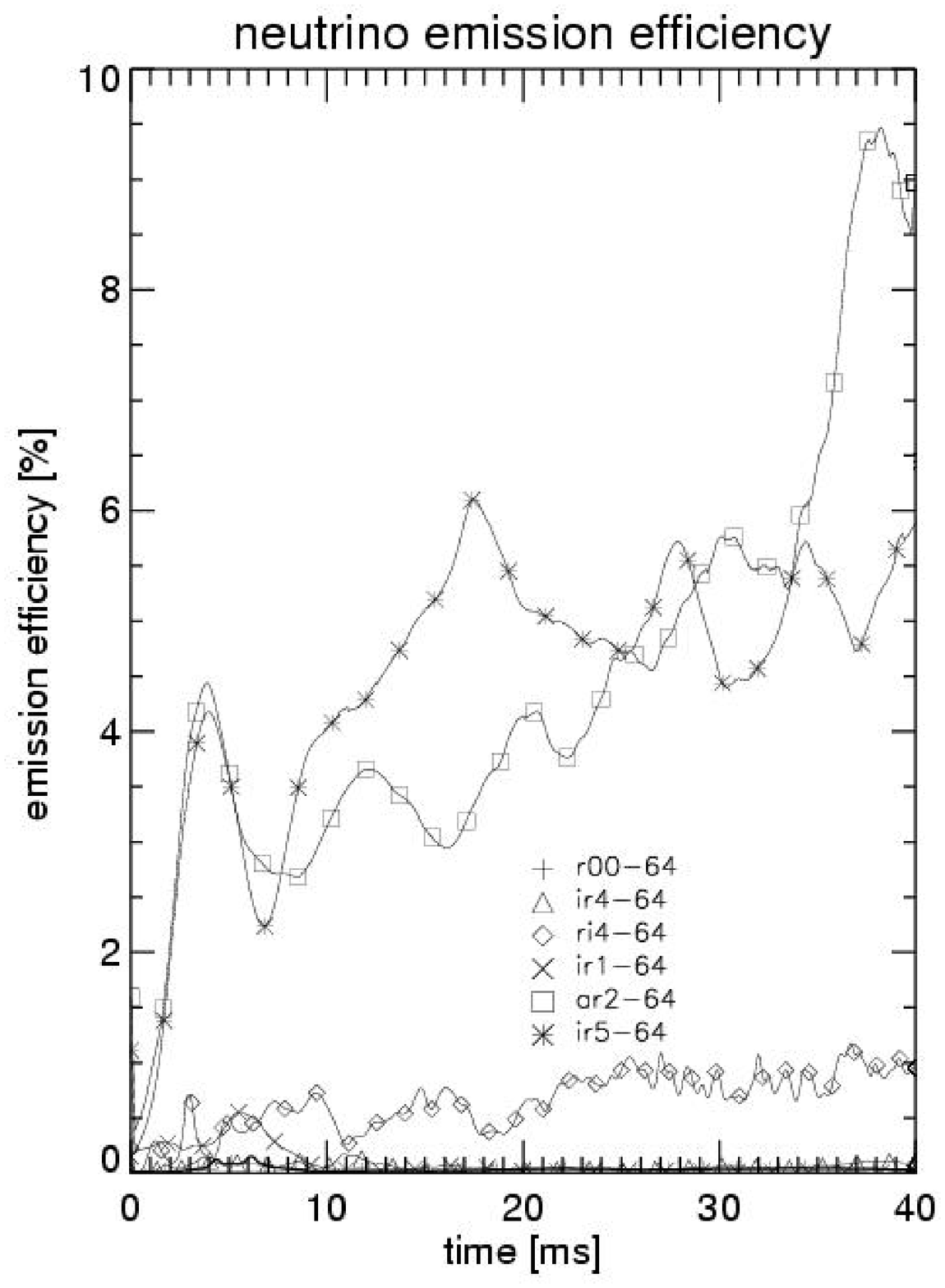,width=8.8cm} &
  \psfig{file=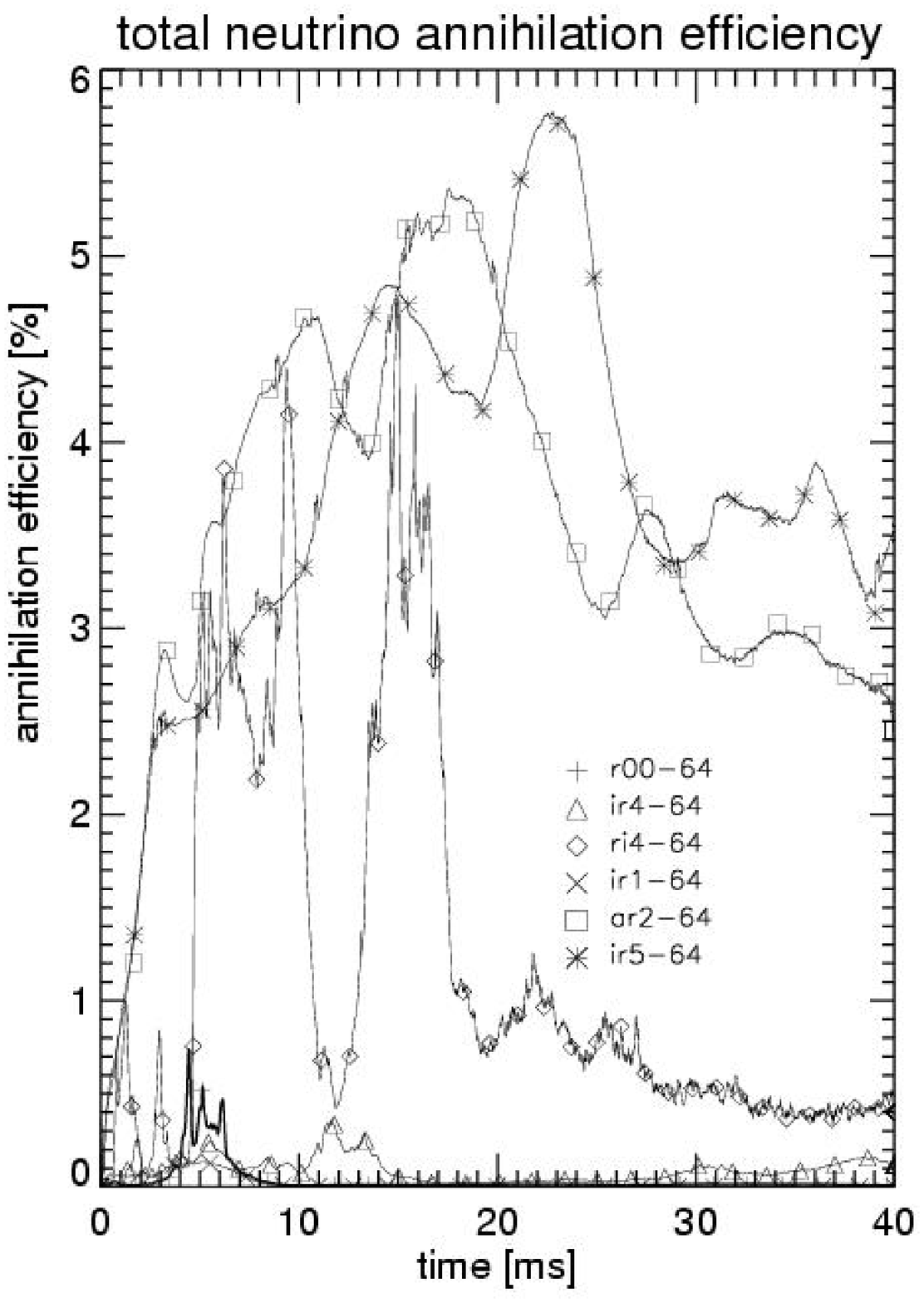,width=8.8cm} \\
  \psfig{file=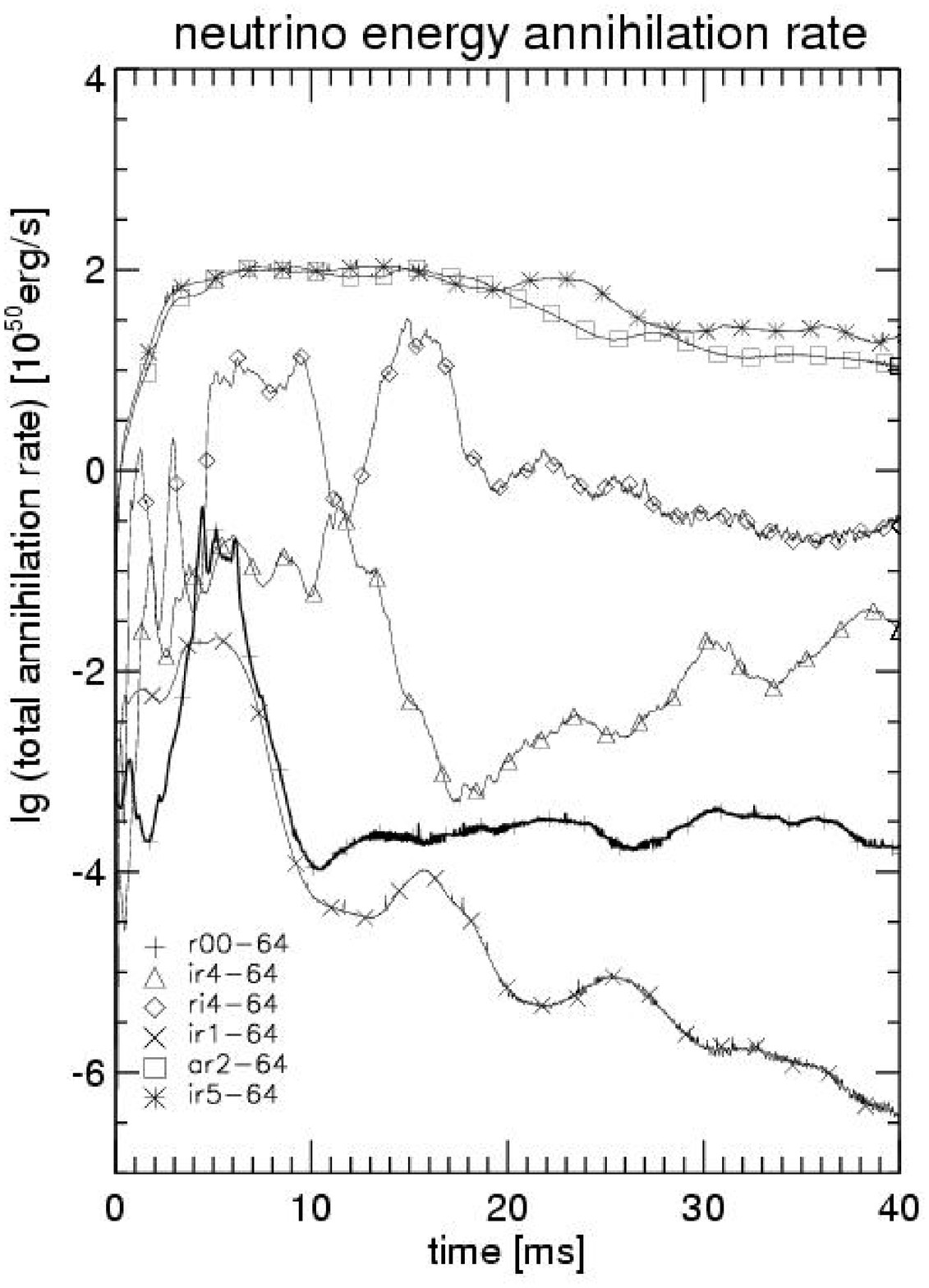,width=8.8cm} &
  \psfig{file=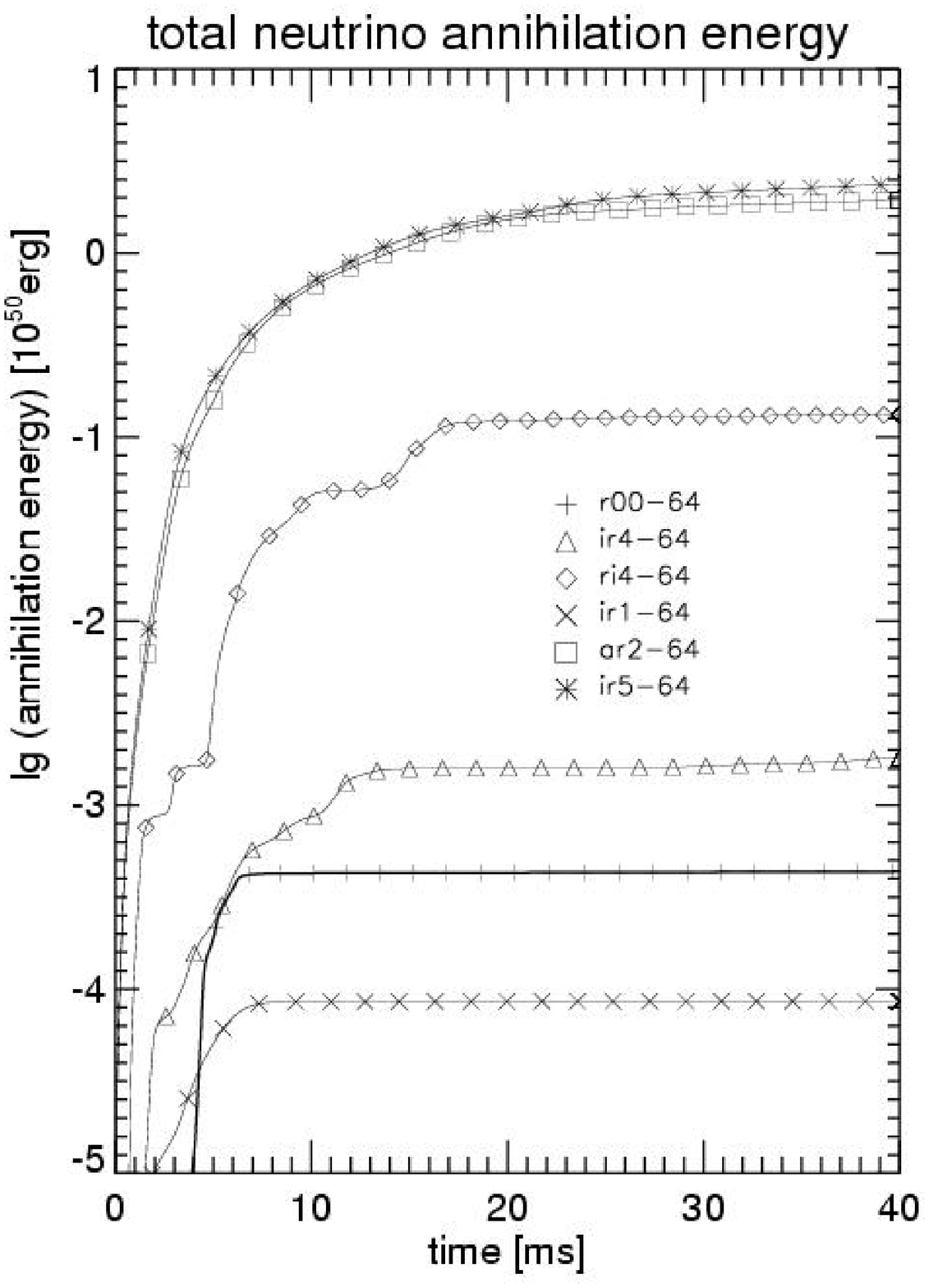,width=8.8cm} 
\end{tabular}
\caption[]{
Conversion efficiency of rest-mass energy to neutrinos,
  $q_{\nu}\equiv L_{\nu}/(\dot Mc^2)$ (top left), conversion efficiency
  of neutrino energy to electron-positron pairs by neutrino-antineutrino
  annihilation, $q_{\nu\bar\nu} \equiv \dot E_{\nu\bar\nu}/L_{\nu}$ (top
  right), integral rate of energy deposition by neutrino-antineutrino
  annihilation around the accretion torus, $\dot{E}_{\nu\bar{\nu}}$
  (bottom left), and cumulative energy deposition by $\nu\bar\nu$
  annihilation, $E_{\nu\bar{\nu}}$ (bottom right), as
  functions of time for the same set of models shown in 
  Figs.~\ref{fig:3a}, and \ref{fig:III45a}.
\label{fig:IIIeff64}
}
\end{figure*}

\begin{figure*}[htp!]
\begin{tabular}{cc}
  \psfig{file=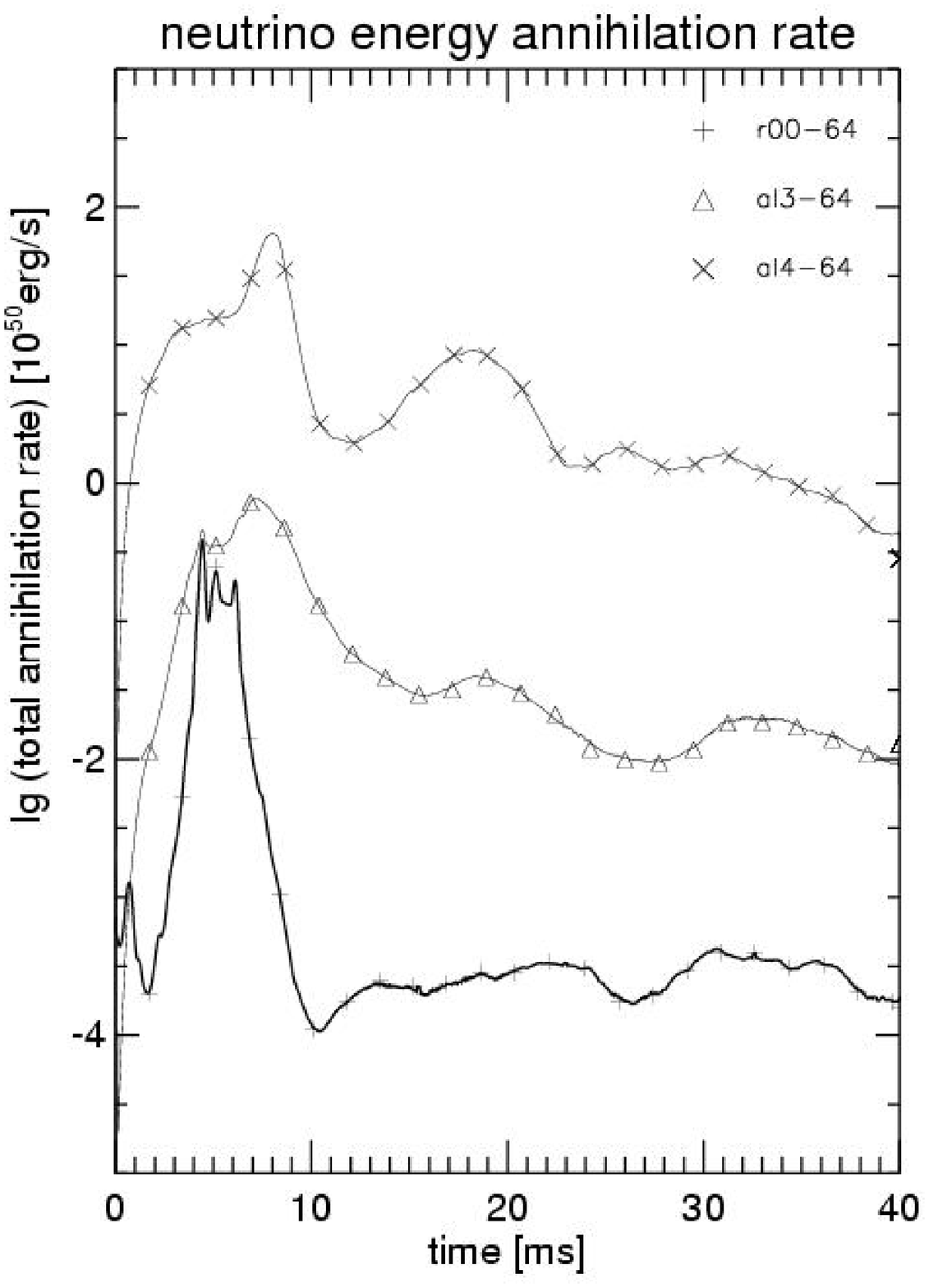,width=8.8cm} &
  \psfig{file=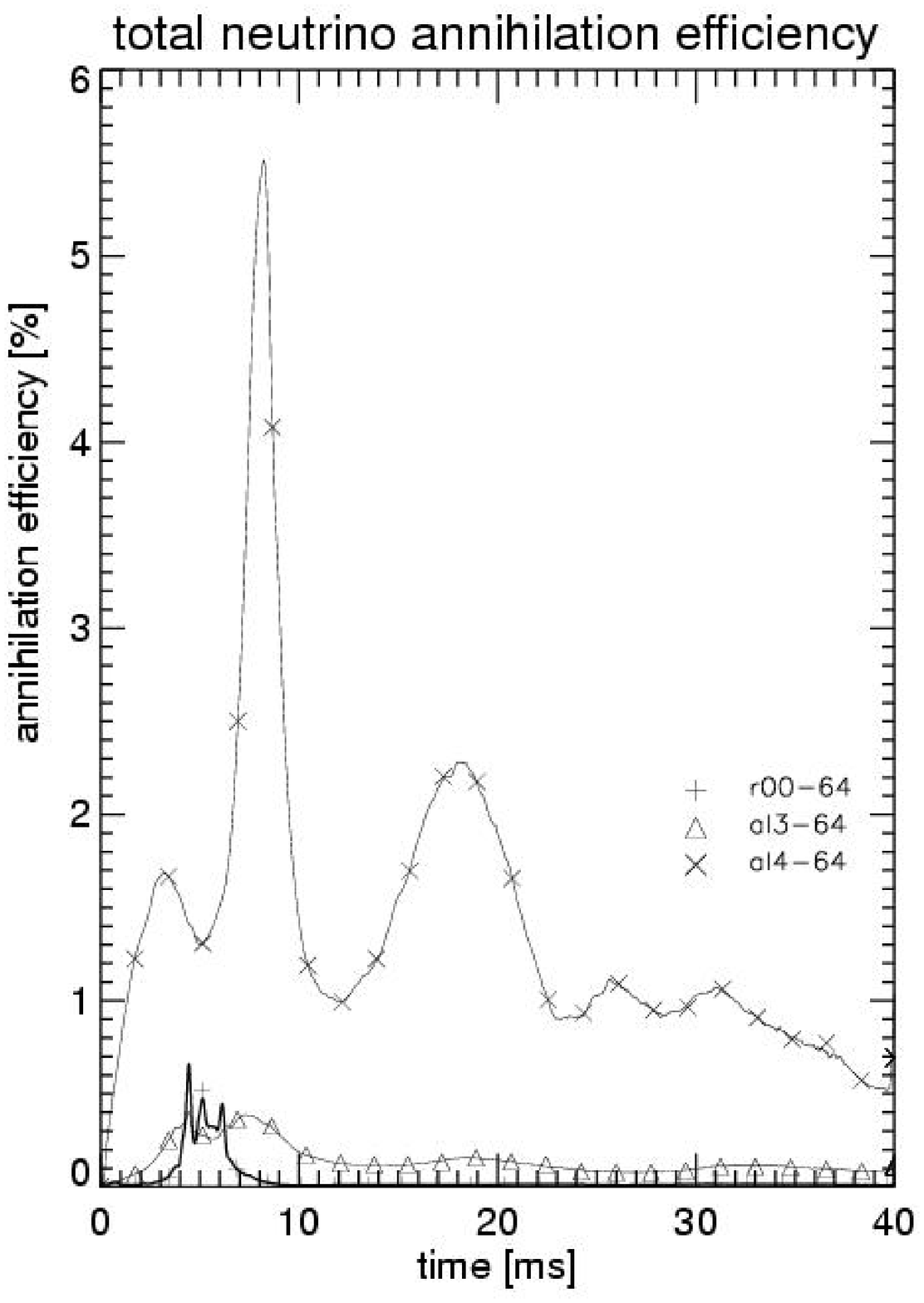,width=8.8cm}   \\[-2ex]
\end{tabular}
\caption[]{
  Integral rate of energy deposition by neutrino-antineutrino
  annihilation around the accretion torus, $\dot{E}_{\nu\bar{\nu}}$
  (left) and conversion efficiency  
  of neutrino energy to electron-positron pairs by
  neutrino-antineutrino annihilation
  $q_{\nu\bar\nu} \equiv \dot E_{\nu\bar\nu}/L_{\nu}$ (right), as
  functions of time for selected 64-resolution models with the same
  torus mass but different torus viscosity. The models are the same 
  as in Figs.~\ref{fig:3a} and \ref{fig:8bnew}.
\label{fig:eff64}
}
\end{figure*}

Because we calculate energy and lepton number loss in our 3D
hydrodynamic simulations of the torus evolution by a 
neutrino trapping scheme, but do not solve the neutrino-transport
problem, the corresponding energy deposition is evaluated
for our hydrodynamic models in a post-processing step.
The numerical procedure for doing this is rather CPU-time
consuming and explained in detail
in Ruffert \& Janka~(\cite{ruffert,ruffert3}) and 
Ruffert et al.~(\cite{ruffert2}). The energy deposition rate
per unit volume, $\dot Q_{\nu\bar\nu}$, 
at a point ${\bf r}$ is calculated as the sum of the 
contributions from neutrinos and antineutrinos emitted from
all regions of the accretion torus, thus taking into account
the specific emission geometry and local variations of the neutrino
loss rates as visible in the emission maps of 
Figs.~\ref{fig:16n1}.
Neutrinos and antineutrinos radiated from
the hot accretion torus interact with each other in the surroundings
with a finite probability, which depends on the number
densities and the energies of these neutrinos and on the
angle between the directions of neutrino and antineutrino propagation
(see, e.g., Goodman et al.~\cite{goo87}, Ruffert et al.~\cite{ruffert2}).
The total energy deposition rate,
%
%
\begin{equation}
\dot E_{\nu\bar\nu} \,=\, C\,L_{\nu} L_{\bar\nu}
\left( \frac{\displaystyle \langle \epsilon_{\nu}^2\rangle
\langle \epsilon_{\bar\nu}\rangle + \langle \epsilon_{\bar\nu}^2\rangle
\langle \epsilon_{\nu} \rangle}
{\displaystyle \langle \epsilon_{\nu}\rangle
\langle \epsilon_{\bar\nu}\rangle} \right) \,\, ,
\label{eq:efneu}
\end{equation}
therefore 
increases with the product of neutrino and antineutrino luminosities
and with the spectrally averaged neutrino energies, times a factor $C$
that accounts for the dependence on the angular distribution of the
neutrinos. The quantities $\langle
\epsilon_\nu \rangle$ and $\langle \epsilon_\nu^2 \rangle$ in
Eq.~(\ref{eq:efneu}) denote the mean energies and mean squared energies
of $\nu$ and $\bar\nu$. The factor $C$ contains the weak interaction
coefficients and terms that depend on the geometry of the
neutrino-emitting torus region.
This factor is a function of the emission geometry and
drops steeply with the distance from the neutrino source. 

\begin{figure*}[htp!]
\begin{tabular}{cc}
  \psfig{file=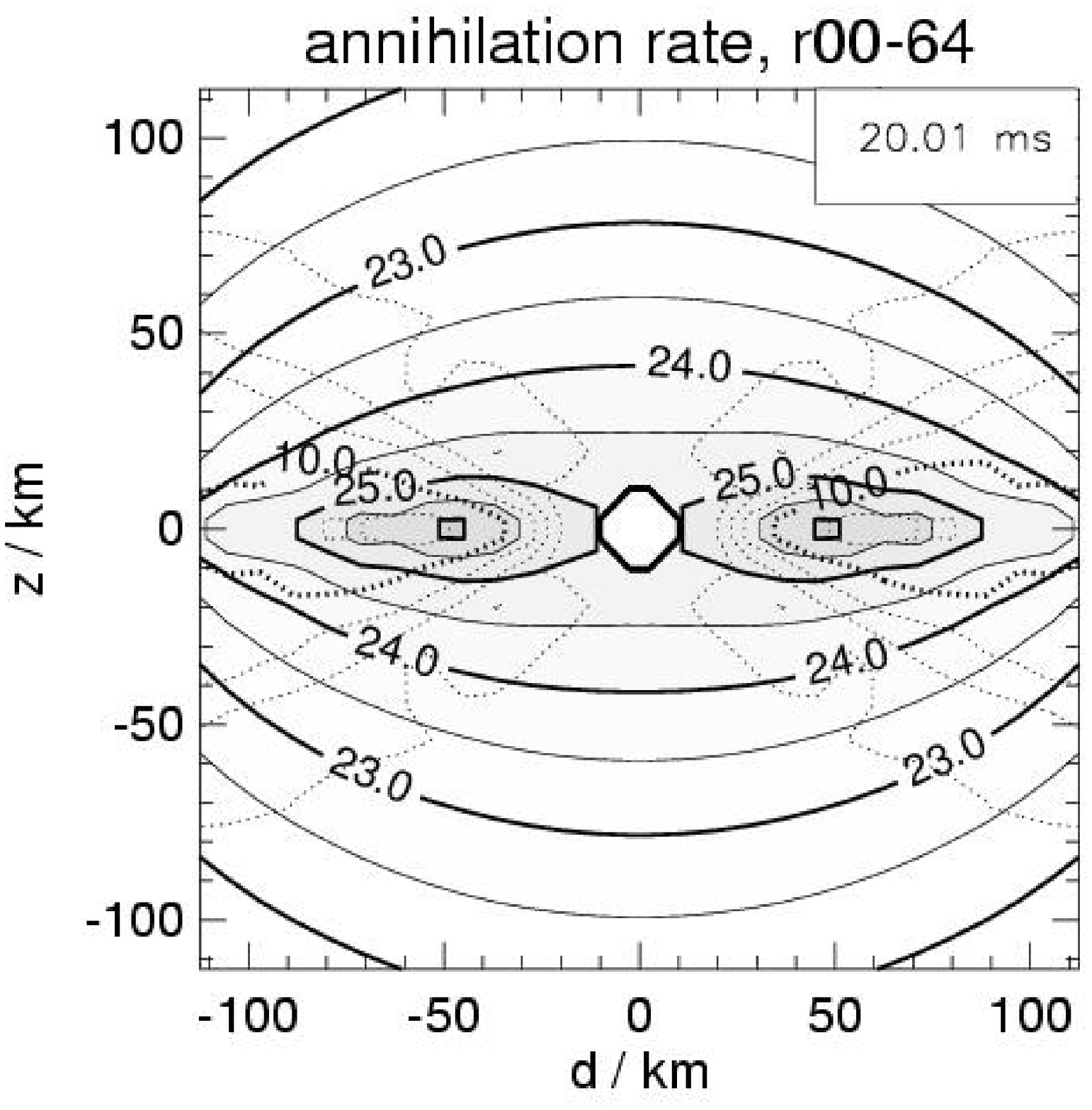,width=8.2cm} &
  \psfig{file=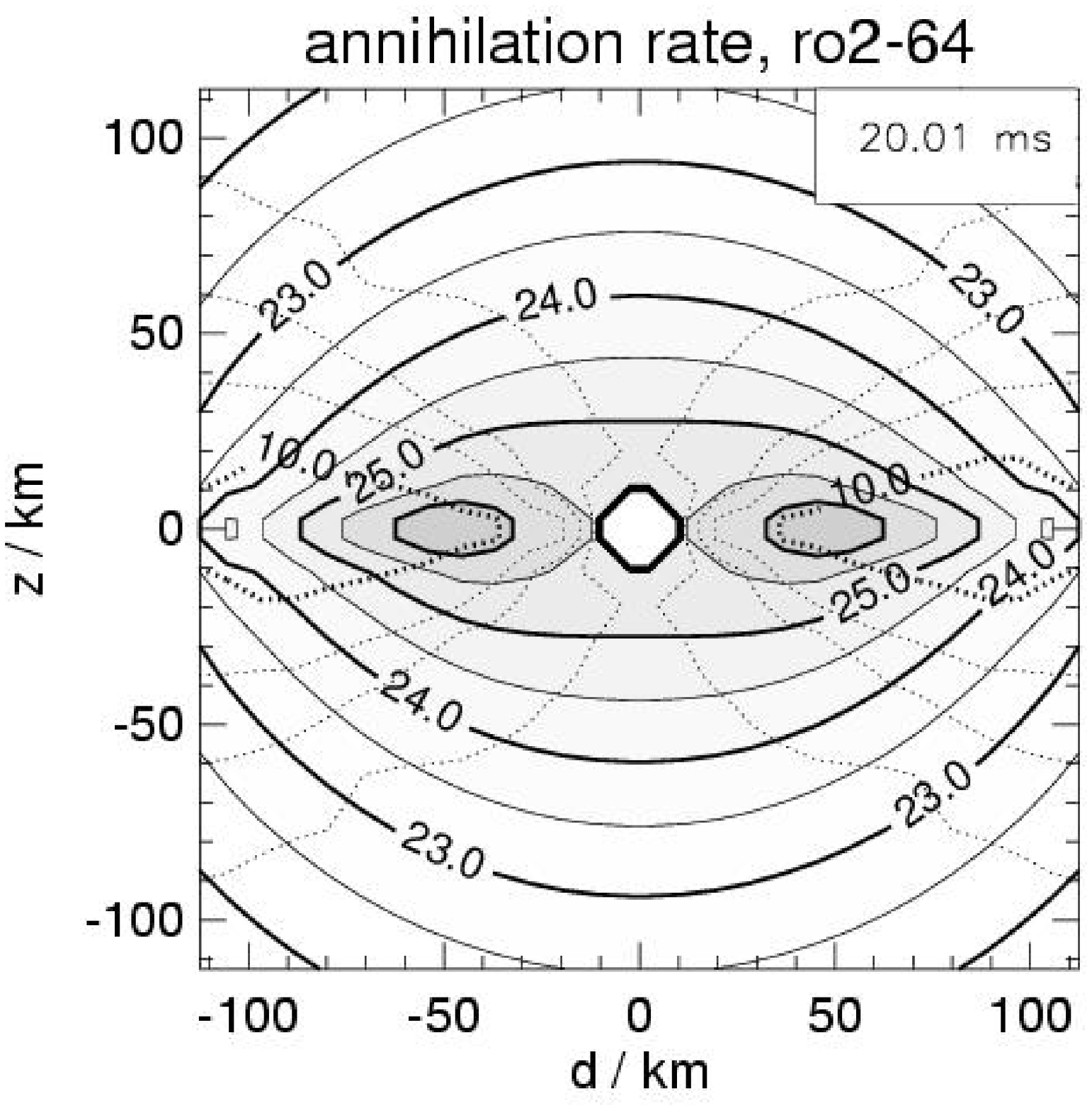,width=8.2cm} \\
  \psfig{file=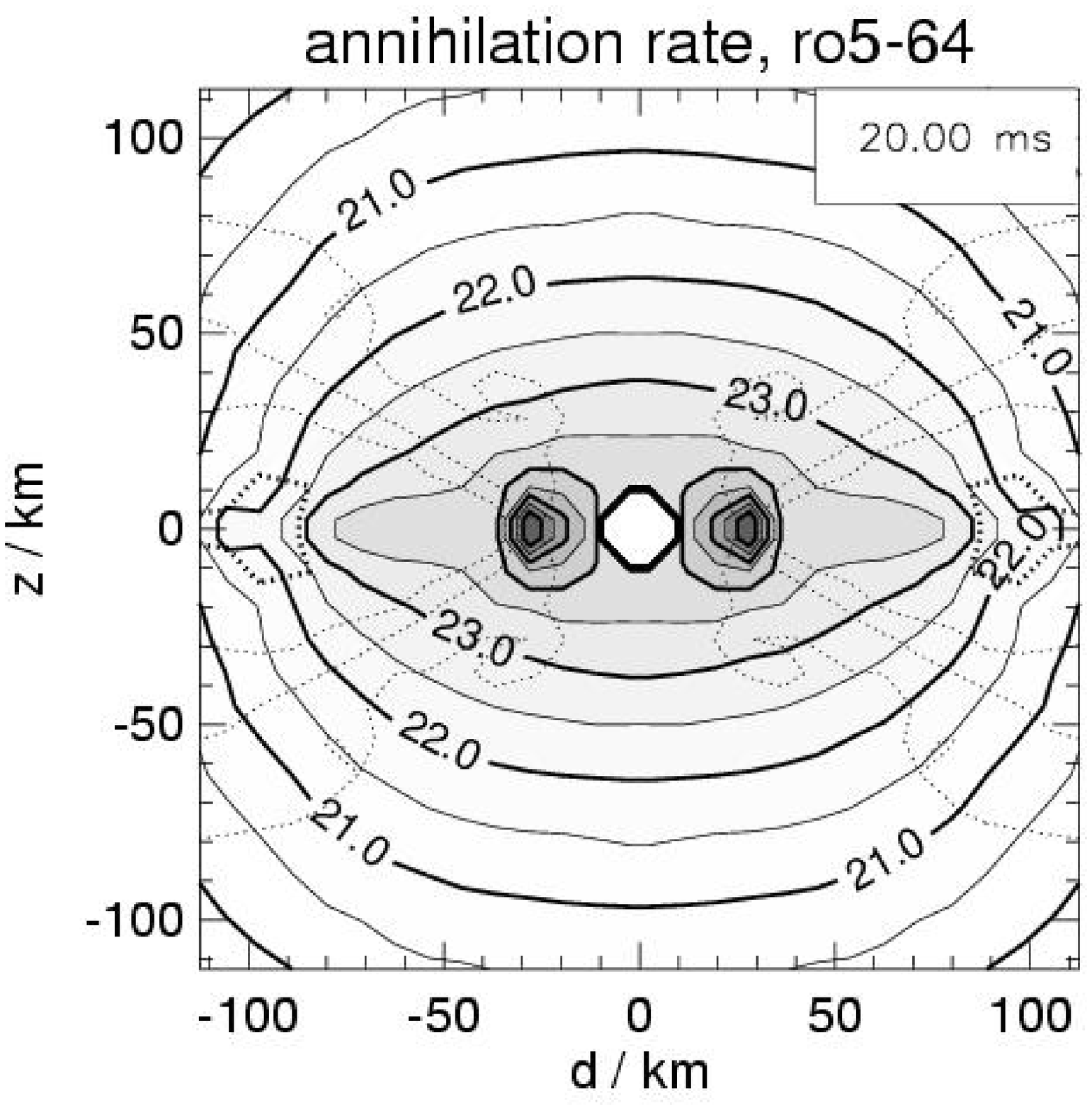,width=8.2cm} &
  \psfig{file=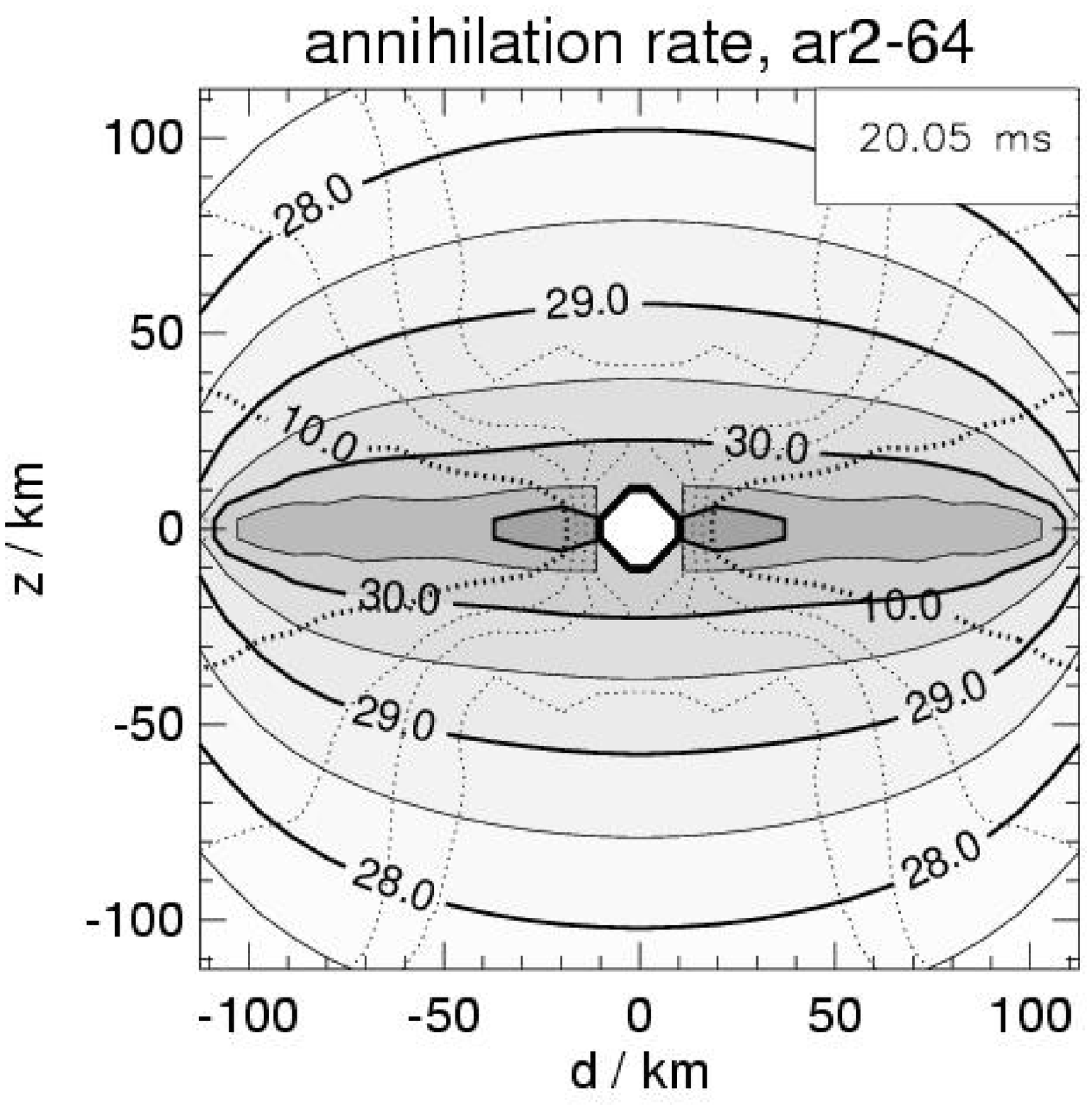,width=8.2cm} \\
\end{tabular}
\caption[]{Maps of the local energy deposition rates
(in~erg$\,$cm$^{-3}\,$s$^{-1}$) by $\nu\bar\nu$ annihilation to
$e^+e^-$ pairs in the surroundings of the accretion torus for
Models~r00-64, ro2-64, ro5-64, and ar2-64 at a time 20$\,$ms after the start of
the simulation. The solid contour lines in a plane perpendicular
to the equatorial plane represent values averaged over azimuthal angles
around the $z$-axis. The contours
are logarithmically spaced in steps of 0.5~dex and the grey shading
emphasises the levels with dark grey meaning high energy
deposition rate. The energy deposition rate was
evaluated only in that region around the torus where the baryon
mass density is below $10^{11}\,{\rm g\,cm}^{-3}$. The white octagonal area 
around the centre with a semidiameter of one Schwarzschild radius
indicates the presence of the central black hole. The dotted contour lines
represent levels of constant values of the azimuthally averaged
mass density, also logarithmically spaced with intervals of 0.5~dex;
the bold dotted line corresponds to $\rho=10^{10}\,{\rm g\,cm}^{-3}$.
\label{fig:16n3}
}
\end{figure*}

Large collision angles between neutrinos and antineutrinos 
are found only close to the source and therefore annihilation
is favoured near the hottest parts of the torus,
close to the equatorial plane. There, however, the gas densities
are still high and the energy deposition by $\nu\bar\nu$ annihilation
is compensated by very rapid cooling due to the charged-current
reactions of Eqs.~(\ref{eq:reac1}) and (\ref{eq:reac2})\footnote{The
energy loss associated with the reemission of neutrinos by 
charged-current reactions
is not taken into account in Fig.~\ref{fig:16n3}.}.
Neutrino-antineutrino annihilation is therefore not
efficient in establishing net energy transfer to
regions with large baryon densities. On the other hand, the
annihilation rate is also higher above the poles of the black hole,
where neutrinos emitted from the torus have a high probability
to collide with large angles in low-density environment.
This can be seen in the four panels of Fig.~\ref{fig:16n3}, 
which show the annihilation
rate in the $d$-$z$-planes perpendicular to the orbital plane 
for Models~r00-64, ro2-64, ro5-64, and ar2-64 at
20$\,$ms after the start of the simulations.

The corresponding integral rate of energy deposition by
neutrino-antineutrino annihilation around the accretion torus,
$\dot E_{\nu\bar\nu}$,
the cumulative energy deposition by $\nu\bar\nu$ annihilation,
$E_{\nu\bar{\nu}}$, and the neutrino conversion efficiency,
\begin{equation}
q_{\nu\bar\nu}\ \equiv\ \dot E_{\nu\bar\nu}/L_{\nu}\ , 
\label{eq:anneff}
\end{equation}
as functions of
time are plotted for some of our computed models in 
Figs.~\ref{fig:IIIeff64} and \ref{fig:eff64}.
Values of $\dot E_{\nu\bar\nu}$ and 
$E_{\nu\bar{\nu}}$ for all 64-resolution models at the end
of the simulations are listed in Table~\ref{tab:efficiencies}.

The contributions from muon and tau neutrinos and antineutrinos 
can safely be ignored in this evaluation, because their luminosities are
much lower (see Table~\ref{tab:models}) and their annihilation
cross sections are also smaller by a factor of about 5 compared to
those of electron neutrinos and antineutrinos. 
The geometry of the neutrino-emitting torus region and therefore
the factor $C$ does not change 
strongly during the quasi-stationary phase of the torus
evolution and therefore $C$ can be calibrated by evaluating 
the $\nu\bar\nu$ annihilation in detail at one or more
representative times. Thus Eq.~(\ref{eq:efneu}) with constant $C$
allows for a good approximation of the time evolution
of $\dot E_{\nu\bar\nu}$ (as shown in Figs.~\ref{fig:IIIeff64} 
and \ref{fig:eff64}) by using the neutrino luminosities and 
mean neutrino energies as determined from the torus simulations,
without the need to perform the CPU-time demanding post-processing
for the $\nu\bar\nu$-annihilation maps of 
Fig.~\ref{fig:16n3} many times for our sample of runs.

Only our most massive tori possess initially neutrino-opaque
``cores'' with densities above $\sim 10^{11}\,$g$\,$cm$^{-1}$, 
whereas in the other cases such regions are absent. When 
the optical depths for neutrinos are larger than unity and 
neutrinospheres exist, these typically have ellipsoidal 
or egg-shaped cross
sections in the $d$-$z$ plane. Also in our massive tori
the opaque cores disappear and the neutrino optical depths
fall below unity everywhere after the tori have lost some 
mass to the black hole and/or thermal inflation by viscous 
heating has decreased the density. All of the tori shown in
Fig.~\ref{fig:16n3} are therefore transparent to neutrinos.
A comparison of the annihilation maps of Models~r00, ro2,
and ro5 in Fig.~\ref{fig:16n3} shows the influence of
the black hole rotation. The values of the energy deposition
rates differ and scale roughly with the neutrino luminosity as 
expected from Eq.~(\ref{eq:efneu}) (cf.\ also 
Table~\ref{tab:efficiencies}), but the morphology of the
contour lines which envelope the black-hole torus system
is very similar. In the counter-rotating case (Model~ro5) 
the two central maxima are more pronounced and represent
a toroidal volume which is more compact than in the
other cases.

Only our most massive tori (Models~ri4, ir4, ir5, and ar2)
and the viscous intermediate-mass Models~al4 and ar1 
reveal $\nu\bar\nu$ annihilation rates $\dot E_{\nu\bar\nu}$
which become larger than $10^{50}\,$erg$\,$s$^{-1}$ at least
for some period of the simulated evolution (see 
Figs.~\ref{fig:IIIeff64} and \ref{fig:eff64}; the non-viscous
massive torus Model~ir4, however, only marginally reaches
this value). The numbers
listed in Table~\ref{tab:efficiencies} clearly show the
correlation with the neutrino luminosities. They also show
that the black hole rotation (up to a rotation parameter
of $a = 0.6$) has hardly any influence on the total 
energy $E_{\nu\bar\nu}$ deposited by $\nu\bar\nu$ annihilation
(compare the model pairs r00 and ro2, al4 and ar1, and
ir5 and ar2). In contrast, the torus mass makes a big effect 
for viscous (model pairs al4 and ir5, ar1 and ar2) as well
as nonviscous cases (model pairs r00 and ir4, ro2 and ri4);
in case of the higher torus masses $E_{\nu\bar\nu}$ is 
larger by factors of 5--40. Again shear viscosity ($\alpha = 0.1$
instead of $\alpha = 0$) makes the biggest difference. For
the latter, viscous models of the pairs (ri4,ar2), (ir4,ir5),
and (r00,al4) $E_{\nu\bar\nu}$ is higher by factors between
15 and 1000. The lower right panel of Fig.~\ref{fig:16n3}
shows the annihilation map for one of the two most extreme 
cases, Model~ar2, at 20$\,$ms after the start of the
simulation when the integral rate of energy conversion by 
$\nu\bar\nu$ annihilation was $\dot E_{\nu\bar\nu} = 
5.8\times 10^{51}\,$erg$\,$s$^{-1}$. The energy deposition 
rates above the poles of the black hole reach 
$10^{30}\,$erg$\,$cm$^{-3}$s$^{-1}$ at this time. Even
at $z$-distances of 50$\,$km they are still more than
$10^{29}\,$erg$\,$cm$^{-3}$s$^{-1}$.
For the most massive of our tori and the viscous tori with
intermediate masses, the annihilation efficiencies $q_{\nu\bar\nu}$
can reach several percent, peaking for short times at more
than 5\% (Figs.~\ref{fig:IIIeff64} and \ref{fig:eff64}).
We note that the results for $\nu\bar\nu$ annihilation which 
we have obtained for the 32-resolution 
runs are in general smaller by about 20\% compared to the
better resolved models. We also point out that the
total energy deposition by $\nu\bar\nu$ annihilation
($E_{\nu\bar{\nu}}$) of Model~ir5-64 at the end of our 
simulations (at 40$\,$ms) is about 25\% higher than that
of Model~ar2-64, but we estimate the remaining torus lifetime
of Model~ar2-64 to be more than two times longer.

At the end of the simulated evolution, the tori have lost
typically about half of their mass to the black hole, and
the mass accretion rates have decreased to 
0.1--0.4$\,M_\odot\,$s$^{-1}$ in most cases. The total 
neutrino luminosities have declined from their peak values by
roughly an order of magnitude for the tori with intermediate
mass and by about a factor of 5 for our most massive cases
(Fig.~\ref{fig:III45a} and Table~\ref{tab:efficiencies}).
Correspondingly, the $\nu\bar\nu$ annihilation
rates have dropped by roughly two orders of magnitude
below their maximum values in the former case
(Fig.~\ref{fig:eff64}), while in the latter case they have 
decreased only by a factor of 5--10
(Models~ar2 and ir5 in Fig.~\ref{fig:IIIeff64}, which have
peak values of $\dot E_{\nu\bar\nu} \ga 10^{52}\,$erg$\,$s$^{-1}$). 
This is less
steep than the square of the total neutrino luminosities
and suggests a time-dependent change of the emission of 
electron neutrinos relative to electron antineutrinos 
(note the dependence of $\dot E_{\nu\bar{\nu}}$ on 
the product $L_\nu L_{\bar\nu}$ in Eq.~\ref{eq:efneu}). This
fact is consistent with the large rise of $Y_e$
relative to the initially low values (e.g.~visible in the comparison
of Fig.~\ref{fig:II1a} with Fig.~\ref{fig:1}). 
When $Y_e$ climbs closer to 0.5 and thus
tries to approach a value where kinetic equilibrium of 
$\nu_e$ and $\bar\nu_e$ production holds, the
initially much stronger emission of $\bar\nu_e$ 
evolves to a more balanced release of electron neutrinos
and antineutrinos\footnote{Note that the product $L_\nu L_{\bar\nu}$
has a maximum for $L_\nu = L_{\bar\nu}$ if $L_\nu+L_{\bar\nu}$
is a constant.}.

\section{Axial low-density funnel and GRB production}
\label{sec:GRBs}

Near the poles of the black hole and along the system axis, the
density in our simulations decreases quickly. 
Material which is swept into these polar regions during and
immediately after the merging of the double neutron stars or
neutron-star black-hole binary falls into the (newly formed) black
hole within a free-fall time scale, because it is not supported by 
a strong negative pressure gradient or centrifugal forces.
Thus a cylindrical funnel along the rotation axis of the black-hole
torus system is rapidly ``evacuated''. 
The continuous entrainment of more gas from the torus into that region
is hindered by the centrifugal barrier which this gas experiences due
to its large angular momentum.

Attempts to estimate the characteristic properties of ultrarelativistic
outflows (e.g., its terminal Lorentz factor or the collimation factor)
on the basis of results from hydrodynamic merger or accretion simulations
like the ones presented here must be taken with great caution and
might be quantitatively meaningless.
The formation of collimated outflows in the vicinity of a
torus-accretor system is a violent, time-dependent hydrodynamic 
process. 
The jet could clean its own funnel through an initially baryon-loaded
environment as happens in the core of a collapsar, where
large amounts of gas envelope the black-hole and accretion disk.
On the other hand, Kelvin-Helmholtz instabilities due to the
interaction of the shear flow with the dense torus may lead to baryon
entrainment and collimation. 
Reliable and quantitatively meaningful answers therefore require the
modelling of the jet formation and evolution by hydrodynamic simulations
(for more details, see Aloy et al.~\cite{aloy05}).

The low-density polar funnels along the rotation axis of the torus are
unlikely to form near the poles of a hot, neutrino-radiating massive
neutron star, if such a merger remnant escapes the collapse to a black
hole. 
At the surface of such a neutrino-emitting compact object, 
a dense baryonic wind is driven by neutrino energy deposition.
This wind produces an expanding cloud of baryonic matter around the
star, a phenomenon which is well-known for the hot proto-neutron stars
born in supernova explosions (Duncan et al.~\cite{dun86};
Woosley~\cite{woo93b}; Qian \& Woosley~\cite{qian96}; Thompson et
al.~\cite{tho01}).  

The hot merger remnant emits neutrinos and antineutrinos with very
high luminosities. 
These energetic neutrinos diffuse and then stream outward to the
surface where they must deposit energy by charged-current absorption
on free nucleons  (the inverse of the reactions Eqs.~(\ref{eq:reac1})
and (\ref{eq:reac2})) in the layers of decreasing temperature
outside of the neutrinosphere. 
This energy deposition drives the subrelativistic, baryonic wind,
whose formation cannot be treated in existing neutron star merger
simulations, because the simulations do not include neutrino
transport. 
Neutrino emission is at best described by a neutrino-trapping scheme,
in which neutrinos are released locally, but the transport of energy
by the neutrinos from layers deep inside the merger remnant to the
surface is not followed.  
Because of the absence of the neutrino-driven wind, the densities in
the polar regions of the massive post-merger neutron stars are 
{\em largely underestimated} by the current hydrodynamical models.

The baryon ``pollution'' by the neutrino-driven wind makes it unlikely
that such a hot neutron star merger remnant can be the central engine
of a cosmic gamma-ray burst, in particular if the energy release that
powers the GRB is considered to happen in the close vicinity of the
neutron star. 
The baryon loading in this region will prevent the formation of a
pair-plasma fireball or jet which can accelerate to ultrarelativistic
velocities. 
Neutrino-antineutrino annihilation, for example, deposits energy
efficiently only very close to the neutrinosphere and the annihilation
rate drops extremely steeply with distance (e.g., like $r^{-8}$ if the
neutrino source is a spherical or ellipsoidal body). 
In case of a hot neutron star the gas density outside of the 
neutrinosphere is very high --- in contrast to the baryon-poor
conditions above the poles of a black hole --- and therefore
$\nu\bar\nu$ annihilation competes with the charged-current absorption
of neutrinos by nucleons (Eqs.~(\ref{eq:reac1}) and (\ref{eq:reac2}))
and with the very efficient cooling due to the inverse of these
reactions. 
In detailed wind models it is therefore seen that $\nu\bar\nu$
annihilation yields only a minor contribution to the energy
transferred to the baryonic wind (see, e.g., Fig.~6 in Thompson et
al.~\cite{tho01}), and this additional energy deposition triggers
enhanced mass loss from the surface of the neutron star instead
of driving ultrarelativistic expansion. Similar arguments also apply
for models where the GRB outflow is considered to be powered by 
magnetohydrodynamic activity of a rapidly, differentially rotating
hot neutron star. Also in this case it has to be demonstrated that
the high baryon densities associated with the neutrino-driven wind 
around the neutron star do not represent a unsurmountable obstacle
for the development of ultrarelativistic fireballs or jets.

\section{Summary}
\label{sec:summary}

We have presented 3D hydrodynamic simulations of the 
time-dependent evolution of 
black-hole accretion-torus systems with parameters typical of the
remnants of binary neutron star or neutron-star black-hole mergers.
The tori which are created from neutron stars disrupted in such 
violent events, and the conditions in the torus gas are not well 
described by steady-state assumptions and the gas evolution
requires time-dependent modelling.

The black hole was described by a vacuum boundary condition and
pseudo-Newtonian gravitational potential, corresponding to an initial
mass of about 4$\,M_\odot$. The effects of black hole rotation were
included according to Artemova et al.~(\cite{art96}), taking into
account the variation of black hole mass and spin by the accretion
of matter from the torus. The torus gas was treated with Newtonian 
dynamics and self-gravity, its shear viscosity was modelled by 
a simple $\alpha$-law, and the initial conditions were assumed to 
be cool (temperatures of 1--2$\,$MeV) and neutron-rich   
(proton-to-baryon ratio $Y_e\sim 0.02$--0.1), decompressed neutron
star matter in nuclear statistical equilibrium. 
The initial spatial distribution of the gas was 
modelled by using results from (Newtonian) merger simulations.

Assuming different initial masses of the torus 
($M_{\mathrm{d,ini}} \sim 0.01$--0.2$\,M_{\odot}$),
black hole spins (spin parameter $a = 0,\,0.3,\,0.6,\,0.8$),
and gas viscosity ($\alpha$-parameter between 0 and 0.1), 
we followed the accretion process and the
associated neutrino emission for 40$\,$ms with high-resolution 
runs, and for 70$\,$ms with lower resolution runs. The global
parameters of both simulations at the same times agree 
satisfactorily well, but the better resolved simulations 
exhibit slightly higher gas temperatures and correspondingly
larger neutrino emission.

Since our initial conditions with chosen torus mass and black
hole potential did not correspond to exact rotational equilibria, the
first 10$\,$ms of our simulations are characterised by a relaxation
phase with high mass accretion rates and a dynamical reconfiguration
of the torus matter, followed by a more ``quiet'', quasi-stationary
phase in which the mass accretion rate of the black hole settles
to a lower, slowly changing value.

The tori start to evolve quickly from their initially 
neutron-rich state to more proton-rich conditions. 
This protonization
proceeds by the emission of electron antineutrinos, which dominates
over the emission of electron neutrinos and heavy-lepton
neutrinos, and is faster for hotter tori. The gas viscosity thus
has an extremely strong influence on the torus evolution. While
simulations without physical viscosity ($\alpha = 0$) show 
only compressional heating and thus stay relatively cool with
$Y_e$ changing rather slowly, shear viscosity ($\alpha = 0.01$ or 0.1)
leads to a rapid rise of the gas temperature, enhanced accretion
of the torus regions close to the black hole, and thermally driven
inflation of the outer parts of the torus with consequently lower
densities. Therefore the mass of viscous tori can decrease initially
more rapidly and then continues to decline more slowly with a 
much reduced mass-loss rate to the black hole. After 40$\,$ms
of evolution the tori have lost about half of their initial mass
to the black hole, which at that time accretes matter at rates 
between about 0.2$\,M_{\odot}\,$s$^{-1}$ and about
0.6$\,M_{\odot}\,$s$^{-1}$ in all simulated
cases except those with $\alpha = 0$ and the highest and lowest
initial torus masses. 

The maximum densities in the tori have dropped to 
$\la 10^{11}\,$g$\,$cm$^{-3}$ after 10$\,$ms in most cases, and
stay around $2\times 10^{11}\,$g$\,$cm$^{-3}$ after the early 
relaxation phase only in our most massive accretion tori.
The tori are therefore optically thin to neutrinos
or develop neutrinospheres just marginally. For this reason
neutrinos, once created, stream off essentially freely without
being hindered by scattering and reabsorption. Due to the reduced
gas densities and relatively high temperatures in viscous tori, 
the protonization there proceeds quickly and produces
$Y_e$-values of 0.3--0.4 within only 20$\,$ms, tending to 
establish kinetic equilibrium conditions for the competing
$\nu_e$ and $\bar\nu_e$ 
emission processes. As a consequence, the $\nu_e$ and $\bar\nu_e$
luminosities become more similar toward the end of our
simulations, although those of electron antineutrinos are 
still a factor of 2--3 higher. Muon and tau neutrinos and 
antineutrinos contribute to the energy loss of the accretion
tori on a negligible level, because their creation by
nucleon-nucleon bremsstrahlung and electron-positron pair
annihilation is much slower than the charged-current production
of $\nu_e$ and $\bar\nu_e$ at the temperatures and
low densities of the tori.

The effect of the protonization on the neutrino luminosities 
has interesting consequences for the rate at which 
neutrinos and antineutrinos deposit energy around the black-hole
torus system by the annihilation to electron-positron pairs.
While the total neutrino luminosity is strongly correlated with 
the torus mass and therefore tends to decay with falling gas
mass, the energy deposition scales with the product of
neutrino and antineutrino luminosities, $L_\nu L_{\bar\nu}$
(see Eq.~\ref{eq:efneu}; electron neutrinos and antineutrinos
make by far the dominant contribution). It therefore decreases
{\em much less steeply} with time than the square of the total 
neutrino luminosities, because the electron neutrino emission 
is initially very small but rises continuously
relative to the one of electron antineutrinos.

As mentioned above, shear viscosity has
the largest influence on enhancing the neutrino production.
We found the highest luminosities and largest $\nu\bar\nu$ 
annihilation rates for our simulations of viscous tori 
($\alpha = 0.1$). Values in excess of $10^{53}\,$erg$\,$s$^{-1}$
and up to $\sim\,$10$^{50}\,$erg$\,$s$^{-1}$ for $L_\nu$ and 
$\dot E_{\nu\bar\nu}$, respectively, were reached even for
relatively low torus masses on only some $10^{-2}\,M_{\odot}$.
The maximum conversion efficiencies of rest-mass energy to
neutrinos could reach 5--10\%, and of neutrinos to electron-positron
pairs 3--5\%. While viscosity, however, tends to decrease the torus
lifetime (measured as the ratio of torus mass to mass accretion rate),
black hole rotation in the direction of the torus spin has the
opposite effect. The higher disk temperatures close to the more
compact ISCO in this case, however, have a noticeable effect on the 
neutrino emission only if the gas viscosity is low or if the 
black hole spin is close to maximal ($a \ga 0.8$).

An evaluation of our results for the growth of nonradial
instabilities, the presence of which might be suggested by the fact
that we found the axial symmetry of our initial models 
destroyed in a few revolutions, is prevented by our
use of a Cartesian grid with three nested levels of 
different resolution. This grid introduces visible numerical 
artifacts, in particular imprints a distinct 4-fold symmetry,
and must be suspected to be also the reason for higher-order
azimuthal modes.

\section{Conclusions and outlook}
\label{sec:akhir}

Our models are another contribution to the ongoing efforts
to theoretically study the viability of accretion tori around
stellar-mass black holes as central engines of cosmic GRBs.
A detailed, critical assessment of previous attempts to link the
results of similar simulations to GRB properties has already
been made in Sect.~\ref{sec:GRBs}. The crucial results of our
studies in this respect are the neutrino luminosities and 
the energy conversion by neutrino-antineutrino annihilation,
the former being the direct output from the neutrino 
trapping/leakage scheme applied in our model calculations, the
latter being obtained by a post-processing of our results.
A direct calculation of the energy release by 
$\nu\bar\nu$-annihilation would require the application of a
3D transport description, which remains a true challenge for the
future.

The results of our studies show that for values of the
torus mass, shear viscosity, and black hole rotation well
in the range of expectations for compact binary mergers, the
neutrino emission is sufficiently powerful and the 
associated efficiencies for converting gravitational binding 
energy to neutrinos as well as
neutrino-antineutrino pairs to a $e^+e^-$-photon plasma,
are sufficiently large to account for the energetics of 
short GRBs, in particular if collimation of the ultrarelativistic
outflow is invoked. Aloy et al.~(\cite{aloy05})
showed by general relativistic calculations that the fireball-torus
interaction can indeed produce collimated jets with 
``beaming fractions''
$f_\Omega = 2\delta \Omega_{\mathrm{jet}}/(4\pi) = 
1-\cos\theta_{\mathrm{jet}}$ between $\sim\,$0.5 and $\sim\,$2\%
of the sky ($\delta \Omega_{\mathrm{jet}}$ is the solid angle
subtended by one jet and $\theta_{\mathrm{jet}}$ its 
semi-opening angle).
This reduces the true energy outputs of the GRBs
compared to the observed ones by a factor of 50--200.
For the recently detected, short, hard bursts GRB~050509b
(Gehrels et al.~\cite{geh05}; Bloom et al.~\cite{blo06};
Hjorth et al.~\cite{hjo05}) and GRB~050724 (Berger et al.~\cite{ber05})
this corresponds to true $\gamma$-energies between
some $10^{46}\,$ergs and some $10^{48}\,$ergs, provided the
collimation factor of about 100 inferred by 
Berger et al.~(\cite{ber05}) for GRB~050724 --- which is in 
agreement with the predictions of Aloy et al.~(\cite{aloy05}) ---
applies to both bursts. Our results suggest that
such energies can be delivered by $\nu\bar\nu$ annihilation
without the need to invoke additional power from mechanisms 
based on the presence of extremely strong magnetic fields.
We therefore refrain from discussing here once more possible 
ways to use such magnetic fields for tapping the rotational energy
of the torus or black hole to drive relativistic motion of small 
amounts of baryonic matter. We feel unable to provide any deeper
insight into such possibilities on grounds of our current 
simulations than that obtained in the discussions of
previous works. These exploit the results of hydrodynamic models
to come up with estimates of equilibration field strengths
in the swirling gas of the merger remnant, or parametrise 
the possible energy conversion due to the Blandford-Znajek 
mechanism in terms of (undetermined or free) system parameters
(see, e.g., Rosswog et al.~\cite{ros03a}; 
Rosswog \& Ramirez-Ruiz~\cite{ros02, ros03c}; Lee et al.~\cite{lee05a}).

Besides limited numerical resolution, another drawback of 3D
simulations is the very limited time over which the evolution of
the accretion torus can be followed with the computer resources
available to us. The models presented here certainly add 
only a small brick to our slowly growing picture of the scenario
and physics that can make up a GRB engine.
The ultimate goal certainly must be simulations
which trace the merger history consistently through all phases,
from the merging of the compact objects, through the formation
of the black hole, to the accretion of the torus matter, until
the completion of the phase of powerful energy release from
this accretion process. Such simulations will require general
relativistic modelling (or a reasonable approximation of it)
and should include the microphysics which describes the gas
properties as well as the processes that establish the 
energy loss of the gas and the conversion of some of this 
energy to relativistic and non-relativistic outflows. 
Further progress in modelling will not
only require adding a spectral treatment of the neutrino emission,
preferably by solving the 3D transport problem. It may at some
stage also be necessary to perform magnetohydrodynamic simulations
for investigating the combined effects of neutrino transport and 
magnetic fields.

\begin{acknowledgements}
We thank the anonymous referee for extensive suggestions which
significantly improved and shortend our paper.
SS highly appreciated continuing encouragement from S.F.\ Gull and is
grateful to G.I.\ Ogilvie for valuable discussions.
SS acknowledged support from the Particle Physics and Astronomy Research
Council (PPARC), MR from the University of Edinburgh Development Trust,
HTJ from the Sonderforschungsbereich 375
``Astro-Teilchenphysik'' und the Sonderforschungsbereich-Transregio 7
``Gravitationswellenastronomie'' of the Deutsche Forschungsgemeinschaft.
Parts of the simulations were performed at the UK Astrophysical Fluids
Facility (UKAFF) and the Edinburgh Parallel Computing Centre (EPCC) of
the University of Edinburgh.
\end{acknowledgements}

{}

\end{document}